%% file: a.tex
\documentclass{CSML}
\pdfoutput=1

\usepackage{lastpage}
\newcommand{\dOi}{13(4:19)2017}
\lmcsheading{}{1--\pageref{LastPage}}{}{}%
{Aug.~12, 2013}{Nov.~27, 2017}{}

\usepackage{tikz}
\usepackage{array}
\usepackage{color}
\usepackage{ifthen}
\usepackage[english]{babel}
\usepackage{amssymb}
\usepackage{amsmath}
\usepackage{amsthm}
\usepackage[all]{xy}
\CompileMatrices
\usepackage{subfiles}
\usepackage{todonotes}
\usepackage{mathtools}

\input{common.tex}

\input{version.tex}
\usepackage[final]{hyperref}
\hypersetup{hidelinks}

\begin{document}
\title{Statman's Hierarchy Theorem}

\author[Westerbaan]{Bram Westerbaan}
\address{Radboud University Nijmegen}
\email{bram@westerbaan.name}
\email{bas@westerbaan.name}
\email{r.kuyper@math.ru.nl}
\email{carst@cs.ru.nl}
\email{remyviehoff@student.ru.nl}
\email{henk@cs.ru.nl
\vspace{-6mm}
}

\author[Westerbaan]{Bas Westerbaan}
\author[Kuyper]{Rutger Kuyper}
\author[Tankink]{Carst Tankink}
\author[Viehoff]{Remy Viehoff}
\author[Barendregt]{Henk Barendregt}


\begin{abstract}
In the Simply Typed $\lambda$-calculus \cite{Hindley,BDS} Statman
investigates the reducibility relation $\leqbe$ between types: for
$A,B \in \mathbb{T}^0$, types freely generated using $\ra$ and a
single ground type~$0$, define $A \leqbe B$ if there exists a
$\lambda$-definable injection from the closed terms of type $A$ into
those of type $B$. Unexpectedly, the induced partial order is the
\emph{(linear) well-ordering (of order type) $\omega + 4$}, see
\cite{stat1980,stat1980a,stat1981,BDS}.

In the proof a finer relation~$\leqh$ is used, where the above
injection is required to be a B\"ohm transformation (\cite{B}), and an
(a posteriori) 
coarser relation~$\leqhp$, requiring a finite family of B\"ohm
transformations that is jointly injective.

We present this result in a self-contained, syntactic, constructive
and simplified manner. \emph{En route} similar results for~$\leqh$
(order type~$\omega + 5$) and~$\leqhp$ (order type~$8$) are
obtained. 
Five of the 
equivalence classes of~$\leqhp$
correspond to canonical term models of Statman, one to the trivial
term model collapsing all elements of the same type, and one does not
even form a model by the lack of closed terms of many types,
\cite{BDS}.
\end{abstract}

\maketitle

{ \subfile{overview.tex} }

{ \subfile{prelim.tex} }
{ \subfile{part1.tex} }      
{ \subfile{intermezzo.tex} } 
{ \subfile{part23.tex} }      
{ \subfile{part4.tex} }      
{ \subfile{conclusions.tex} }

\bibliography{main}{}
\bibliographystyle{amsalpha}

\end{document}

%% file: common.tex
\newcommand{\iref}[1]{(\ref{#1})}

\theoremstyle{definition}

\newtheorem{cnv}[thm]{Convention}

\newcommand{\T}[1]{\ensuremath{\mathbb{T}_{#1}}}
\newcommand{\HH}[1]{\ensuremath{\mathbb{H}_{#1}}}
\newcommand{\BB}[1]{\ensuremath{\mathbb{B}_{#1}}}

\newcommand{\kernedBe}{{\beta\kern-.5pt\eta}}
\newcommand{\kernedHp}{{h^{\kern-.8pt+}}}

\newcommand{\eqbe}{\ensuremath{=_\kernedBe}}
\newcommand{\eqb}{\ensuremath{=_\beta}}
\newcommand{\obseq}[1]{\ensuremath{\approx^{\mathrm{Ob}}_{#1}}}
\newcommand{\exeq}[1]{\ensuremath{\approx^{\mathrm{Ex}}_{\smash{#1}}}}
\newcommand{\neqbe}{\ensuremath{\neq_\kernedBe}}
\newcommand{\sleqbe}{\,\leqbe} 
\newcommand{\snleqbe}{\,\nleqbe}
\newcommand{\leqbe}{\ensuremath{\leq_\kernedBe}}

\newcommand{\leqs}{\ensuremath{\leq^{s}}}
\newcommand{\leqa}{\ensuremath{\leq^{a}}}
\newcommand{\nleqa}{\ensuremath{\nleq^{a}}}

\newcommand{\leqh}{\ensuremath{\leq_{h}}}
\newcommand{\nleqh}{\ensuremath{\nleq_{h}}}
\newcommand{\nleqbe}{\ensuremath{\nleq_\kernedBe}}
\newcommand{\leqsh}{\ensuremath{\leq_{h}^{s}}}
\newcommand{\leqhp}{\ensuremath{\leq_\kernedHp}}
\newcommand{\nleqhp}{\ensuremath{\nleq_\kernedHp}}
\newcommand{\nleqhpbeh}{\ensuremath{\nleq_{h,\kernedBe,\kernedHp}}}
\newcommand{\nleqbeh}{\ensuremath{\nleq_{h,\kernedBe}}}
\newcommand{\isbe}{\ensuremath{\sim_\kernedBe}}
\newcommand{\ish}{\ensuremath{\sim_{h}}}
\newcommand{\iss}{\ensuremath{\sim^s}}

\newcommand{\ishp}{\ensuremath{\sim_\kernedHp}}

\newcommand{\ra}{\rightarrow}
\newcommand{\radots}{\rightarrow \cdots \rightarrow}

\newcommand{\Msep}[1][]{\ensuremath{/_{\kern-2pt#1}}}

\newcommand{\rk}{\ensuremath{\operatorname{rk}}}

\newcommand{\htam}[1]{\text{#1}}

\newcommand{\lm}[1]{\lambda#1.\;}

\newcommand{\ots}[1]{\Lambda(#1)}
\newcommand{\ts}[2][\ectx]{\Lambda^{#1}(#2)}
\newcommand{\ectx}{\ensuremath{\varepsilon}}
\newcommand{\ct}[2]{{^{#1}\!#2}}

\newcommand{\svec}[1]{\smash{\vec{#1}}}
\newcommand{\subtup}[3][\!]{\smash{\vec{#2}^{\,#1}_{#3}}}

\newcommand{\ctx}[1]{{{}^*\kern-2.2pt{#1}}}

\DeclareMathAlphabet{\mathpzc}{T1}{pzc}{m}{it}

\newcommand{\keyword}[1]{\textbf{#1}}

\newcommand{\myvcenter}[1]{\ensuremath{\vcenter{\hbox{#1}}}}

\newcommand\lbb{\mathchoice{[\kern-1.42pt[}%
                    {[\kern-1.42pt[}%
                    {[\kern-1.2pt[}%
                    {[\kern-1.15pt[}}
\newcommand\rbb{\mathchoice{]\kern-1.42pt]}%
                    {]\kern-1.42pt]}%
                    {]\kern-1.2pt]}%
                    {]\kern-1.15pt]}}

\newcommand{\bnf}{\ensuremath{\mathrel{::=}}}

\newcommand{\subs}{\ensuremath{{:=}}}

\newcommand{\scare}[1]{`#1'}

\newcommand{\rsub}[2]{\phantom{#2}\mathllap{#1}}

\newcommand{\gray}[1]{\textcolor{gray}{#1}}

\newcommand{\grayparop}{\mathrlap{\!\!\gray{(}}}
\newcommand{\grayparcl}{\mathrlap{\gray{)}}}

\newcommand\weg[1]{{}}
\newcommand\eqdf{\,\smash\triangleq\,}

\newcommand\set[1]{\{#1\}}
\newcommand\FV{{\rm FV}}


%% file: overview.tex
\section{Hierarchy of types}
We work in simply typed lambda calculus over a single base type 0.
The set of open terms of (simple) type~$A$ is written~$\ots{A}$,
while the set of closed terms of type~$A$
is denoted by~$\ts{A}$ (for reasons which become clear
in Section~\ref{S:prelim}).

For types $A,B$ one defines $A\leqbe B$ if there is a closed term
$\Phi\colon A\to B$ that is an injection on closed terms modulo
$\beta\eta$-equality.

To get some feeling for the  relation $\leqbe$ we begin by observing
\begingroup
\allowdisplaybreaks
\begin{alignat*}{4}
B\ra \grayparop A\ra C \grayparcl
&\sleqbe A\ra \grayparop B\ra C \grayparcl
  &\quad& \htam{via $\lm{mab}{mba}$ (see Corollary~\ref{L:sr-glt});}
\\
A\ra C&\sleqbe A\ra B\ra C&& \htam{via $\lm{mab}{ma}$ (see Lemma~\ref{L:ar-glp});}
\\
A &\sleqbe (A\ra 0)\ra 0&& \htam{via $\lm{mf}{fm}$ (see Lemma~\ref{L:sr-c});}\\
[0,0]\,\eqdf\,0\ra 0\ra 0&\snleqbe  0\ra0&& \htam{by counting closed inhabitants.}
\intertext{Less intuitively clear is that for all simple types $A$ over~$0$ one has}
A&\sleqbe [0,0]\ra 0\ra 0 &&\htam{(see Lemma \ref{top})}.
\intertext{Also, one might ponder (writing $1\eqdf0\ra 0$) whether}
1\ra 1 \ra 0\ra  0&\sleqbe 1 \ra 0\ra 0
&&\htam{(no);}
\\
1\ra1\ra 1 \ra 0\ra  0&\sleqbe 1\ra1 \ra 0\ra 0\quad
&&\htam{(yes!);}
\\
[0,0]\ra 0\ra 0&\sleqbe 1 \ra 1 \ra 0 \ra 0
&&\htam{(no).}
\end{alignat*}
\endgroup

The general problem whether $A\leqbe B$
for given types~$A$ and~$B$ is
solved by the Hierarchy Theorem (stated on page~\pageref{T:Hierarchy},
due to Richard Statman~\cite{stat1980,stat1980a}),
which describes among other things the equivalence classes of~$\leqbe$
in terms of (relatively) simple syntactic properties.

We give a new proof,
 which is self-contained, syntactic and constructive.
We assume only basic knowledge of the simply typed lambda calculus
(long normal form, rank, \dots),
and recall the most important notions before using them.
Roughly speaking, the proof is one long syntactic analysis
of inhabitants of simple types
and reductions between them; 
we make no use of
term models and the like.
The proof is constructive in the sense that we do not use the law of 
the excluded middle,
and so one may easily ignore this feature of the proof
(except perhaps when reading Theorem~\ref{T:inhabitation}).

Applications of the Hierarchy Theorem
include 
the polynomial decidability of the Decidable Unification
    Problem,\cite{stat1981}
the 1-Section Theorem\cite{statman85}
(which itself has many uses, see~\cite{onesection}),
and the construction of the five canonical term models
(see \S3.5 of~\cite{BDS}).
For more details on the Hierarchy Theorem,
its applications, and another proof, we refer the reader 
to Section~3.4 of~\cite{BDS}.

\subsection{Hierarchy Theorem}
\label{SS:hierarchy}
To formulate the theorem
we first recall a few notions and some notation
from the simply typed lambda calculus, 
see Section~1.1 and Section~3.4 of~\cite{BDS}.
\begin{defi}
\label{D:e}\hfill
\begin{enumerate}[label=(\roman*),ref=\roman*]
\item 
Let~$A$ be a type.  The \keyword{components} of~$A$
are the unique types $A_1,\dotsc,A_n$
such that 
\begin{equation*}
A\,=\,A_1\ra \dotsb \ra A_n\ra 0.
\end{equation*}
\item
\label{D:e-rank}
Each type~$A$ has a \keyword{rank} denoted by~$\rk(A)$;
it is defined recursively by
\begin{equation*}
\rk (0)  \,\eqdf\, 0\htam;\qquad 
\rk(A\ra B)\,\eqdf\, \max\{\,\rk (A) +1,\ \rk (B)\,\}\htam.
\end{equation*}
\item
\label{D:e-fat}
A type~$A \equiv A_1 \ra \dotsb \ra A_n \ra 0$ is \keyword{fat}
when $n\geq 2$.
\item
\label{D:e-large}
A type~$A \equiv A_1 \ra \dotsb \ra A_n \ra 0$ is \keyword{large}
if either~$A$ has a fat component~$A_i$,
or one of~$A$'s components $A_i \equiv A_{i1}\ra \dotsb \ra A_{im}\ra 0$
has a large component~$A_{ij}$.
\item
\label{D:e-small}
A type which is not large,
is called~\keyword{small}.
\item Let $A$, $B$ be types and $k$, $n$ natural numbers.
The following notation is used.
\begin{equation*}
n+1 \,\eqdf\, n\ra 0\htam; 
\qquad\qquad
\biggl[\ 
\begin{aligned}
A^0\ra B \,&\eqdf\, B \\
A^{k+1} \ra B\,&\eqdf\, A\ra A^{k}\ra B\htam.
\end{aligned}
\end{equation*}
\end{enumerate}
\end{defi}
\begin{defi}[Reducibility relations]
\label{defn:reductions}
\hfill

\noindent Let $A\equiv A_1 \radots A_n \ra 0$ 
and $B \equiv B_1 \radots B_m \ra 0$ be types.
\begin{enumerate}[label=(\roman*),ref=\roman*]
\item
\label{defn:reductions-be}
\keyword{$A$ $\beta\eta$-reduces to $B$},
notation $A \leqbe B$,
if for some $R\in\ts{A\to B}$
\begin{equation*}
\rsub{R M_1\eqbe R M_2}{R^{(i)} M_1 \eqbe R^{(i)} M_2 }
\implies M_1\eqbe M_2
\qquad ( M_1, M_2 \in \ts{A}\,)\htam{.}
\end{equation*} 
This $R$  is then called a \keyword{reducing term}
from $A$ to $B$.  
\item \label{defn:reductions-h}
\keyword{$A$ head reduces  to $B$}, notation $A \leqh B$, if
$A\leqbe B$ with a reducing term of the form
$R\,\equiv\,\lm{m^A b_1^{B_1} \cdots b_m^{B_m}} m \varrho_{a_1}
\cdots \varrho_{a_n}$
where
$\varrho_{a_i}{:}A_i$ are open terms 
with free variables from ${b_1^{B_1}, \ldots,
b_m^{B_m}}$.
We call a term of this form a \emph{B\"ohm term}.
\item \label{defn:reductions-hp}
\keyword{$A$ reduces multi-head to $B$}, notation $A \leqhp B$,
provided  there exist
B\"ohm terms $R^{(1)},\ldots,R^{(\ell)}$
which
are \emph{jointly injective}, that is,
for $M_1,M_2 \in \ts{A}$,
\begin{equation*}
\forall i \,[ \ \  R^{(i)} M_1 \eqbe R^{(i)} M_2\ \ ]
 \quad  \implies\quad   M_1 \eqbe M_2.
\end{equation*}
\end{enumerate}
\end{defi}

\begin{thm}[Statman Hierarchy]
\label{T:Hierarchy}
The relations $\leqhp$, $\leqbe$ and $\leqh$ are increasingly fine.\footnote{%
\emph{Viz.} 
$A\leqhp B \implies A \leqbe B$
and $A \leqbe B\implies A\leqh B$
for all types $A$ and $B$.}
Their equivalence classes 
are listed below vertically
in descending order.
The types~$H_\alpha$ in the last column
(called \keyword{canonical types})
are representatives for the equivalence classes
of~$\leqh$.
\input {figures/hasse-h-be-hp.tex}
\vspace{.8em}
\noindent Moreover,
the equivalence classes~$\HH{\alpha}$ of~$\leqh$ have the following
syntactic description,
and the relations $\leqhp$, $\leqbe$ and $\leqh$ are hence decidable.
\begingroup
  \allowdisplaybreaks
\begin{align*}
\HH{\omega+4} & \,=\, \{\, A  \colon  \htam{$A$ is inhabited and large} \,\}; \\
\HH{\omega+3} & \,=\, \{\, A  \colon  \htam{$A$ is inhabited, small and $\rk(A)>3$} \,\}; \\
\HH{\omega+2} & \,=\, \{\, A ;
\begin{aligned}[t]
& \htam{$A$ is inhabited, small, $\rk(A)\in\{2,3\}$} \\
& \htam{and $A$ has at least two components of rank $\geq 1$}\, \};
\end{aligned} \\
\HH{\omega+1} & \,=\, \{\, A \colon
\begin{aligned}[t]
& \htam{$A$ is inhabited, small, $\rk(A)=3$} \\
& \htam{and $A$ has exactly one component of rank $\geq 1$} \,\};
\end{aligned} \\
\HH{\omega} & \,=\, \{\, A \colon
\begin{aligned}[t]
& \htam{$A$ is inhabited, small, $\rk(A)=2$} \\
& \htam{and $A$ has exactly one component of rank $1$} \,\};
\end{aligned} \\
\HH{k} & \,=\, \{ \, A \colon
\begin{aligned}[t]
& \htam{$A$ is inhabited, small, $\rk(A)=1$} \\
& \htam{and $A$ has exactly $k$ components of rank $0$} \,\};
\end{aligned} \\
\HH0 & \,=\, \{\, A  \colon  \htam{$A$ is not inhabited} \,\}.
\end{align*}
\endgroup
\end{thm}


We give an overview of the proof of Theorem~\ref{T:Hierarchy}
in Subsection~\ref{SS-sketch-of-the-proof}.
To be able to do this,
we first expose the precise relation
between the syntactic structure of a type and
the shape of its (long normal form) inhabitants 
in Subsection~\ref{S:structure}, and we  examine the inhabitants of
the canonical types $H_0$, $H_1$, \dots 
in Subsection~\ref{S:inhabitants}.

While technical details
are unavoidable in a paper like this,
we make them
more palatable 
by introducing some syntactic sugar in Section~\ref{S:prelim}.
With it we can already prove the inequalities between 
the canonical types (such as $[0,0] \nleqh [0]$)
in Section~\ref{S:part1}.
We proceed by developing a general theory about reductions 
in Section~\ref{S:intermezzo}
to establish the order type of~$\leqh$ in Section~\ref{S:part23},
and the order types of~$\leqbe$ and~$\leqhp$ in Section~\ref{S:part4}.

Since we are in the fortunate position to have strong normalization,
every term has a \emph{long normal form} (lnf), which is the
$\beta\eta^{-1}$-normal form. As default we will only consider terms
in lnf. The few exceptions will not pose a problem to the reader.

\subsection{Syntactic structure and inhabitants}
\label{S:structure}
Recall that for any type~$A$,
there are unique types $A_1,\dotsc,A_n$
such that $A\,=\,A_1\ra\dotsc\ra A_n\ra 0$.
Hence it is natural to write 
\begin{defi}\quad
$[A_1,\dotsc,A_n]\ \eqdf\ A_1 \ra \dotsb \ra A_n \ra0\htam{.}$
\end{defi}
\begin{obs}
\label{O:lnf-type}
Any (lnf-)inhabitant $M$ of a type~$A$
is of the form
\begin{equation*}
\label{eq:lnf-example}
\lm{a_1 \dotsb a_n}{b^B\, N_1 \dotsb N_m}\htam{.}
\end{equation*}
Writing $A=[A_1,\dotsc,A_n]$ and $B=[B_1,\dotsc,B_m]$ we have that
\begin{enumerate}[label=(\roman*),ref=\roman*]
\item
the types of the variables  $a_1,\dotsc,a_n$ 
are respectively $A_1,\dotsc,A_n$ and
\item
the types of the terms $N_1,\dotsc, N_m$
are respectively $B_1,\dotsc,B_m$.
\end{enumerate}
\end{obs}
\begin{obs}
We can write every type~$A$
using only the operation $[\ \ ]$
in a unique way.
For example, $0= [\,]$ and $n+1 = [n]$.
In this way we can consider
types to be finite trees.
For instance, 
the canonical
types are represented by the following trees.
\begin{equation*}
\input{figures/tree_Cs.tex}
\end{equation*}
From this we see that 
given a type~$A$,
the nodes on odd height of the tree~$A$
are the types of the variables which might occur in
closed terms of type~$A$,
while the possible types of the subterms are those on even height.
(E.g.,
in a closed term of type~$[3,0]$ ---
such as 
$\lm{\Phi^3 c^0} \Phi \lm{f_1^1} f_1 \Phi \lm{f_2^1} f_1 f_2\,c$ --- 
the introduced variables are of type~$3$ and~$1$,
while the subterms are of type $2$ and~$0$.)
\end{obs}
\begin{obs}
\label{O:treesyn}
The syntactic properties mentioned in Theorem~\ref{T:Hierarchy}
are more easily defined and understood 
when considering a type to be a tree.
\begin{enumerate}[label=(\roman*),ref=\roman*]
\item 
The \emph{rank} of a type~$A$ is its height as a tree.
If the rank of~$A$ is restricted from above to, say~$2$,
the variables occurring in a closed term~$M$ of type~$A$
are of rank~$0$ or~$1$
so that all the variables in~$M$ are introduced at the head,
contrary to types like $[3,0]$.

\item
\label{O:treesyn-large}
A type~$A$ is \emph{fat} if it has more than one component.
Fat types are important,
because a variable of fat type can be used to construct a pairing
(see Lemma~\ref{L:large-pairing}).

Moreover, $A$ is \emph{large} if $A$ as tree has a fat type on odd height.
One can show that 
if $A$ is inhabited,
then $A$ is large if and only if
there is closed inhabitant~$M$ of~$A$
which contains a (bound) variable of fat type.

In particular,
if a type is \emph{small} (= not large),
then its inhabitants are ``strings'' of a variable
followed by the mandatory abstractions
\begin{equation*}
\lm{a_1^{A_1} \dotsc a_n^{A_n}}
 \underline{G\,\lm{b_1^{B_1} \dotsc b_m^{B_m}}}
 H\,\lm{c_1^{C_1} \dotsc c_k^{C_k}}  \quad \dotsb\quad\htam{.}
\end{equation*}
\end{enumerate}
\end{obs}

\subsection{\texorpdfstring%
        {Inhabitants of the canonical types $H_\alpha$}
        {Inhabitants of the canonical types H(alpha)}}
\label{S:inhabitants}
By the preceding observations we can determine the (lnf-)inhabitants
of a given type.  As an example (and also since we will need them), we
list by an iconic shorthand (explained by 
$\rightsquigarrow$) 
the inhabitants of the canonical types
below.  The verification is left to the reader.
(The graphical ``inhabitation machines''
from Examples~1.3.8 of~\cite{BDS} 
might prove illuminating here,
see also~\cite{schubert}.)

To make terms
more readable,
we leave out parentheses.
There is only one way to place parentheses
to get a term obeying the typing rules.
E.g.,
we read
\begin{alignat*}{2}
\lm{f^1 g^1 c^0} fggf c
\qquad&\htam{as}\qquad
\lm{f^1g^1 c^0} f(g(g(fc)))\htam;\\
\lm{b^{[0,0]} c^0} bcbcc
\qquad&\htam{as}\qquad
\lm{b^{[0,0]} c^0} (bc)((bc)c)\htam.
\end{alignat*}

We abbreviate $[A,A,A,C,C]$ to~$[A^3,C^2]$,
etcetera.%
\subsubsection{$H_{k}=[0^k]$}
\label{SSS:inh-0k}
The inhabitants (of $[0^k]$) are the projections on $k$ elements:
\begin{equation*}
U^k_i \ \rightsquigarrow\ 
\lm{x_1^0 \cdots x_k^0} x_i \qquad \htam{for }0<i\leq k
\end{equation*}%
\vspace{-1.5em}
%
\subsubsection{$H_{\omega}=[1,0]$}
\label{SSS:inh-10}
The inhabitants are the Church-numerals,
\begin{equation*}
c_n \ \rightsquigarrow\  \lm{f^1c^0}{f^{(n)} c}\htam{,}
\end{equation*}	
where 
$f^{(0)}c=c$ and $f^{(n+1)}c=f(f^{(n)}c)$.
As a warm-up for what is coming,
note that the inhabitants of $[1,0]$ 
are produced by the following two-level grammar.
\begin{equation*}
\lm{f^1c^0} \mathsf{N}
\quad\htam{ where }\quad
\mathsf{N}\bnf (f \mathsf{N}) \ | \ c
\end{equation*}%
\vspace{-1.5em}
\subsubsection{$H_{\omega+1}=[2]$}
\label{SSS:inh-2}
An inhabitant can be identified by a pair of natural numbers
\begin{equation*}
\left< i, j \right> \ \rightsquigarrow\ 
  \lm{F^2}F\lm{x_1^0}\cdots F\,\lm{x_i^0} x_j
\qquad\htam{ where } j\leq i\htam{.}
\end{equation*}
These terms are produced by the following grammar.
\begin{equation*}
\lm{F^2} \mathsf{P}_0
\quad\htam{where}\quad
\mathsf{P}_{n} 
  \bnf (F\lm{x^0_{n+1}} \mathsf{P}_{n+1}) \ |\ x_1\ |\  \dotsb\ |\  x_{n}
\end{equation*}%
\vspace{-1.5em}
%
%
\subsubsection{$H_{\omega+2}=[1,1,0]$}
\label{SSS:inh-110}
The inhabitants are essentially `words over a two element alphabet',
\begin{equation*}
\lm{f^1g^1c^0} \mathsf{W} \quad\htam{where}\quad \mathsf{W}
	\bnf (f \mathsf{W}) \ |\  (g \mathsf{W}) \ |\  c\htam{.}
\end{equation*}
Hence we use words over $\{f,g\}$ as shorthands.
For instance,
\begin{equation*}
ffggf \ \rightsquigarrow\  \lm{f^1g^1c^0}ffggfc \htam.
\end{equation*}%
\vspace{-1.5em}
%
%
\subsubsection{$H_{\omega+3}=[3,0]$}
\label{SSS:inh-30}
The inhabitants are produced by the following grammar.
\begin{equation*}
\lm{\Phi^3c^0}\mathsf{M}_1 \quad\htam{where}\quad
\begin{aligned}[t]
	\mathsf{M}_n \bnf \
		& (\Phi\,\lm{f^1_{n}}\mathsf{W}_{n})\ \ |\  c \\
        \mathsf{W}_n \bnf \ 
                & (f_1 \mathsf{W}_n) \ |\ \dotsb\ |\ (f_n \mathsf{W}_n)
			\ \  |\ \ \mathsf{M}_{n+1}
\end{aligned}
\end{equation*}
By replacing ``$\Phi(\lm{f^1_i}\cdots)$''
with ``$\Msep\,$'',
hiding ``$\lm{\Phi^3c^0}$'' and hiding the~``$c$'' at the end,
we obtain a shorthand for the inhabitants of~$[3,0]$.
For instance,
\begin{equation*}
\Msep 1 \Msep\Msep 23\  \rightsquigarrow\ 
		\lm{\Phi^3c^0}\, \Phi\lm{f_1^1} f_1 \Phi\lm{f_2^1} 
			\Phi\lm{f_3} f_2f_3c \htam.
\end{equation*}
So we identify an inhabitant
of~$[3,0]$
with a list of words $w_1,\ldots,w_n$ with
$w_i\in\{1,\dotsc,i\}^*$.%
%
\subsubsection{$H_{\omega+4}=[[0,0],0]$}
\label{SSS:inh-000}
The inhabitants are (like) binary trees:
\begin{equation*}
\lm{b^{[0,0]}c^0} \mathsf{T} \quad \htam{where} \quad
\mathsf{T} \bnf (b\mathsf{T}\mathsf{T})\  |\  c\htam.
\end{equation*}
We will denote them as such.  For instance,
\begin{equation*}
\myvcenter{\input{figures/tree_bbccc.tex}} \ \rightsquigarrow\ 
	\lm{b^{[0,0]}c^0}bbccc \qquad \htam{and} \qquad
\myvcenter{\input{figures/tree_bcbcc.tex}}\  \rightsquigarrow\ 
	\lm{b^{[0,0]}c^0} bcbcc \htam.
\end{equation*}

\subsection{Structure of the proof}
\label{SS-sketch-of-the-proof}
In this Subsection, we present the proof of the Hierarchy Theorem.
We delegate most of the work to the remainder of
this article
by using statements proved later on.
What is left is the compact skeleton of the proof.

\proof [Proof of Theorem~\ref{T:Hierarchy}]
We need to prove the following.
\begin{enumerate}[label=(\Roman*),ref=\Roman*]
\item\label{i:proof-I}
The relations~$\leqh$, $\leqbe$ and $\leqhp$
are as displayed on page~\pageref{T:Hierarchy}.

\item\label{i:proof-II}
The relations~$\leqh$, $\leqbe$ and~$\leqhp$ are decidable.
\end{enumerate}
Concerning~\iref{i:proof-I}.
We first consider the relation~$\leqh$.
Let the sets~$\HH\alpha$
be defined as on page~\pageref{T:Hierarchy}.
One easily verifies that the $\HH\alpha$
form a partition of $\mathbb{T}^0$,
and that $H_\alpha \in \HH\alpha$ for all~$\alpha$.

To show that~$\leqh$ is of the form
as on page~\pageref{T:Hierarchy} it suffices to show that
for all $A,B\in\mathbb{T}^0$ 
and $\alpha,\beta\in\omega+5$
with $A\in\HH\alpha$ and $B\in\HH\beta$,
we have that
\begin{equation}
\label{eq:H-to-prove}
A\leqh B 
\quad\iff\quad H_\alpha \leqh H_\beta
\quad \iff \quad \alpha\leq\beta
\end{equation}
For this we use the following four facts
proved later on.
\begin{enumerate}[label=(\roman*),ref=\roman*]
\item
\label{program-1}
$A\in \HH\alpha\implies H_\alpha\leqh A$
(see \textbf{Subsection~\ref{S:part2}}).
\item
\label{program-2}
$A\in \HH\alpha\implies A\leqh H_\alpha$
(see \textbf{Subsection~\ref{S:part3}}).
\item
\label{program-3}
$\alpha\leq\beta \implies H_\alpha \leqh H_\beta$
(see \textbf{Subsection~\ref{SS:HaleqHb}}).
\item 
\label{program-4}
$\alpha\nleq\beta\implies H_\alpha\nleqh H_\beta$
(see \textbf{Section~\ref{S:part1}}).  
\end{enumerate}
Before we prove Statement~\eqref{eq:H-to-prove}
let us spend some words on fact~\iref{program-4}.
In Section~\ref{S:part1}
we do not directly prove that
 $\alpha\nleq \beta\implies H_\alpha \nleqh H_\beta$.
Instead we show
the inequalities listed below in Statement~\eqref{eq:ineq} 
(writing $A\nleqbeh\!B$ for $A\nleqbe\!B$ \&
$A\nleqh B$, etcetera).
Together with fact~\iref{program-3},
this is sufficient to establish fact~\iref{program-4}.
Indeed suppose that $\alpha\nleq \beta$
and $H_\alpha \leqh H_\beta$
for some~$\alpha$ and~$\beta$
in order to obtain a contradiction.
Then $\beta<\alpha$, so $\beta+1\leq \alpha$.
Thus
\begin{equation*}
H_{\beta+1} \ \leqh\  H_{\alpha}\ \leqh\  H_{\beta}
\end{equation*}
by fact~\iref{program-3}.
This contradicts the inequality $H_{\beta+1}\nleqh H_{\beta}$
from Statement~\eqref{eq:ineq}.

It is interesting to note that we will prove the inequalities 
from Statement~\eqref{eq:ineq} of the form $H_\beta \nleqhpbeh\!\!H_\alpha$
(except  one)
by showing that there are distinct terms $N_1, N_2\in\ts{H_\beta}$
such that $R\,N_1 \eqbe R\,N_2$ for all  $R\in\ts{A\ra B}$
(see Lemma~\ref{lem1}).
These terms $N_1,N_2$ are listed on the right in Statement~\eqref{eq:ineq}
using the notation from Subsection~\ref{S:inhabitants}.
\vspace{-.5em}
\begin{equation}
\label{eq:ineq}
\begin{alignedat}{5}
[[0,0],0]  \,& \nleqhpbeh\  &\,& [3,0] &\qquad& 
N_1=\myvcenter{\input{figures/tree_ce1.tex}} \quad
&N_2&=\myvcenter{\input{figures/tree_ce2.tex}}\\
 [3,0]\, & \nleqhpbeh && [1,1,0] &\qquad&
	N_1=\Msep1\Msep21&N_2&=\Msep1\Msep22 \\
 [1,1,0]\,& \nleqhpbeh && [2] && N_1=fgfg &N_2&=fggf \\
 [2]\,& \nleqh  && [1,0] &&  \\
 [1,0]\,& \nleqhpbeh && [0^{k+2}] && N_1=c_1&N_2&=c_2 \\
 [0^{k+2}]\,& \nleqbeh && [0^{k+1}] && \\
 [0^{k+1}]\,& \nleqhpbeh && 0 && 
\end{alignedat}
\end{equation}

Let us prove Statement~\eqref{eq:H-to-prove}.
The first equivalence
follows  from facts~\iref{program-1} and~\iref{program-2},
the second equivalence follows directly
from facts~\iref{program-3} and~\iref{program-4}.

We now turn to the 
order type of the reducibility relation~$\leqbe$.
To show that~$\leqbe$ is of the
form as depicted on page~\pageref{T:Hierarchy},
we need to prove that
\begin{equation}
\label{eq:B-to-prove-minus-one}
A \leqbe B
\qquad\iff\qquad 
\alpha\leq\beta\quad\htam{ or }
\quad
\alpha,\beta\in\{\omega,\,\omega+1\}
\end{equation}
for all~$\alpha,\beta\in\omega+5$
and all~$A\in\HH{\alpha}$ and~$B\in\HH{\beta}$.
Note that
\begin{alignat}{7}
\label{eq:h-implies-be}
A&\leqh B&\quad&\implies&\quad A &\leqbe B&
\qquad\quad&\htam{for all types }A\htam{ and }B.
\end{alignat}
Hence $A\isbe H_\alpha$ for all~$A\in\HH{\alpha}$,
since $A\ish H_\alpha$ for $A\in\HH{\alpha}$ 
by facts~\iref{program-1} \& \iref{program-2}.
So to prove Statement~\eqref{eq:B-to-prove-minus-one},
it suffices to show that 
\begin{equation}
\label{eq:B-to-prove}
H_\alpha \leqbe H_\beta 
\qquad\iff\qquad 
\alpha\leq\beta\quad\htam{ or }
\quad
\alpha,\beta\in\{\omega,\,\omega+1\}
\htam.
\end{equation}
The implication ``$\Longleftarrow$''
follows from Statements~\eqref{eq:H-to-prove},
Statement~\eqref{eq:h-implies-be}
and 
\begin{equation*}
H_{\omega+1}\ \leqbe\  H_{\omega}
\qquad \htam{(see \textbf{Subsection~\ref{SS:part4-be}}).}
\end{equation*}
Concerning  ``$\Longrightarrow$''.
Let $\alpha,\beta\in\omega+5$ be given. Suppose that $H_\alpha\leqbe H_\beta$.
Then since $\leq$ and~$=$ on~$\omega+5$ are decidable,
it suffices to show that the negation of the right-hand side of
Statement~\eqref{eq:B-to-prove} leads to a contradiction.
Suppose that $\beta<\alpha$
and not $\alpha,\beta\in \{\omega,\,\omega+1\}$.
Then $\beta\leq\gamma<\gamma+1\leq\alpha$
for some $\gamma \in \omega+5$ with~$\gamma\neq\omega$.
(Pick $\gamma = \beta$ if $\beta\neq\omega$,
or pick $\gamma=\omega+1$ otherwise.)
Then 
$H_{\gamma+1}\nleqbe H_\gamma$
by Statement~\eqref{eq:ineq},
but we also have that 
$H_{\gamma+1}\leqbe H_\alpha
\leqbe H_\beta
\leqbe H_\gamma$,
a contradiction.
We have proven Statement~\eqref{eq:B-to-prove-minus-one}.

We continue with the order type of~$\leqhp$.
We need to prove that
\begin{equation}
\label{eq:T-to-prove}
\begin{split}
A \leqhp B
\qquad\iff\qquad
\alpha\leq\beta\quad&\htam{ or }\quad
\alpha,\beta\in\{\omega,\,\omega+1\}\ \\
\quad&\htam{ or }\quad 
\alpha,\beta\in\{2,\dotsc,\omega\}
\end{split}
\end{equation}
for all~$\alpha,\beta\in\omega+5$
and all $A\in\HH{\alpha}$ and $B\in\HH{\beta}$.
Again, we have
\begin{alignat}{7}
\label{eq:h-implies-hp}
A&\leqh B&\quad&\implies&\quad A &\leqhp B&
\qquad\quad&\htam{for all types }A\htam{ and }B,
\end{alignat}
and $A\ishp H_\alpha$ for all~$A\in\HH{\alpha}$.
So it suffices to show that
\begin{equation}
\label{eq:T-to-prove-2}
\begin{split}
H_\alpha \leqhp H_\beta 
\qquad\iff\qquad
\alpha\leq\beta\quad&\htam{ or }\quad
\alpha,\beta\in\{\omega,\,\omega+1\}\ \\
\quad&\htam{ or }\quad 
\alpha,\beta\in\{2,\dotsc,\omega\}.
\end{split}
\end{equation}
The implication ``$\Longleftarrow$''
follows from Statement~\eqref{eq:H-to-prove}, 
and Statement~\eqref{eq:h-implies-hp} and
\begin{equation*}
H_{\omega+1} \leqhp H_\omega
\qquad\text{and}\qquad
H_{k+1} \leqhp H_k\quad(k\geq 2)
\qquad
\text{(see \textbf{Subsection~\ref{SS:part4-hp}})}.
\end{equation*}
The implication ``$\Longrightarrow$''
can be proven
using the inequalities of Stat.~\eqref{eq:ineq}
in a similar fashion
as the implication ``$\Longrightarrow$''
of Stat.~\eqref{eq:B-to-prove}
was proven above.
We leave this to the reader.

\noindent
Concerning~\iref{i:proof-II}.
To show that the reducibility relations $\leqh$, $\leqbe$ and $\leqhp$
are decidable,
we prove that for every type~$A$ an $\alpha\in\omega+5$
can be computed with $A\in\HH\alpha$.
(This is sufficient because
if $A\in\HH\alpha$
and
$B\in\HH\beta$
then $A\leqh B$ can be decided using Statement~\eqref{eq:H-to-prove};
$A\leqbe B$ using Statement~\eqref{eq:B-to-prove-minus-one};
$A\leqhp B$ using Statement~\eqref{eq:T-to-prove}.)
Of course,
algorithms to determine the rank of a type,
the number of components and
whether the type is large or small
are defined easily enough;
the difficulty here is how to decide whether
a given type is inhabited.

By Proposition~2.4.4 of~\cite{BDS}
(which is proven using the law of the excluded middle)
a type $A\equiv [A_1,\dots,A_n]$ is inhabited
iff $A_i$ is uninhabited for some~$i$.
From this fact
a recursive algorithm to determine whether a type~$A$ is inhabited
is easily concocted.
Since we want constructive proof of the Hierarchy Theorem,
we have formulated and proven a constructive variant of Proposition~2.4.4
of~\cite{BDS},
see Theorem~\ref{T:inhabitation} below.
Note that the algorithm to determine inhabitation
in the constructive case is the same as in the classical case;
only the proof that the algorithm is correct is different.

This concludes the proof of the Hierarchy Theorem.\qed

\begin{thm}
\label{T:inhabitation}
Let $A=[A_1,\dotsc,A_n]$ be a type.
Then either $A$ is inhabited or not, and
\begin{equation}
\label{eq:inh}
\htam{$A$ is uninhabited}\quad\iff\quad\htam{ all $A_i$ are inhabited.}
\end{equation}
\end{thm}
\proof 
Concerning ``$\Longleftarrow$''.
Suppose towards a contradiction
that all~$A_i$ are inhabited,
and~$A$ is inhabited too.
Pick~$M\in\ts{A}$ and~$N_i\in\ts{A_i}$ for each~$i$.
Then $N\,M_1\dotsb M_n$ is a closed inhabitant of~$0$,
which is impossible.

We prove ``$\Longrightarrow$''
and ``either~$A$ is inhabited or not''
by induction on the buildup of types as `tuples' using the operation $[\ \ ]$.
Let $A=[A_1,\dotsc,A_n]$ with~$A_i=[A_{i1},\dotsc,A_{im_i}]$ be given.
For all~$i\in\{1,\dotsc,n\}$,
suppose the following.
\begin{enumerate}[label=(\roman*),ref=\roman*]
\item \label{inhI}
Either~$A_i$ is inhabited or~$A_i$ is uninhabited.
\item \label{inhII}
If $A_i$ is uninhabited then~$A_{ij}$ is inhabited for all~$j$.
\end{enumerate}
We need to prove that all~$A_i$ are
inhabited provided that~$A$ is uninhabited,
and that either~$A$ is inhabited or $A$ is uninhabited.

Assume that~$A$ is uninhabited in order to show that
all~$A_i$ are inhabited.
By~\iref{inhI},
either all~$A_i$ are inhabited
or some~$A_i$ is uninhabited.
In the former case we are done;
so let us prove the latter case leads to a contradiction.
Assume~$A_i$ is uninhabited for some~$i$.
By~\iref{inhII}
$A_{ij}$ is inhabited for all~$j$.
Pick $N_j\in\ts{A_{ij}}$ for all~$j$.
Then $\lm{a^{A_1}_1 \dotsb a^{A_n}_n} a_i N_1 \dotsb N_{m_i}$
is a closed inhabitant of~$A$,
contradicting that~$A$ is uninhabited.
Therefore,
all~$A_i$ are inhabited.

Consequently,
$A$ is inhabited iff not all~$A_i$ are inhabited.
Since (by~\iref{inhI}) either all~$A_i$ are inhabited or not,
it follows~$A$ is either inhabited or not.\qed

%% file: figures/hasse-h-be-hp.tex
{\newcommand{\PS}{\mathstrut} 
\[ \begin{xy}
(24,0)*[.]!(1.3,0){\PS\HH{\omega+4}} ="h0"+
(0,-6)*[.]!(1.3,0){\PS\HH{\omega+3}} ="h1"+
(0,-6)*[.]!(1.3,0){\PS\HH{\omega+2}} ="h2"+
(0,-6)*[.]!(1.3,0){\PS\HH{\omega+1}} ="h3"+
(0,-6)*[.]!(1.3,0){\PS\HH{\omega}} ="h4"+
(0,-6)*{\vdots} ="hv1"+
(0,-6)*[.]!(1.3,0){\PS\HH{k}} ="h5"+
(0,-6)*{\vdots} ="hv2"+
(0,-6)*[.]!(1.3,0){\PS\HH{2}} ="h6"+
(0,-6)*[.]!(1.3,0){\PS\HH{1}} ="h7"+
(0,-6)*[.]!(1.3,0){\PS\HH{0}} ="h8"+
(0,-6)*[.]!(2.3,0){\PS\leqh} ="h9",
%
(12,0)*[.]!(1.3,0){\PS\BB{\omega+3}} ="b0"+
(0,-6)*[.]!(1.3,0){\PS\BB{\omega+2}} ="b1"+
 (0,-6)*[.]!(1.3,0){\PS\BB{\omega+1}} ="b2"+
 (0,-9)*[.]!(1.3,0){\PS\BB{\omega}} ="b3"+
(0,-9)*{\vdots} ="bv1"+
 (0,-6)*[.]!(1.3,0){\PS\BB{k}} ="b5"+
(0,-6)*{\vdots} ="bv2"+
 (0,-6)*[.]!(1.3,0){\PS\BB{2}} ="b6"+
 (0,-6)*[.]!(1.3,0){\PS\BB{1}} ="b7"+
 (0,-6)*[.]!(1.3,0){\PS\BB{0}} ="b8"+
 (0,-6)*[.]!(2.3,0){\PS\leqbe} ="b9",
%
  (0,0)*[.]!(1.3,0){\PS\T{5}} ="t0"+
 (0,-6)*[.]!(1.3,0){\PS\T{4}} ="t1"+
 (0,-6)*[.]!(1.3,0){\PS\T{3}} ="t2"+
 (0,-9)*[.]!(1.3,0){\PS\T{2}} ="t3"+
(0,-15)*[.]!(1.3,0){\PS\T{1}} ="t5"+
(0,-18)*[.]!(1.3,0){\PS\T{0}} ="t7"+
 (0,-6)*[.]!(1.3,0){\PS\T{-1}} ="t8"+
 (0,-6)*[.]!(2.3,0){\PS\leqhp} ="t9",
%
(40, 0)*[.]{\PS H_{\omega+4} \,\eqdf\ (0 \ra 0 \ra 0) \ra 0 \ra 0 } +
(0,-6)*[.]{\PS H_{\omega+3} \,\eqdf\ 3 \ra 0 \ra 0} +
(0,-6)*[.]{\PS H_{\omega+2} \,\eqdf\ 1 \ra 1 \ra 0 \ra 0} +
(0,-6)*[.]{\PS H_{\omega+1} \,\eqdf\ 2 \ra 0} +
(0,-6)*[.]{\PS H_{\omega\phantom{+1}}\,\eqdf\ 1 \ra 0 \ra 0} +
(0,-12)*[.]{\PS H_{k\phantom{+1}}\,\eqdf\ 0^k \ra 0} +
(0,-12)*[.]{\PS H_{2\phantom{+1}}\,\eqdf\ 0^2 \ra 0} +
(0,-6)*[.]{\PS H_{1\phantom{+1}} \,\eqdf\ 0 \ra 0} +
(0,-6)*[.]{\PS H_{0\phantom{+1}} \,\eqdf\ 0} ,
(20,-21)*+=(0,10)\frm{\{},
(8,-36)*+=(0,22)!(0,3)\frm{\{},
%
(60,0)*{}
%
\ar @{-} (24,-1.6); (24,-3.8)   
\ar @{-} (24,-7.6); (24,-9.8)  
\ar @{-} (24,-13.6);(24,-15.8) 
\ar @{-} (24,-19.6);(24,-21.8) 
\ar @{-} (24,-33.6);(24,-33.8) 
\ar @{-} (24,-37.6);(24,-39.8)
\ar @{-} (24,-45.6);(24,-45.8) 
\ar @{-} (24,-49.6);(24,-51.8)
\ar @{-} (24,-55.6);(24,-57.8)

\ar @{-} (12,-1.6); (12,-3.8)
\ar @{-} (12,-7.6); (12,-9.8)
\ar @{-} (12,-13.6);(12,-18.8)
\ar @{-} (12,-33.6);(12,-33.8) 
\ar @{-} (12,-37.6);(12,-39.8)
\ar @{-} (12,-45.6);(12,-45.8) 
\ar @{-} (12,-49.6);(12,-51.8)
\ar @{-} (12,-55.6);(12,-57.8)

\ar @{-} (0,-1.6); (0,-3.8)
\ar @{-} (0,-7.6); (0,-9.8)
\ar @{-} (0,-13.6);(0,-18.8)
\ar @{-} (0,-22.6);(0,-33.8) 
\ar @{-} (0,-37.6);(0,-51.8)
\ar @{-} (0,-55.6);(0,-57.8)

\end{xy} \]}%

%% file: figures/tree_Cs.tex
\begin{tikzpicture}[
    level distance=0.7cm, 
    sibling distance=.7cm, 
    grow=up,
    nodes={draw=none, fill=white}]

\draw[help lines] (-.25,0.0) -- (11.25,0.0);
\draw[help lines,densely dashed] (-.25,0.7) -- (11.25,0.7);
\draw[help lines] (-.25,1.4) -- (11.25,1.4);
\draw[help lines,densely dashed] (-.25,2.1) -- (11.25,2.1);
\draw[help lines] (-.25,2.8) -- (11.25,2.8);
\draw (11.25,0.0) node[right]{\footnotesize$\mathsf{0}$};
\draw (11.25,0.7) node[right]{\footnotesize$\mathsf{1}$};
\draw (11.25,1.4) node[right]{\footnotesize$\mathsf{2}$};
\draw (11.25,2.1) node[right]{\footnotesize$\mathsf{3}$};
\draw (11.25,2.8) node[right]{\footnotesize$\mathsf{4}$};

\begin{scope}[xshift=0cm]
\node{$0$};
\end{scope}

\begin{scope}[xshift=.75cm]
\node{$[0]$}
child { node{$0$} };
\end{scope}

\begin{scope}[xshift=1.75cm]
\node{$[0,0]$}
child { node{$0$} }
child { node{$0$} };
\end{scope}

\begin{scope}[xshift=2.75cm,yshift=.7cm]
\node{$\cdots$};
\end{scope}
\begin{scope}[xshift=2.75cm,yshift=0cm]
\node{$\cdots$};
\end{scope}

\begin{scope}[xshift=4cm]
\node{$[1,0]$}
child { node{$0$} }
child { node{$1$} 
  child { node{$0$} }
};
\end{scope}

\begin{scope}[xshift=5.25cm]
\node{$[2]$}
child { node{$2$} 
  child { node{$1$} 
    child { node{$0$} }
  }
};
\end{scope}

\begin{scope}[xshift=6.75cm]
\node{$[1,1,0]$}
child { node{$0$} }
child { node{$1$} 
  child { node{$0$} }
}
child { node{$1$} 
  child { node{$0$} }
};
\end{scope}

\begin{scope}[xshift=8.5cm]
\node{$[3,0]$}
child { node{$0$} }
child { node{$3$} 
  child { node{$2$} 
    child { node{$1$}
      child { node{$0$} }
    }
  }
};
\end{scope}

\begin{scope}[xshift=10.25cm]
\node{$[[0,0],0]$}
child { node{$0$} }
child { node{$[0,0]$} 
  child { node{$0$} }
  child { node{$0$} }
};
\end{scope}
\end{tikzpicture}

%% file: figures/tree_bbccc.tex
\begin{tikzpicture}[level distance=0.3cm, sibling distance=0.9cm]
\coordinate 
child {
	child
	child } 
child;
\end{tikzpicture}

%% file: figures/tree_bcbcc.tex
\begin{tikzpicture}[level distance=0.3cm, sibling distance=0.9cm]
\coordinate 
child
child {
	child
	child };
\end{tikzpicture}

%% file: figures/tree_ce1.tex
\begin{tikzpicture}[level distance=0.3cm, sibling distance=0.9cm]
\coordinate
child  {
	child[sibling distance=0.6cm]
	child[sibling distance=0.6cm]{
		child
		child } }
child {
	child[sibling distance=0.6cm]
	child[sibling distance=0.6cm] };
\end{tikzpicture}

%% file: figures/tree_ce2.tex
\begin{tikzpicture}[level distance=0.3cm, sibling distance=0.9cm]
\coordinate 
child {
	child[sibling distance=0.6cm]
	child[sibling distance=0.6cm] }
child {
	child[sibling distance=0.6cm] {
		child
		child }
	child[sibling distance=0.6cm] };
\end{tikzpicture}

%% file: prelim.tex
\section{Reductions and contexts}
In this section
we introduce some syntactic sugar
that
will save ink later on.
\label{S:prelim}
%
%
\begin{defi}
\label{D:ctx}\hfill
\begin{enumerate}[label=(\roman*),ref=\roman*]
\item A \keyword{context} is a sequence of distinct typed variables,
$c_1^{C_1},\dotsc,c_k^{C_k}$.

The letters $\Gamma$, $\Delta$, $\Theta$, and~$\Xi$ denote contexts.  The
empty context is denoted by~$\ectx$; concatenation of contexts
is written as $\Gamma,\Delta$. 
\item \label{D:ctx-ts}
For a context $\Gamma = {c_1^{C_1}, \cdots, c_k^{C_k}}$
write 
\begin{align*}
\lm\Gamma N\ &\eqdf\ \lm{c_1^{C_1} \ldots c_k^{C_k}} N\\
\{\Gamma\} & \eqdf\ \{c_1^{C_1},\dotsc,c_k^{C_k}\} \\
\ts[\Gamma]A\ &\eqdf\ \set{\,M\in\ots{A}\,\mid\,
\FV(M)\subseteq\{\Gamma\}\,}\\
[\Gamma]\ &\eqdf\ [C_1,\dotsc,C_k].
\end{align*}
\item 
\label{D:ctx-fits}
Let $\Gamma=c_1^{C_1},\dotsc,c_k^{C_k}$ be a context.
We say $\svec{P}$ \keyword{fits} in $\Gamma$ if
$\svec{P}=P_1,\dotsc,P_k$ is a tuple of (open) terms,
and~$P_i:C_i$ for every~$i$.
In that case we write 
\begin{equation*}
M[\Gamma{{\subs}}\svec{P}]\ \eqdf\  M[c_1{{\subs}}P_1]\ldots[c_k{{\subs}}P_k]\htam.
\end{equation*}
\end{enumerate}\end{defi}
\begin{rem}
Recall that we have assumed that all
terms are in long normal form.\\
In particular,
if $M\in\ts[\Gamma]{[\Delta]}$, then~$M$ is of the form
 $M\equiv\lambda\Delta.N$ where
$\FV(N)\subseteq\set{\Gamma,\Delta}$.
\end{rem}
Using contexts one can formulate statements such as
$N \in \ts[\Gamma]0\implies \lm{\Gamma}N \in \ts{[\Gamma]}$,
and $(\lm\Gamma N) \svec{P} \eqb N[\Gamma{\subs}\svec{P}]$
for any term~$N$ and~$\svec{P}$ which fits in~$\Gamma$.
Also contexts lighten the study of reductions
as will be shown in the following.

We study the relation $A_1\leqh A_2$ for types~$A_1,A_2$
(see Definition~\ref{defn:reductions}\iref{defn:reductions-h}).
Note that $A_1 \leqh A_2$
if and only if
 there is a \emph{B\"ohm transformation}~$\Phi\colon\ts{A_1}\ra\ts{A_2}$,
which is injective (on lnf-terms).
That is, 
$\Phi$ should be of the form
$\Phi (M) \eqbe R M$
where~$R$ is some B\"ohm term
(see Definition~\ref{defn:reductions}\iref{defn:reductions-h}).
More explicitly,
writing $A_1 \equiv [\Delta_1]$ and~$A_2 \equiv [\Delta_2]$,
the map $\Phi$ should be of 
the form $\Phi(M)=\lm{\Delta_2}M \svec{P}$
where $\svec{P}$ fits in~$\Delta_1$
and~$\FV(\vec{P})\subseteq\{\Delta_2\}$.

Let~$\Phi$ be such a B\"ohm transformation, then it transforms
\begin{equation*}
\lm{\Delta_1} N\quad\text{to}\quad
\lm{\Delta_2} N[\Delta_1{\subs}\svec{P}]\htam.
\end{equation*}
To see if~$\Phi$ is injective,
we only need to focus on the transformation mapping
\begin{equation*}
N\quad\htam{to}\quad N[\Delta_1{\subs}\svec{P}]\htam.
\end{equation*}
A map $\ts[\Delta_1]0\ra\ts[\Delta_2]0$ of this form is also called a
B\"ohm transformation.

In order to construct these B\"ohm transformations
it pays off to consider the  more general B\"ohm transformations
from~$\ts[\Gamma_1]{[\Delta_1]}\ra\ts[\Gamma_2]{[\Delta_2]}$
which map
\begin{equation*}
\lm{\Delta_1} N
\quad\htam{to}\quad
\lm{\Delta_2} N\,[\Delta_1{\subs}\svec{P}]\,
[\Gamma_1 {\subs}\svec{Q}]\htam,
\end{equation*}
where $\svec{P}$, $\svec{Q}$
fit in~$\Delta_1$, $\Gamma_1$, respectively,
having free variables from~$\Gamma_2,\Delta_2$.
Note that the core of these transformations
is the \emph{substitution} of~$\svec{P},\svec{Q}$
for~$\Delta_1,\Gamma_1$.

These considerations lead to the next set of definitions.
%
\begin{defi}
\label{D:ct}\hfill
\begin{enumerate}[label=(\roman*),ref=\roman*]
\item 
A pair $\ct\Gamma{A}$ is called a \keyword{context--type}
and has as intended meaning the set $\ts[\Gamma]{A}$
of terms $M:A$ with $\FV(M)\subseteq\{\Gamma\}$
(see Definition~\ref{D:ctx}\iref{D:ctx-ts}).
\item\label{D:ct-ar}
Define for such $\ct\Gamma{A}$ the type
\begin{equation*}
\Gamma\ra A\ \ \eqdf\ \  C_1\to\ldots\to C_k\to A
\ \ =\ \ [\Gamma,\Delta],
\end{equation*}
if $\Gamma\equiv c_1^{C_1},\ldots,c_k^{C_k}$ and  $A\equiv [\Delta]$.
\item\label{D:ct-red} We say $\ct{\Gamma_1}{A_1}$ \keyword{reduces} to
$\ct{\Gamma_2}{A_2}$ and write $\ct{\Gamma_1}{A_1}\leq \ct{\Gamma_2}{A_2}$
provided that
\begin{equation*}
\Gamma_1\ra {A_1}  \,\leqh\, \Gamma_2\ra {A_2}\htam.
\end{equation*}
\item
We write $\ct{\Gamma_1}{A_1}\sim\ct{\Gamma_2}{A_2}$
provided that $\ct{\Gamma_1}{A_1}\leq \ct{\Gamma_2}{A_2}$
and $\ct{\Gamma_1}{A_1} \geq \ct{\Gamma_2}{A_2}$.
\item\label{D:ct-sub-ctx}
Let $\Gamma$,~$\Delta$ be contexts.
A \keyword{substitution}
from~$\Gamma$ to~$\Delta$
is a map~$\varrho$ from~$\{\Gamma\}$ to terms such that
$\varrho_{c} \eqdf \varrho(c) \in \ts[\Delta]C$
for all~$c^C\in\{\Gamma\}$.
\item\label{D:ct-sub}
Let~$\ct{\Gamma_1}{[\Delta_1]}$ and $\ct{\Gamma_2}{[\Delta_2]}$
be given.
A \keyword{substitution}~$\varrho$ from~$\ct{\Gamma_1}{[\Delta_1]}$ 
to~$\ct{\Gamma_2}{[\Delta_2]}$
is a substitution from~$\Gamma_1,\Delta_1$ to~$\Gamma_2,\Delta_2$.
\item\label{D:ct-subtup}
Let~$\varrho$ be as in \iref{D:ct-sub-ctx}.
For every context~$\Theta\equiv d_1^{D_1},\dotsc,d_k^{D_\ell}$
with $\{\Theta\}\subseteq\{\Gamma\}$,
define $\subtup\varrho\Theta\eqdf \varrho_{d_1},\dotsc,\varrho_{d_\ell}$.
(Then $\subtup\varrho\Theta$ fits in~$\Theta$,
see Definition~\ref{D:ctx}\iref{D:ctx-fits}.)
\item
\label{D:ct-bohm}
Let~$\varrho$ be as in \iref{D:ct-sub}.
Writing $A_1=[\Delta_1]$ \& $A_2 = [\Delta_2]$,
define~$\hat\varrho\colon\ts[\Gamma_1]{A_1}\ra \ts[\Gamma_2]{A_2}$
by
\begin{alignat*}{4}
\hat\varrho(\,\lm{\Delta_1} N\,)
\ &\eqdf\ 
&&\lm{\Delta_2} N\,[\Gamma_1 {\subs} \subtup\varrho{\Gamma_1}]
         \,[\Delta_1{\subs}\subtup\varrho{\Delta_1}]\\
 &\!\eqb \ 
&&\lm{\Delta_2}(\lm{\Delta_1}N)[\Gamma_1{\subs}\subtup\varrho{\Gamma_1}]
         \,\subtup\varrho{\Delta_1}\htam.
\end{alignat*}
Such a map $\hat\varrho$ is called a \keyword{(B\"ohm-)transformation}
from~$\ct{\Gamma_1}{A_1}$ to~$\ct{\Gamma_2}{A_2}$.
\end{enumerate}
\end{defi}

\begin{prop}
\label{P:redsub}
Let $\ct{\Gamma_1}{A_1}$ and $\ct{\Gamma_2}{A_2}$
be context--types.
Then
\begin{equation*}
\ct{\Gamma_1}{A_1}\leq\ct{\Gamma_2}{A_2}
\quad\iff\quad
\left[\ \begin{minipage}{0.60\columnwidth}
There is a substitution $\varrho$ 
from $\ct{\Gamma_1}{A_1}$ to $\ct{\Gamma_2}{A_2}$
such that the transformation
$\hat\varrho\colon\ts[{\Gamma_1}]{A_1}\ra\ts[\Gamma_2]{A_2}$
is injective.
\end{minipage}
\right.
\end{equation*}
\end{prop}
\proof 
Just unfold the definitions.\qed

\noindent Hence, if 
$\ct{\Gamma_1}{A_1}$ reduces to~$\ct{\Gamma_2}{A_2}$, then there is
an injective B\"ohm transformation $\Phi=\hat\varrho$ 
from $\ts[\Gamma_1]{A_1}$
to~$\ts[\Gamma_2]{A_2}$. 
We will focus on $\varrho$
instead of~$R$
as the following convention shows.
(The benefit of this becomes clear later, see Remark~\ref{rem:reduction-def}.)
\begin{cnv}
\label{V:red}
A \keyword{reduction} from $\ct{\Gamma_1}{A_1}$
to~$\ct{\Gamma_2}{A_2}$
is a \emph{substitution}~$\varrho$
from~$\ct{\Gamma_1}{A_1}$ to~$\ct{\Gamma_2}{A_2}$
such that the B\"ohm transformation $\hat\varrho$ is injective.
\end{cnv}

Since $A\leqh B\iff \ct\ectx{A}\leq \ct\ectx{B}$
for all types~$A$ and~$B$,
it is natural to regard the types part of the
context--types by identifying~$A$ with~$\ct\ectx{A}$.
As such,
any notion defined for context--types
can be applied to types as well.

For notational brevity,
we also identify~$\Gamma$ and~$\ct\Gamma0$
for any context~$\Gamma$.
In this way 
we also regard the contexts as part of the context--types.
As such, any notion defined for context--types
is applicable to contexts.
In particular,
we obtain a notion of reduction between contexts;
$\Gamma\leq \Delta
\iff
\ct\Gamma0\leq\ct\Delta0$.

Note that 
with these identifications
we have 
$\Gamma,\Delta \sim \ct\Gamma{[\Delta]} \sim [\Gamma,\Delta]$
for all~$\Gamma$, $\Delta$.

%% file: part1.tex
\section{Inequalities between canonical types}\label{S:part1}
In this section we will prove the
 inequalities listed in Statement~\eqref{eq:ineq} on page~\pageref{eq:ineq}.
This is one of the bits left out of the proof of the Hierarchy Theorem
in Subsection~\ref{SS-sketch-of-the-proof}.

We start with two of the simpler inequalities.


\subsection{\texorpdfstring%
	{Ad $H_{k+1} \nleqbe H_{k}$}
	{Ad H(k+1) does not reduce to H(k)}}
As $H_k$ has exactly~$k$ inhabitants,
there is no injection from~$\ts{H_{k+1}}$ to~$\ts{H_k}$,
and hence no $\beta\eta$-reduction
from $H_{k+1}$ to~$H_k$
(see Definition~\ref{defn:reductions}\iref{defn:reductions-be}).

\subsection{\texorpdfstring%
	{Ad $H_{\omega+1} \nleqh H_\omega$}
	{Ad H(omega+1) does not reduce to H(omega)}}
\label{SS:wp1nrw}
We need to prove that $[2]\nleqh [1,0]$.
Or in other words,
we must prove that $[2]\nleq f^1,c^0$ 
(see Definition~\ref{D:ct}\iref{D:ct-red}).
Let~$\varrho$ be a substitution from~$[2]$ to~$f^1,c^0$;
we will prove
that the B\"ohm transformation~$\hat\varrho$ is not injective.
(Hence $[2]\nleq f^1,c^0$ by Proposition~\ref{P:redsub}.)
To this end we simply calculate $\hat\varrho (M)$
for~$M\in \ts{[2]}$.

Recall that an inhabitant of~$[2]$ is of the following form
(see~\ref{SSS:inh-2}).
\begin{equation*}
\left<i,j\right> \ \eqdf\ \lm{F^2} F\lm{x^0_1} \dotsb F\lm{x^0_i} x_j
\end{equation*}
So we have 
\begin{equation}
\label{eq:w1-1}
\hat\varrho (\,\left< i,j \right>\,) 
\ \equiv\  \left<i,j\right> \varrho_F 
\ \eqb\  \varrho_F \lm{x^0_1} \dotsb \varrho_F \lm{x^0_i} x_j\htam.
\end{equation}
Further, note that since $\varrho_F$ 
is an element of~$\ts[f^1,c^0]{2}$ it must be of the following form.
\begin{equation*}
\varrho_F\ \equiv\ \lm{g^1} f^{(k_0)}g f^{(k_1)} \cdots g  f^{(k_n)} c\htam.
\end{equation*}
We first exclude a pathological case. If $n=0$,
then $\hat\varrho(\,\left< i, j\right>\,) \eqb f^{(k_0)} c$.
So $\hat\varrho$ is constant and hence not injective.
So let us assume that $n>0$.

To reduce Equation~\eqref{eq:w1-1},
note the following.
\begin{enumerate}[label=(\roman*),ref=\roman*]
\item
Let $G\eqdf \lm{y^0} f^{(m)} x$; we calculate $\varrho_F\,G$.
To start, $G M \eqb f^{(m)} x$ for all terms~$M$.
So if $M\eqdf f^{(k_1)} G f^{(k_2)} \dotsb G f^{(k_n)} c$,
then 
\begin{equation*}
\varrho_F G
\ \equiv\ 
\varrho_F\, \lm{y^0} f^{(m)} x
\ \eqb\ f^{(k_0)} G M
\ \eqb\ f^{(k_0)}f^{(m)} x
\end{equation*}
\item
Let $G\eqdf \lm{y^0} f^{(m)} y$. We have
\begin{align*}
\varrho_F\lm{y^0} f^{(m)} y 
\ &\eqb\ f^{(k_0)} G f^{(k_1)} \dotsb G f^{(k_n)} c \\
\ &\eqb\ f^{(k_0 + m + k_1 + \dotsb + m + k_n)} c
\ \eqb\ f^{(m\cdot n + \sum_i k_i)}c\htam.
\end{align*}
\end{enumerate}
So if we apply (i) and (ii) 
to Equation~\eqref{eq:w1-1},
in this order, (i), (ii), (i),
we obtain
\begin{align*}
\varrho_F (\, \left< i,j \right>\,)
\ &\eqb\ \varrho_F \lm{x_1^0} \dotsb \varrho_F \lm{x_j^0}
       \  f^{((i-j)k_0)}\  x_j\\
\ &\eqb\ \varrho_F \lm{x_1^0} \dotsb \varrho_F \lm{x_{i-1}^0} 
       \ f^{(n(i-j)k_0+\sum_i k_i)} \ c\\
\ &\eqb\ f^{((i-1)k_0 + n(i-j)k_0 + \sum_i k_i)} \ c\htam.
\end{align*}
Consequently, $\left<3,1\right>$
and $\left<n+3,n+2\right>$ are both sent to 
$\smash{f^{(2(n+1)k_0 + \sum_i k_i)} }c$ by~$\hat\varrho$.
\subsection{Indiscernibility}
\label{SS:indiscernibility}
The remaining inequalities are of the form $H_\alpha \nleqhpbeh H_\beta$.
In this subsection we develop some general theory to prove them.
In fact, we prove a stronger statement:
there are  terms $M_1\neq M_2$ in $\HH\alpha$ (listed 
in Statement~\eqref{eq:ineq}
on page~\pageref{eq:ineq})
such that
\begin{equation}
\label{eq:obseq}
R M_1 \eqbe R M_2
\qquad\htam{ for all }\qquad
R\colon H_\alpha \ra H_\beta\htam.
\end{equation}
That is, $M_1$ and~$M_2$ are 
indiscernible for any term~$R\colon H_\alpha \ra H_\beta$.
(This is called \emph{observational equivalence} in the literature\cite{BDS}.)
As Proposition~\ref{P:part1-main} shows,
instead of proving that $M_1$ and~$M_2$
are indiscernible for every~$R\colon H_\alpha \ra H_\beta$,
it suffices to prove
that for certain variants~$H_\beta'$ of~$H_\beta$,
the terms 
$M_1$ and~$M_2$ are indiscernible for any B\"ohm transformation
from $H_\alpha$ to~$H_\beta'$. (This is called \emph{existential equivalence}.)
This general method of proving that $H_\alpha \nleqhpbeh H_\beta$
has been extracted from the proof in~\cite{Dekkers-1988}
of $[3,0]\nleqh [1,1,0]$.
%
%
\begin{defi}
\label{D:eq}
Let~$\ct\Gamma{A}$ and $\Delta$ be given.
 For $M_1,M_2\in\ts[\Gamma]{A}$
define
\begin{alignat*}{4}
M_1 &\obseq\Delta M_2& &\quad\iff\quad&
\forall R\,[&\ R(\lm\Gamma M_1) \eqbe R(\lm\Gamma M_2)\ ]\htam,\\
M_1 &\exeq\Delta M_2& &\quad\iff\quad&
\forall \varrho\,[&\ \hat\varrho M_1 = \hat\varrho M_2\ ]\htam,
\end{alignat*}
where~$R$ ranges over the closed terms~$R\colon (\Gamma \ra A)\ra[\Delta]$
and~$\varrho$ ranges over the substitutions
from~$\ct\Gamma{A}$ to~$\Delta$.
(So $\hat\varrho$ is a B\"ohm transformation.)
\end{defi}
%
%
\begin{lem}\label{lem1}
Let~$A$ and $B\equiv[\Delta]$ be types
and $M_1,M_2 \in \ts{A}$ with $M_1 \neq M_2$.
\begin{enumerate}[label=(\roman*),ref=\roman*]
\item\label{obseqeqbe}%
$M_1 \obseq\Delta M_2$ implies $A\nleqbe B$.
\item\label{exeqeqh}%
$M_1 \exeq\Delta M_2$ implies $A\nleqh B$ and $A\nleqhp B$.
\end{enumerate}
\end{lem}
\proof 
\iref{obseqeqbe}.\ 
Suppose $M_1 \obseq\Delta M_2$ and $A\leqbe B$
towards a contradiction.
Since  $A\leqbe B$,
there a reducing term~$R\in\ts{A\ra B}$
(see Definition~\ref{defn:reductions}\iref{defn:reductions-be}).
Since $M_1 \obseq\Delta M_2$,
we get $RM_1 \eqbe RM_2$ 
(see Definition~\ref{D:eq}).
But 
this implies that $M_1 = M_2$
(as~$R$ is a reducing term),
contradicting $M_1 \neq M_2$.
Hence $A \nleqbe B$.

\iref{exeqeqh}.\ 
Assume that $M_1 \exeq\Delta M_2$.
We will that prove~$A\nleqhp B$,
and hence a fortiori $A\nleqh B$ (see Definition~\ref{defn:reductions}).
Suppose  that $A\leqhp B$ towards a contradiction.
Pick a family of substitutions~$\varrho^1,\dotsc,\varrho^n$
from~$A$ to~$[\Delta]$ such that
\begin{equation}
\label{hpexobseq}
\forall i\,[\ \hat\varrho^i ( M ) = \hat\varrho^i ( N )\ ]
\ \implies\  M = N \qquad(M,N\in \ts{A})\htam.
\end{equation}
Let~$i$ and~$M\in\ts{A}$ be given.
Then we know that  $\hat\varrho^i(M) = \lm\Delta M \subtup[i]\varrho\Gamma$
and $M \subtup[i]\varrho\Gamma \in \ts[\Delta]0$
where $A=[\Gamma]$.
Hence  $M\mapsto M \subtup[i]\varrho\Gamma$
is a B\"ohm transformation from~$A$ to~$\Delta$.
Thus we get $M_1\subtup[i]\varrho\Gamma = M_2\subtup[i]\varrho\Gamma$ 
  by $M_1 \exeq\Delta M_2$
(see Definition~\ref{D:eq}).
Then $\hat\varrho^i (M_1) = \hat\varrho^i (M_2)$.
So Statement~\eqref{hpexobseq}
implies that $M_1 = M_2$,
contradicting~$M_1\neq M_2$.
Hence $A\nleqhp B$.\qed

\noindent To formulate Proposition~\ref{P:part1-main}, 
we need one more notion.
%
%
\begin{obs}
\label{O:der}
An inhabitant~$M$ of a context~$\Gamma$,
i.e. $M\in\ts[\Gamma]0$, is of the form
\begin{equation*}
M\,\equiv\, a^A \,(\lm{\Gamma_1} M_1) \dotsb (\lm{\Gamma_k} M_k)\htam,
\end{equation*}
where $a^A\in\{\Gamma\}$,  $A=[[\Gamma_1],\dotsc,[\Gamma_k]]$
and $M_i$ is an inhabitant of $\Gamma,\Gamma_i$.
\end{obs}
%
%
\begin{defi}
\label{D:der}
Let $\Gamma$ be a context 
and  $a^A \in\{\Gamma\}$ with $A\equiv[[\Gamma_1],\dotsc,[\Gamma_k]]$.
Then for all~$i$ the 
context~$\Gamma,\Gamma_i$ is said to be a \emph{direct derivative}
of~$\Gamma$.
A context $\Gamma'$ is \keyword{a derivative} of~$\Gamma$
if there is a chain of direct derivatives from~$\Gamma$ to~$\Delta$.%
\footnote{Equivalently,
the relation on contexts of being a derivative
is the \emph{transitive--reflexive closure} of
the relation on contexts of being a direct derivative.}
\end{defi}
\begin{exas}
\begin{enumerate}
\item
The only derivative of $x^0,f^1$ is $x^0,f^1$ itself.
In fact,
a context $\Gamma$ has only one derivative (c.q.~itself)
iff  $\rk ([\Gamma]) \leq 2$.

\item
Any derivative of $F^2$ is of the form $F^2,x_1^0,\dotsc,x_n^0$
for some $n$,
and any derivative of $\Omega^3$ is 
of the form $\Omega^3,f_1^1,\dotsc,f^1_n$ for some~$n$.

\item
The context
$\Phi^4,F^2,x^0,G^2$ is a derivative of~$\Phi^4$.
Any derivative of $\Phi^4$ is of the form $\Phi^4,\Delta'$,
where~$\Delta'$ is a context
with  $\{\Delta'\}=\{F_1^2,\dotsc,F_m^2,x_1^0,\dotsc,x_n^0\}$
for some $n,m$ with $n\neq 0\implies m\neq 0$.
\end{enumerate}
\end{exas}

%
%
\begin{prop}
\label{P:part1-main}
Given a type~$A$,
terms 
$M_1,M_2 \in\ts{A}$ and a context~$\Delta$,
\begin{equation*}
\forall \Delta'\  [\ M_1 \exeq{\Delta'} M_2 \ ]
\quad\implies\quad
M_1 \obseq{\Delta} M_2\htam.
\end{equation*}
Here $\Delta'$ ranges over
contexts such that 
$m^{A},\Delta'$ is a derivative of $m^{A},\Delta$.
\end{prop}
\proof 
Assume $M_1 \exeq{\Delta'} M_2$
for every~$\Delta'$.
We will prove that
\begin{equation}
\label{eq:prop1}
N[m{\subs}M_1]=N[m{\subs}M_2]
\qquad\htam{ for each }\Delta'\htam{ and }N\in\ts[m^{A},\Delta']{0}\htam.
\end{equation}
This is sufficient.
Indeed,
suppose that~$R\colon A\ra [\Delta]$
with $R\equiv \lm{m^A} N$.
Then we have that 
$RM_1 \eqb N[m{\subs}M_1] = N[m{\subs}M_2] \eqb RM_2$.
Hence $M_1 \obseq{\Delta} M_2$, as required.

To prove Statement~\eqref{eq:prop1},
we use induction
(over the long normal form of~$N$).
Let $N\in\ts[m^A,\Delta']0$ be given
for some~$\Delta'$ such that
 $m^A,\Delta'$ is a derivative of~$m^A,\Delta$.
We have
\begin{equation*}
N \,\equiv\, c\,(\lm{\Gamma_1} N_1) \cdots(\lm{\Gamma_k} N_k),
\end{equation*}
where $c\in\{m^A,\Delta'\}$ 
and $N_j \in\ts[m^A,\Delta',\Gamma_j]0$ for all~$j$
(see Observation~\ref{O:der}).
To use induction over~$N$,
we need to prove that every~$N_j$ 
falls in the scope of Statement~\eqref{eq:prop1},
i.e.~that $\smash{N_j\in\ts[m^A,\Delta'']{0}}$
for some~$\Delta''$ such that
$m^A,\Delta''$ is a derivative of~$m^A,\Delta$.
That is,
we need to have that
 $m^A,\Delta',\Gamma_j$ is a derivative of~$m^A,\Delta$.
This is indeed the case
because~$m^A,\Delta',\Gamma_j$ is a direct derivative of~$m^A,\Delta'$,
and $m^A,\Delta'$ itself is a derivative of~$m^A,\Delta$
(see Definition~\ref{D:der}).

We need to prove that
$N[m\subs M_1] = N[m\subs M_2]$,
and by induction we may assume 
that  $N_i[m {\subs} M_1] = N_i [m{\subs}M_2]$ for all~$i$.
Note that either $c\in \{\Delta'\}$ or $c=m^A$.

In the former case, $m\neq c$, so
\begin{align*}
N[m{\subs}M_j]
  \,&=\, c\,(\lm{\Gamma_1} N_1[m{\subs}M_j]) \cdots (\lm{\Gamma_k} N_k[m{\subs}M_j])
\htam{,}
\end{align*}
hence $N[m{\subs}M_1]=N[m{\subs}M_2]$, as required.

In the latter case,
we have $c=m^A$ (and thus $A=[[\Gamma_1],\dotsc,[\Gamma_k]]$), so
\begin{align*}
N[m{\subs}M_j]
  \,&=\, M_j (\lm{\Gamma_1} N_1[m{\subs}M_j]) \cdots (\lm{\Gamma_k} N_k[m{\subs}M_j])
  = \hat{\varrho} M_j
\htam{,}
\end{align*}
where~$\varrho$ is the substitution
from~$A=[\smash{a_1^{[\Gamma_1]},\dotsc,a_k^{[\Gamma_k]}}]$
to~$\Delta'$
given by
\begin{equation*}
\varrho_{a_i} \ \eqdf\  \lm{\Gamma_i} N_i [m {\subs} M_j]
\htam{,}
\end{equation*}
but since $M_1 \exeq{\Delta'} M_2$,
we have~$\hat{\varrho}M_1 = \hat{\varrho} M_2$
and~$N[m{\subs}M_1]=N[m{\subs}M_2]$.\qed

\begin{cor}\label{cor1}
Let~$A$ and~$B=[\Delta]$ be types.
Let $M_1,M_2 \in\ts{A}$ with
\begin{equation*}
M_1 \neq M_2\qquad\htam{ and }\qquad
\forall \Delta'\  [\ M_1 \exeq{\Delta'} M_2 \ ]\htam,
\end{equation*}
where $\Delta'$ ranges over
contexts such that 
$m^{A},\Delta'$ is a derivative of $m^{A},\Delta$.
Then
\begin{align*}
A\nleqbe B; &&A\nleqh B;&& A\nleqhp B.
\end{align*}
\end{cor}
\proof 
Combine Proposition~\ref{P:part1-main} and Lemma~\ref{lem1}.\qed


\subsection{\texorpdfstring%
	{Ad $H_{\omega} \nleqhp H_{k+1}$ and $H_{\omega} \nleqbe H_{k+1}$}
	{Ad H(omega) does not reduce to H(k+1)}}
We need to prove that $[1,0]\nleqhp [0^{k+1}]$ and $[1,0]\nleqbe [0^{k+1}]$.
We use Corollary~\ref{cor1}
with $M_i \eqdf c_i$
(see~\ref{SSS:inh-10})
and $\Delta\eqdf x_1^0,\dotsc,x_{k+1}^0$.
Let~$m^{[0,1]},\Delta'$ be a derivative
of~$m^{[0,1]},\Delta$.
We need to prove that $c_1 \exeq{\Delta'} c_2$.
Note that $\Delta'\equiv x^0_1,\ldots,x^0_m$
for some~$m\geq k+1$.
Let~$\varrho$ be a substitution
from~$[f^1,c^0]$ to~$\Delta'$.
In order to show that  $M_1 \exeq{\Delta'} M_2$,
we need to prove that $\hat\varrho c_1 = \hat\varrho c_2$.

The term~$\varrho_{f}\in\smash{\ts[\Delta']{1}}$
is either~$\lm{y^0}y$
or~$\lm{y^0}x_i$ for some $i$.
\begin{enumerate}[label=(\roman*),ref=\roman*]
\item
In the former case,
$\hat{\varrho} c_i
  = \smash{\varrho_f^{(i)}} \varrho_c
  = \varrho_c
\htam{,}$
so $\hat{\varrho} c_1 = \hat{\varrho} c_2$.

\item
In the latter,
$\varrho_f M \eqb x_i$ for each $M$, so
in particular $\hat{\varrho} c_1 = x_i = \hat{\varrho} c_2$.
\end{enumerate}


\subsection{\texorpdfstring%
	{Ad $H_{\omega+2} \nleqhp H_{\omega+1}$ and
        $H_{\omega+2} \nleqbe H_{\omega+1}$}
	{Ad H(omega+2) does not reduce to H(omega+1)}}
\label{SS:ineq-w2}
Again we use Corollary~\ref{cor1}, but
now with $M_1 = fgfg$ and $M_2 = fggf$
(see~\ref{SSS:inh-110}).

Let $\varrho$ be a substitution from $[f^1,g^1,c^0]$ 
to a context~$\Delta'$,
such that~$m^{[1,1,0]},\Delta'$ 
is a derivative of~$m^{[1,1,0]},F^2$;
we need to show that
 $\hat{\varrho} M_1 = \hat{\varrho} M_2$.
Note that $\Delta' = F^2,d_1^0,\dotsc,d_\ell^0$ for some $\ell$.
Let us first study~$\varrho_f\in\smash{\ts[\Delta']1}$;
it is of the form 
\begin{equation*}
\varrho_f \equiv \lm{z^0} F \lm{x_1^0} \cdots F\lm{x_i^0} e \htam{,}
\end{equation*}
for some $i$, where either  $e=z$, $e=d_k$ or $e=x_k$ for some $k$.
So for any term~$M$,
\begin{equation*}
\varrho_f M \ \eqb \ 
\begin{cases}
 F\lm{x_1^0} \cdots F\lm{x_i^0}M & \htam{if }e = z \\
 F\lm{x_1^0} \cdots F\lm{x_i^0}e & \htam{otherwise. }
\end{cases}
\end{equation*}
In the latter case, 
$\hat{\varrho}M_1\equiv\hat{\varrho} (fgfg) 
  \eqbe \varrho_f \hat{\varrho} (gfg)
  \eqb F\lm{x_1^0} \cdots F\lm{x_i^0}e$
and similarly $\hat\varrho M_2 \eqbe F\lm{x_1^0} \cdots F\lm{x_i^0}e$,
so $\hat\varrho M_1 = \hat\varrho M_2$.
So let us instead assume that
\begin{equation*}
\rho_f \,=\, \lm{z^0} F\lm{x_1^0} \cdots F\lm{x_i^0} z.
\end{equation*}
By similar reasoning for~$g$, 
we are left with the case that, for some~$j$,
\begin{equation*}
\varrho_g = \lm{z^0} F\lm{x_1^0} \cdots F\lm{x_j^0} z.
\end{equation*}
Abusing notation, one could
set $h\eqdf \text{``}F\lm{z^0}\text{''}$ and 
write~$\varrho_f = h^{i} z$.
Then
\begin{equation*}
\hat\varrho M_1 \equiv
\hat{\varrho} (fgfg)
   \,\eqb\, h^{i} h^{j} 
      h^{i} h^{j} \varrho_c 
   \,=\, h^{2(i+j)} \varrho_c 
   \,\eqb\, \hat{\varrho} (f g g f ) \equiv \hat\varrho M_2
\htam{.}
\end{equation*}

\subsection{Ad \texorpdfstring{%
$H_{\omega + 3} \nleqhp H_{\omega + 2}$
and
$H_{\omega + 3} \nleqbe H_{\omega + 2}$}{%
H(omega+3) does not reduce to H(omega+2)}}
\label{SS:w3nrw2}%
We use Corollary~\ref{cor1} 
with 
(see~\ref{SSS:inh-30})
\begin{equation*}
M_i\eqdf\Msep1\Msep2i.
\end{equation*}
Let~$m^{[3,0]},\Delta'$
be a derivative of~$m^{[3,0]},f^1,g^1,d^0$,
and
$\varrho$ a substitution
from~$[\Phi^3,c^0]$
to~$\Delta'$.
One easily verifies that
$\{\Delta'\} = \{f,g,d,G^2_1,\dotsc,G^2_\nu,d_1^0,\dotsc,d^0_\mu\}$
for some~$\nu,\mu$.

We need to prove the following equality.
\begin{equation}
\label{eq:w3}
\hat{\varrho} M_1 = \hat{\varrho} M_2
\end{equation}
To this end, we first calculate $\varrho_\Phi M$
for $M\in\ts[\Xi ]2$
where $\Xi \eqdf h_1^1,\dotsc,h_m^1,\Delta'$.
The result is recorded in Lemma~\ref{L:w3}.
We start with two remarks.

First, note that $\varrho_\Phi\in\smash{\ts[\Delta']3}$ 
is of the form
\begin{equation}
\label{eq:PhiFFF}
\varrho_\Phi
  \,\equiv\, \lm{F^2} w_0 F \lm{z^0_1} w_1 \cdots F \lm{z^0_n} w_n e
\htam{,}
\end{equation}
where 
$w_i$ are words on the  alphabet
\begin{equation*}
\mathcal A = \{f,g,G_1\lm{y^0},\ldots,G_\nu\lm{y^0}\}
\end{equation*}
and $e$ is a variable of type~$0$,
so either $e=d$,
$e=z_i$ for some $i\in\{1,\ldots,n\}$,
$e=d_i$ for some $i\in \{1,\ldots,m\}$,
or $e=y$ for some $y$ introduced by a $G_j$ in one of the $w_k$.

Secondly, we know that any~$M\in\ts[\Xi]2$ is of the form
\begin{equation*}
M\equiv \lm{h^1} H_1 h H_2 h \dotsb  H_\ell h R
\end{equation*}
where $H_i \in \ts[\Xi]1$ and $R\in\ts[\Xi]0$.

\begin{lem}
\label{L:w3}
Let $M\equiv \lm{h^1} H_1 h H_2 h \dotsb  H_\ell h R$
from~$\ts[\Xi]2$ be given.
\begin{enumerate}[label=(\roman*),ref=\roman*]
\item
\label{L:w3-1}
If $h$ does not occur in~$M$ (i.e. $M\equiv\lm{h^1} R$),
then $\varrho_\Phi M \eqb w_0 R$.
\item\label{L:w3-2}
If $h$ occurs in~$M$ then
\begin{equation}
\label{eq:L:w3--0}
\varrho_\Phi M
    \eqb 
\begin{cases}
W^H (T P^H_i)&
  \text{if } e = z_i,\\
    W^H e&
    \text{otherwise,}
\end{cases}
\end{equation}
where $H\eqdf H_1$ and $T\eqdf \lm{h^1} H_2 h \dotsb H_\ell h R$ and
\begin{equation*}
W^H \,\eqdf\, w_0 H w_1  \cdots H w_n
\quad\text{ and }\quad
P^H_i\,\eqdf\,\lm{z^0_i} w_i H w_{i+1}  \cdots H w_n z_i\htam.
\end{equation*}%
\end{enumerate}%
\end{lem}%
\proof 
\iref{L:w3-1}.\ 
Assume~$h$ does not occur in~$M$.
Then $MK\eqb R$ for every~$K:1$.
We apply this to Equation~\eqref{eq:PhiFFF}.
Writing $K\eqdf \lm{z_1^0} w_1 \dotsb M\lm{z_n^0} w_n e$,
we have
\begin{equation*}
\varrho_\Phi M \,\eqb\,
w_0 M (\,\lm{z_1^0} w_1 \dotsb M\lm{z_n^0} w_n e\,)
\,\equiv\,w_0 M K
\,\eqb\, w_0 R.
\end{equation*}

\iref{L:w3-2}.\ 
Assume~$h$ occurs in~$M$.  Note that
by definition of~$H$ and~$T$,
\begin{equation}
\label{eq:MHT}
M \eqb \lm{h^1} H h (Th)
\htam.
\end{equation}
In particular,
for any term~$K:1$ in which~$z_j$ does not occur,
we have
\begin{equation}
\label{eq:w3-*}
M (\lm{z_j^0} K) \,\eqb\, H K\htam.
\end{equation}

Either $e=z_i$ for some~$i$ or not.
If $e\neq z_i$, then
\begin{alignat*}{4}
\varrho_\Phi M
  & \eqb \ &&w_0 M\lm{z^0_1} w_1  \dotsb M\lm{z_n^0} w_n e
      \qquad &&\htam{by Eq. \eqref{eq:PhiFFF}} \\
  & \eqb &&w_0 H w_1 \dotsb H  w_n e
      \qquad &&\htam{by Eq. \eqref{eq:w3-*}} \\
  & = && W^He
      \qquad &&\htam{by def. of }W^H.
\end{alignat*}
If $e=z_i$, then 
\begin{alignat*}{4}
\varrho_\Phi M
  & \eqb \ &&w_0 M\lm{z^0_1} w_1  \dotsb M\lm{z_n^0} w_n z_i
      \qquad &&\htam{by Eq. \eqref{eq:PhiFFF}} \\
  & \eqb &&w_0 H w_1  \dotsb M \lm{z^0_i} w_i \dotsb H  w_n z_i
      \qquad &&\htam{by Eq. \eqref{eq:w3-*}} \\
  & = &&w_0 H w_1  \cdots M P_i^H 
      \qquad &&\htam{by def. of }P_i^H \\
  & \eqb &&w_0 H w_1  \cdots HP_i^H (TP_i^H) 
      \qquad &&\htam{by Eq. \eqref{eq:MHT}}\\
  & \eqb &&w_0 H w_1  \cdots Hw_i \dotsb H w_n (TP_i^H) 
      \qquad &&\htam{by def. of }P_i^H\\
  & = && W^H(TP_i^H)
      \qquad &&\htam{by def. of }W^H\htam{.}
\end{alignat*}
We have proven Statement~\eqref{eq:L:w3--0} and so we are done.\qed

\noindent We will use the special case of Lemma~\ref{L:w3} where~$H=\lm{x^0}x$.
\begin{cor}
\label{C:w3}
Define $W\eqdf w_0 w_1 \dotsb w_n$
and $P_i\eqdf \lm{z_i^0} w_i w_{i+1} \dotsb w_n z_i$.
Then for any term $M\in\ts[\Xi]2$ of the
form $M\eqb \lm{h^1} h(Th)$
with $T\in\ts[\Xi]2$
we have 
\begin{equation*}
\varrho_\Phi M
    \eqb 
\begin{cases}
W (T P_i)&
  \htam{if } e = z_i,\\
    W e&
    \htam{otherwise.}
\end{cases}
\end{equation*}
\end{cor}
\proof 
Follows immediately from Lemma~\ref{L:w3}.\qed

\noindent We are now ready
to prove Equation~\eqref{eq:w3}.
\begin{cor}
$\hat{\varrho} M_1 = \hat{\varrho} M_2$.
\end{cor}
\proof 
For brevity, let  $K_j\eqdf \lm{x_1 x_2} x_j$.
We have 
\begin{equation}
\label{eq:MiKi}
M_j = \lm{\Phi^3c^0} \Phi \lm{f_1^1} f_1 \Phi\lm{f_2^1} f_2
(K_j f_1 f_2) c\htam.
\end{equation}
We distinguish two cases:
either $e=z_i$ or not.

Assume $e=z_i$ for some~$i$.
We apply Corollary~\ref{C:w3} twice, 
first to 
$M^{\mathrm{I}}\eqdf\lm{f_2^1} f_2 (K_j f_1 f_2) \varrho_c$
and then to 
$M^{\mathrm{II}}\eqdf \lm{f_1^1} f_1 W (K_j f_1 P_i) \varrho_c$.
Indeed,
\begin{alignat*}{4}
\hat{\varrho} M_j
  & \eqb \ &&\varrho_\Phi \lm{f^1_1} f_1
      \varrho_\Phi \lm{f^1_2} f_2 \, (K_j f_1 f_2)  \varrho_c 
    \qquad&&\htam{by Equation \eqref{eq:MiKi}}\\
  & = &&\varrho_\Phi \lm{f^1_1} f_1
      \varrho_\Phi M^{\mathrm{I}}
    \qquad&&\htam{by def. of }M^{\mathrm{I}} \\
  & \eqb && \varrho_\Phi \lm{f^1_1} f_1 W (K_j f_1 P_i) \varrho_c 
    \qquad&&\htam{by Corollary \ref{C:w3}} \\
  & \eqb && \varrho_\Phi M^{\mathrm{II}}
    \qquad&&\htam{by def. of }M^{\mathrm{II}} \\
  & \eqb && W W (K_j P_i P_i)\varrho_c 
    \qquad&&\htam{by Corollary \ref{C:w3}} \\
  & \eqb && WWP_i\varrho_c
    \qquad&&\htam{by def. of }K_j\htam.
\end{alignat*}

Assume $e\neq z_i$.
By Corollary~\ref{C:w3}
applied to $M\eqdf\lm{f_1^1} f_1 \varrho_\Phi \lm{f_2^1} f_2 f_i \varrho_c$,
\begin{equation*}
\hat{\varrho} M_j
  \eqb \varrho_\Phi 
      (\, \lm{f_1^1} f_1 \varrho_\Phi \lm{f_2^1} f_2 f_i \varrho_c\,)
  \equiv \varrho_\Phi M 
  \eqb We
\htam{.}
\end{equation*}

So in both cases
the value of~$\hat\varrho M_j$
does not depend on~$j$.\qed

%
%
%

\subsection{Ad \texorpdfstring{%
$H_{\omega + 4} \nleqhp H_{\omega + 3}$ and
$H_{\omega + 4} \nleqbe H_{\omega + 3}$}{%
H(omega+4) does not reduce to H(omega+3)}}

We use Corollary~\ref{cor1}
with (see~\ref{SSS:inh-000})
\begin{align*}
M_1 & = \myvcenter{\input{figures/tree_ce1.tex}} &
M_2 & = \myvcenter{\input{figures/tree_ce2.tex}}.
\end{align*}
That is,
$M_1 \eqdf \lm{b^{[0,0]}c^0}{bbcbccbcc}$ and
$M_2 \eqdf \lm{b^{[0,0]}c^0}{bbccbbccc}$.
Let~$m^{[[0,0],0]},\Delta'$ be
a derivative of~$m^{[[0,0],0]},\Phi^3,c^0$,
and let~$\varrho$ be a substitution
from~$[b^{[0,0]},c^0]$ to~$\Delta'$.
Note that
\begin{equation*}
\{\Delta'\}
\,=\,\{\Phi^3,c^0,f_1^1,\ldots,f_\ell^1,d_1^0,\ldots,d_\nu^0\}.
\end{equation*}
We need to prove that
\begin{equation*}
\hat\varrho M_1 = \hat\varrho M_2\htam.
\end{equation*}
Consider~$\varrho_b \in\ts[\Delta']{[0,0]}$.
It is of the form
\begin{equation*}
\varrho_b \,\equiv\,
\lm{x^0y^0} w_0 \,\Phi\lm{g_1^1} w_1 \cdots \Phi\lm{g_\mu^1} w_\mu e,
\end{equation*}
where~$e\in\{ x, y, c, d_1, \ldots, d_\nu \}$ and~$w_i$ is a word
over~$\{ f_1, \ldots, f_\ell, g_1, \ldots, g_i \}$.

We see that either $e\in\{ x,c,d_1,\dotsc,d_\nu\}$
or $e=y$.

In the former case,
we have $e\neq y$.
Then $y$ is not used in~$\varrho_b$,
so
we have
$\varrho_b M N \eqb \varrho_b M N'$ for all terms $M,N,N':0$.
In particular,
\begin{alignat*}{4}
\hat\varrho M_1
	& \equiv && \varrho_b (\varrho_b \varrho_c (\varrho_b
	    \varrho_c \varrho_c)) (\varrho_b \varrho_c \varrho_c) \\
	& \eqb&& \varrho_b (\varrho_b \varrho_c N) N'
	&\quad& \htam{ for any $N$, $N'$} \\
	& \eqb\ && \varrho_b (\varrho_b \varrho_c \varrho_c)
	    (\varrho_b (\varrho_b \varrho_c \varrho_c) \varrho_c) \\
	& \equiv && \hat\varrho M_2 \htam.
\end{alignat*}
Similarly,
if $e=y$, then $e\neq x$,
so $\varrho_b M N \eqb \varrho_b M' N$ for all $M,M',N$,
and
\begin{alignat*}{4}
\hat\varrho M_1
	& \eqb\ && \varrho_b M (\varrho_b M' \varrho_c) \,\eqb\, \hat\varrho M_2
	&\quad& \htam{ for all  $M$, $M'$.}
\end{alignat*}

%% file: intermezzo.tex
\section{Calculus of reductions}
\label{S:intermezzo}
Before we proceed,
we establish some general calculation rules
for  reducibility.
Although one can find some trivial rules for~$\leqh$
such as
$[A_1,A_2] \leqh [A_2,A_1]$,
the notion of head reduction
 is otherwise uncooperative.
Therefore,
we work with  \emph{strong reductions}
(Subsection~\ref{SS:strongred})
and \emph{atomic reductions}
(Subsection~\ref{SS:atomic-reductions}) instead,
yielding the
more tractable relations~$\leqs$ and~$\leqa$, respectively.
We prove
later on that for types $A$ and~$B$ we have
\begin{equation*}
A \leqa B \quad\implies\quad A\leqs B \quad\implies\quad A\leqh B\htam.
\end{equation*}
So to show that $A \leqh B$
it suffices to prove that either $A\leqs B$
or~$A\leqa B$.

One of the calculation rules provided in this section
concerns types~$A$ with~$[1,1]\leqa A$.
It states that for such~$A$ and any contexts $\Gamma_1$, $\Gamma_2$,
we have
\begin{equation*}
[\Gamma_1]\leqs A\quad\htam{ and } \quad[\Gamma_2]\leqs A
\quad\implies\quad
[\Gamma_1,\Gamma_2]\leqs A.
\end{equation*}
We will call these types \emph{atomic types}
and study them in Subsection~\ref{SS:atomic-types}.

\subsection{Strong reductions}
\label{SS:strongred}
For the sake of familiarity
we begin with  strong reductions between
\emph{types}.
Let $A_1$ and~$A_2$ be types.
Recall 
that a reducing term from~$A_1$ to~$A_2$
is a closed term $R$ of type $A_1 \ra A_2$
which is injective on closed terms
(see Definition~\ref{defn:reductions}),
that is, 
the map~$\Phi\colon \ts{A_1}\ra\ts{A_2}$
given by $\Phi(M)\eqb RM$ is injective.

If $R$ is also injective
on open terms, then~$R$ is  called \emph{strong}:

\begin{defi}\leavevmode\begin{enumerate}[label*=(\roman*),ref=\roman*]
\label{D:s}
\item Let~$A_1$ and~$A_2$ be types.  
A \keyword{strong reducing term}
from~$A_1$ to~$A_2$
is a closed term~$R\colon A_1\ra A_2$ 
that is injective on open terms,
that is,
for every context~$\Xi$
 the term~$R$ is injective
on open terms with free variables from $\set{\Xi}$,
that is,
the term~$R$ is injective considered as a map
 $\ts[\Xi]{A_1}\ra\ts[\Xi]{A_2}$.
\item
If
there is a strong reducing term~$R\colon A_1\ra A_2$
that is a B\"ohm term
(see Definition~\ref{defn:reductions}\iref{defn:reductions-h}),
we say that $A_1$~\keyword{strongly head reduces
  to}~$A_2$, notation~$A_1\leqsh A_2$.
\item
\label{D:s-red}
For context--types (see Definition~\ref{D:ct})
$\ct{\Gamma_1}{A_1}$ \keyword{strongly reduces}
to~$\ct{\Gamma_2}{A_2}$ if
\begin{equation*}
\Gamma_1 \ra A_1 \ \leqsh\ \Gamma_2 \ra A_2,
\end{equation*}
and we write $\ct{\Gamma_1}{A_1}\leqs\ct{\Gamma_2}{A_2}$.
If in addition $\ct{\Gamma_2}{A_2}\leqs\ct{\Gamma_1}{A_1}$,
we write $\ct{\Gamma_2}{A_2}\iss\ct{\Gamma_1}{A_1}$.
\item
\label{D:s-xi}
Let~$\varrho$ be a substitution
from~$\ct{\Gamma_1}{A_1}$ to~$\ct{\Gamma_2}{A_2}$, and let $\Xi$ be
a fresh context.\\
With~$\varrho^\Xi$ we denote the natural extension of~$\varrho$
to a substitution
from~$\ct{\Xi,\Gamma_1}{A}$ to~$\ct{\Xi,\Gamma_2}{B}$
given by $\varrho^\Xi_c = c$ for all~$c^C\in\{\Xi\}$.
Then $\hat\varrho^\Xi\colon\ts[\Xi,\Gamma_1]{A_1}\ra\ts[\Xi,\Gamma_2]{A_2}$. 
\end{enumerate}
\end{defi}
\begin{prop}
\label{P:strongsubs}
Let $\ct{\Gamma_1}{A_1}$ and $\ct{\Gamma_2}{A_2}$ be context--types.
Then 
\begin{equation*}
\ct{\Gamma_1}{A_1} \leqs \ct{\Gamma_2}{A_2}
\quad\iff\quad
\left[\ \begin{minipage}{0.50\columnwidth}
There is a substitution $\varrho$ 
from $\ct{\Gamma_1}{A_1}$ to $\ct{\Gamma_2}{A_2}$
such that
$\hat\varrho^\Xi\colon\ts[\Xi,\Gamma_1]{A_1}\ra\ts[\Xi,\Gamma_2]{A_2}$
is injective
for every context~$\Xi$.
\end{minipage}
\right.
\end{equation*}
\end{prop}
\proof 
Just unfold the definitions.\qed

\begin{defi}
\label{D:strongred}
Let~$\ct{\Gamma_1}{A_1}$, $\ct{\Gamma_2}{A_2}$ be context--types.
A \keyword{strong reduction from~$\ct{\Gamma_1}{A_1}$
to~$\ct{\Gamma_2}{A_2}$} is a substitution~$\varrho$
from~$\ct{\Gamma_1}{A_1}$
to~$\ct{\Gamma_2}{A_2}$
such that all~$\hat\varrho^\Xi$ are injective.
We write $\varrho\colon \ct{\Gamma_1}{A_1}\leqs\ct{\Gamma_2}{A_2}$.
\end{defi}
\begin{rem}
\label{rem:reduction-def}
It would not make sense to define a strong reduction
to be the B\"ohm transformation~$\Phi\equiv\hat\varrho$
because one can not always reconstruct~$\varrho$---and hence
the~$\hat\varrho^\Xi$s---from~$\hat\varrho$,
which acts only on closed terms.
\end{rem}

The merit of strong reductions
(over regular ones)
is that it is easy to built complex strong reductions from simpler ones.
Moreover, almost all reductions
encountered in this text are strong.

\begin{rems} \leavevmode
  \begin{enumerate}[label=(\roman*),ref=\roman*]
\item
Not every reduction is also a strong reduction: 
the substitution~$\varrho$ 
from the context~$f^1$
to the empty context~$\ectx$ given by $\varrho_f = \lm{x^0}x$ is a
reduction,
because~$\smash{\ts[f^1]0}$ is empty,
and thus $\hat\varrho\colon\smash{\ts[f^1]0 \ra \ts0}$ is injective
(see Convention~\ref{V:red});
but~$\varrho$ is not a strong reduction
since~$\smash{\hat\varrho^{\Xi}}$ with $\Xi\eqdf d^0$ 
maps both $fd$ and $ffd$ to~$\lm{x^0}x$
and is hence not injective (see Definition~\ref{D:strongred}).
\item
Note that~$\Gamma\leqs \Delta$
implies $\Xi,\Gamma \leq \Xi,\Delta$ for all contexts~$\Xi$.
It is not evident whether the reverse implication holds as well.
If $\Gamma\leqs \Delta$ then there is \emph{one} substitution
$\varrho$ 
which yields a family of similar reductions
$\varrho^\Xi\colon\Xi,\Gamma\leq\Xi,\Delta$;
on the other hand,
if $\Xi,\Gamma \leq \Xi,\Delta$ for all~$\Xi$,
we only know there is a family of
(potentially quite dissimilar reductions)
$\varrho_\Xi \colon \Xi,\Gamma\leq \Xi,\Delta$. 
As it turns out,
the reverse implication does hold;
we will not prove this in this article.

\item
As $\Gamma\leqs \Delta \implies [\Gamma]\leqs[\Delta]$
(by Definition~\ref{D:s}\iref{D:s-red})
one could conjecture that we also have that $A\leqs
B\implies [A]\leqs [B]$.  These however are quite different statements.
 By the Hierarchy Theorem, the conjecture is false.
Indeed: later on we will see that $[1,0]\leqs[2]$. If one had
$[[1,0]]\leqs[[2]]$, then also $[0,[1,0]]\leqh [0,[2]]$, quod non as
$[0,[1,0]]\in\HH{\omega+4}$, while $[0,[2]]\in\HH{\omega+3}$.

\item
Nevertheless we do have $A\leqs B\implies [[A]]\leqs[[B]]$ (see
Lemma~\ref{L:sr-ci}).

\item
Similarly,
we have $\Gamma \iss [\Gamma]$ for every context~$\Gamma$
(by Definition~\ref{D:s}\iref{D:s-red}),
but \emph{never} $A \iss [A]$ for a type~$A$.
Indeed,
if $A\iss [A]$, then
$[A]$ is inhabited iff $A$ is inhabited,
while by Theorem~\ref{T:inhabitation},
$[A]$ is inhabited iff~$A$ is uninhabited.
\end{enumerate}
\end{rems}
\begin{lem}
\label{L:xi-func}
$(\varrho^{\Xi_1})^{\Xi_2} = \varrho^{\Xi_2,\Xi_1}$
for every substitution~$\varrho$
and contexts~$\Xi_1$, $\Xi_2$.
\end{lem}
\proof 
By Definition~\ref{D:ct}\iref{D:ct-sub}
we may assume that~$\varrho$ is a substitution between contexts,
say from~$\Gamma$ to~$\Delta$.
Recall that $\varrho^{\Xi_1}$ is a substitution
from~$\Xi_1,\Gamma$ to~$\Xi_1,\Delta$
with $\varrho^{\Xi_1}_c = \varrho_c$ for all~$c\in\{\Gamma\}$
and $\varrho^{\Xi_1}_d = d$ for all~$d\in\{\Xi_1\}$
(see Definition~\ref{D:s}\iref{D:s-xi}).
So both~$(\varrho^{\Xi_1})^{\Xi_2}$
and~$\varrho^{\Xi_2,\Xi_1}$
are a substitution~$\sigma$ from~$\Xi_2,\Xi_1,\Gamma$
to~$\Xi_2,\Xi_1,\Delta$
such that $\sigma_c=\varrho_c$ for all~$c\in\{\Gamma\}$
and~$\sigma_d = d$ for all~$d\in\{\Xi_1,\Xi_2\}$.
Hence~$(\varrho^{\Xi_1})^{\Xi_2}$
and~$\varrho^{\Xi_2,\Xi_1}$
are the same.\qed

%
%
\begin{lem}
\label{L:sr-glp}
Given contexts $\Theta$, $\Gamma$ and $\Delta$,
we have
\begin{equation*}
\Gamma\leqs\Delta
\quad\implies\quad
\Theta,\Gamma\,\leqs\,\Theta,\Delta\htam.
\end{equation*}
\end{lem}
\proof 
Assume that $\Gamma\leqs\Delta$,
that is,
that there is some strong reduction~$\varrho$
from~$\Gamma$ to~$\Delta$ 
(see Proposition~\ref{P:strongsubs} and Definition~\ref{D:strongred}).
To show that~$\Theta,\Gamma\leqs \Theta,\Delta$,
we prove that $\varrho^\Theta$
is a strong reduction from~$\Theta,\Gamma$ to~$\Theta,\Delta$.
Writing $\varrho^{\Theta\,\Xi}\eqdf (\varrho^\Theta)^\Xi
= \varrho^{\Xi,\Theta}$
(see Lemma~\ref{L:xi-func}),
we need to prove that the map
 $\hat\varrho^{\Theta\,\Xi}\colon \ts[\Xi,\Theta,\Gamma]0\ra\ts[\Xi,\Theta,\Delta]0$
is injective for every context~$\Xi$
(see Proposition~\ref{P:strongsubs}).
Since~$\varrho$ is a strong reduction,
we know that 
$\hat\varrho^{\Xi'}\colon\ts[\Xi',\Gamma]0 \ra \ts[\Xi',\Delta]0$
is injective for every~$\Xi'$.
Now, pick $\Xi'=\Xi,\Theta$.\qed

\noindent Before we get to the more
serious reductions,
we study
the workings of a B\"ohm transformation~$\hat\varrho$ 
(see Definition~\ref{D:ct}\iref{D:ct-bohm})
more closely
in Proposition~\ref{P:bohm}.
\begin{defi}
\label{D:rhosc}
Let $\varrho$ be a substitution
from~$\ct{\Gamma_1}{[\Delta_1]}$
to~$\ct{\Gamma_2}{[\Delta_2]}$
(see Definition~\ref{D:ct}\iref{D:ct-sub}).\\
For every type~$A\equiv[\Xi]$,
let $\varrho^A$ 
denote the natural extension of~$\varrho$
to a substitution from $\ct{\Gamma_1}{[\Xi,\Delta_1]}$
to $\ct{\Gamma_2}{[\Xi,\Delta_2]}$
given 
by $\hat\varrho^A(c)=c$ for all~$c\in\{\Xi\}$.
\end{defi}
\begin{rems}
\label{R:rhosc}
Let $\Xi$ be a context
and let $\varrho$ a substitution
from~$\ct{\Gamma_1}{[\Delta_1]}$
to~$\ct{\Gamma_2}{[\Delta_2]}$.
\begin{enumerate}[label=(\roman*),ref=\roman*]
\item
$\varrho^{[\Xi]}$ (Definition~\ref{D:rhosc})
and~$\varrho^\Xi$ (Definition~\ref{D:s}\iref{D:s-xi})
are essentially the same substitution since
we have 
$\varrho^{[\Xi]}_c = \varrho^{\Xi}_c$ for all~$c\in\{\Xi,\Gamma_1,\Delta_1\}$
(see Definition~\ref{D:ct}\iref{D:ct-sub}).
\item
\label{R:rhosc-working}
Using Definition~\ref{D:ct}\iref{D:ct-bohm}
we see that
for all contexts~$\Xi$, $\Theta$
and 
~$M\in\ts[\Xi,\Gamma_1]{[\Theta,\Delta_1]}$,
\begin{equation*}
\varrho^{\Xi\, [\Theta]} M
\ \eqbe\  M\,[\Gamma_1,\Delta_1\subs\subtup\varrho{\Gamma_1,\Delta_1}].
\end{equation*}
\item
\label{R:rhosc-0}
We have $\varrho^0 = \varrho = \varrho^\ectx$.
\end{enumerate}
\end{rems}
\begin{prop}
\label{P:bohm}
A substitution~$\varrho$ from~$\Gamma$ to~$\Delta$
satisfies the `recursion':
\begin{equation*}
\left[\ 
\begin{minipage}{0.95\columnwidth}
Given contexts~$\Xi$, $\Theta$ and $a^A\in \{\Xi,\Gamma\}$
with $A\equiv[A_1,\dotsc,A_n]$, we have
\begin{equation*}
\begin{alignedat}{4}
\hat\varrho^{\Xi}(\, a\,M_1\dotsc M_n \,) 
    & \eqbe\ &&  \varrho^\Xi_a\ 
     (\,\hat\varrho^{\Xi\,A_1} M_1\,) \dotsb 
               (\,\hat\varrho^{\Xi\, A_n} M_n\,), \\
\hat\varrho^{\Xi\,[\Theta]}(\, \lm{\Theta} M \,)
    & = && \lm{\Theta}(\,\hat\varrho^{\Theta,\Xi} M\,)\htam,
\end{alignedat}
\end{equation*}
for all $M_i\in \ts[\Xi,\Gamma]{A_i}$
and~$M\in\ts[\Theta,\Xi,\Gamma]{0}$.
\end{minipage}
\right.
\end{equation*}
\end{prop}
\proof 
It is only a matter of expanding definitions.
Indeed,
\begin{alignat*}{4}
&&& \hat\varrho^\Xi(\,a\,M_1\dotsb M_n\,) \\
& \eqbe\  && 
  (\,a\,M_1\dotsb M_n\,) 
  [\,\Gamma\subs\subtup\varrho{\Gamma}\,]
\quad && 
\htam{by Rem.~\ref{R:rhosc}\iref{R:rhosc-working},\iref{R:rhosc-0}}\\
& \eqbe\   &&
  \varrho^\Xi_a \  M_1[\,\Gamma\subs\subtup\varrho{\Gamma}\,]
  \ \dotsb\ M_n[\,\Gamma\subs\subtup\varrho{\Gamma}\,]\qquad\\
& \eqbe  &&
 \varrho^\Xi_a\ 
  (\,\hat\varrho^{\Xi\,A_1} M_1\,) \dotsb (\,\hat\varrho^{\Xi\,A_n} M_n\,)
\qquad &&\htam{by Rem.~\ref{R:rhosc}\iref{R:rhosc-working},}
\end{alignat*}
where $a^A\in\{\Xi,\Gamma\}$
with $A\equiv[A_1,\dotsc,A_n]$
and $M_i\in\ts[\Xi,\Gamma]{A_i}$.
Similarly,
\begin{alignat*}{4}
\hat\varrho^{\Xi\,[\Theta]}(\,\lm{\Theta}M\,)& \eqbe\  && 
(\lm{\Theta} M)[\Gamma\subs\subtup\varrho{\Gamma}]
     && \qquad\htam{by Rem.~\ref{R:rhosc}\iref{R:rhosc-working}} \\
&=&&\lm{\Theta} (\,M[\Gamma\subs\subtup\varrho{\Gamma}]\,)
     && \qquad\htam{} \\
&\eqbe&&\lm{\Theta} (\,\hat\varrho^{\Theta,\Xi} M \,)
     && \qquad\htam{by 
          Rem.~\ref{R:rhosc}\iref{R:rhosc-working},\iref{R:rhosc-0}} 
\end{alignat*}
for every term~$M\in\ts[\Theta,\Xi,\Gamma]0$.\qed

\noindent
We now give an important condition
for a substitution to be a strong reduction.
\begin{thm}
\label{T:atomic-strong}
Let~$\Gamma$ and $\Delta$ be contexts.
Let $\varrho$ be a substitution from~$\Gamma$
to~$\Delta$.\\
If $\varrho$ has the following property,
then~$\varrho$ is a strong reduction.
\begin{equation*}
\left[\ 
\begin{minipage}{0.95\columnwidth}
Given~$a^A,b^B\in\{\Xi,\Gamma\}$
with $A\equiv[A_1,\dotsc,A_n]$
and $B\equiv[B_1,\dotsc,B_m]$. Then
\begin{equation}
\label{eq:atomic}
\varrho_a^\Xi\,M_1\dotsb M_n \,\eqbe\,\varrho_b^\Xi\,N_1\dotsb N_m
\quad\implies\quad a=b\ \text{ and } \ M_i = N_i
\end{equation}
for all  $M_i\in\ts[\Xi,\Delta]{A_i}$ and $N_i\in\ts[\Xi,\Delta]{B_i}$
and every context~$\Xi$.
\end{minipage}
\right.
\end{equation*}
\end{thm}
\proof 
To prove that~$\varrho$ is a strong reduction,
we need to show that for each context~$\Xi$,
the
B\"ohm
transformation $\hat\varrho^\Xi\colon\ts[\Xi,\Gamma]0\ra\ts[\Xi,\Delta]0$ 
is injective 
(see Definition~\ref{D:strongred}).
So,
consider for each context--type~$\ct{\Xi}{C}$ 
(see Definition~\ref{D:ct})
and 
$M\in\ts[\Xi,\Gamma]{C}$
the property $P(M)$:
\begin{alignat*}{3}
\hat\varrho^{\Xi\,C}(M) &= \hat\varrho^{\Xi\,C}(N)&
\quad&\implies&\quad M&=N\qquad\qquad
\htam{for all }N\in\ts[\Xi,\Gamma]{C}.
\intertext{It suffices to prove that~$P(M)$ for all~$M$,
because then (taking $C=0$),}
\hat\varrho^{\Xi}(M) &= \hat\varrho^{\Xi}(N) &
\quad&\implies&\quad M&=N\qquad\qquad
\htam{for all }M,N\in\ts[\Xi,\Gamma]{0}
\end{alignat*}
for each context~$\Xi$,
so each B\"ohm transformation~$\hat\varrho^\Xi$ is injective.

To prove that~$P(M)$ for all~$M$, we use induction on~$M$.
There are two cases.
\begin{equation*}
\text{\iref{it:atomic-strong-I}}\ \,M\equiv a^A\,M_1 \dotsb M_n\htam{;}
\qquad\quad
\text{\iref{it:atomic-strong-II}}\ \,M\equiv \lm{\Theta} M'\htam.
\end{equation*}
\begin{enumerate}[label=(\Roman*),ref=\Roman*]
\item\label{it:atomic-strong-I}
We have
$M\equiv a^A\,M_1 \dotsb M_n$
where~$a^A\in\{\Xi,\Gamma\}$
and where
writing $A\equiv[A_1,\dotsc,A_n]$
we have  $M_i\in\ts[\Xi,\Gamma]{A_i}$ .
Assume~$P(M_i)$ in order to show that~$P(M)$.

Let $N\in\ts[\Xi,\Gamma]{0}$ with~$\hat\varrho^\Xi{M}=\hat\varrho^\Xi{N}$
be given.  We need to prove that~$M=N$.
We~have~$N\equiv b^B\,N_1\dotsb N_m$
for some~$b^B\in\{\Xi,\Gamma\}$
with $B\equiv[B_1,\dotsc,B_m]$
and $N_i \in \ts[\Xi,\Gamma]{B_i}$.
Then by 
Proposition~\ref{P:bohm},
$\hat\varrho^\Xi{M}=\hat\varrho^\Xi{N}$ implies
\begin{equation*}
\varrho^\Xi_a\ 
  (\,\hat\varrho^{\Xi\,A_1} M_1\,) \dotsb (\,\hat\varrho^{\Xi\,A_n} M_n\,)
\ \eqbe\ 
\varrho^\Xi_b\ 
  (\,\hat\varrho^{\Xi\,B_1} N_1\,) \dotsb (\,\hat\varrho^{\Xi\,B_m} N_m\,)\htam.
\end{equation*}
Now, $\hat\varrho^{\Xi\,A_i} M_i \in \ts[\Xi,\Delta]{A_i}$
and $\hat\varrho^{\Xi\,B_i} N_i \in\ts[\Xi,\Delta]{B_i}$,
so by Statement~\eqref{eq:atomic}, $a=b$ and
$\hat\varrho^{\Xi\,A_i} M_i= \hat\varrho^{\Xi\,B_i} N_i$.
Then~$M_i=N_i$ by~$P(M_i)$,
so~$M=N$. Hence~$P(M)$.

\item\label{it:atomic-strong-II}
We have $M\equiv \lm{\Theta} M'$
where~$\Theta$ is some context and $M'\in\ts[\Theta,\Xi,\Gamma]{0}$.
Assume that~$P(M')$ in order to show that~$P(M)$.

Let~$N\in\ts[\Xi,\Gamma]{[\Theta]}$
with $\hat\varrho^{\Xi\,[\Theta]} M = \hat\varrho^{\Xi\,[\Theta]} N$
be given. We need to prove that~$M=N$.
Write $N\equiv\lm{\Theta}N'$
where~$N'\in\ts[\Theta,\Xi,\Gamma]{0}$.
Then 
by Proposition~\ref{P:bohm}
we have
\begin{equation*}
\lm{\Theta} (\, \hat\varrho^{\Theta,\Xi} M'\,)
\ = \  \lm{\Theta} (\, \hat\varrho^{\Theta,\Xi} N' \,).
\end{equation*}
Then
$\hat\varrho^{\Theta,\Xi} M' = \hat\varrho^{\Theta,\Xi} N'$,
and thus $M'=N'$ by~$P(M')$.
Hence~$M=N$ and so~$P(M)$.
\end{enumerate}
So we see that $P(M)$ for all~$M$.
Hence $\varrho$ is a strong reduction.\qed

%
%
\subsection{Atomic reductions}
\label{SS:atomic-reductions}
\begin{defi}\label{D:ar}\leavevmode
\begin{enumerate}[label=(\roman*),ref=\roman*]
\item\label{D:ar-ctx-red}
Let~$\Gamma$ and~$\Delta$ be contexts.
A substitution from~$\Gamma$ to~$\Delta$
is called an \keyword{atomic reduction}
if it satisfies condition~\eqref{eq:atomic} of Theorem~\ref{T:atomic-strong}.
\item\label{D:ar-red}
A substitution~$\varrho$
from~$\ct{\Gamma_1}{[\Delta_1]}$ to~$\ct{\Gamma_2}{[\Delta_2]}$
is called an \keyword{atomic reduction}
if~$\varrho$,
considered as substitution from~$\Gamma_1,\Delta_1$ to~$\Gamma_2,\Delta_2$
(see Definition~\ref{D:ct}\iref{D:ct-sub}),
is an atomic reduction.
In that case we 
write~$\varrho\colon \ct{\Gamma_1}{[\Delta_1]}
\leqa \ct{\Gamma_2}{[\Delta_2]}$.
\item\label{D:ar-reduces}
We say that~\keyword{$\ct{\Gamma_1}{[\Delta_1]}$
atomically reduces to~$\ct{\Gamma_2}{[\Delta_2]}$}
if there is an atomic reduction
from~$\ct{\Gamma_1}{[\Delta_1]}$ to~$\ct{\Gamma_2}{[\Delta_2]}$.
In that case we write
$\ct{\Gamma_1}{[\Delta_1]}\leqa \ct{\Gamma_2}{[\Delta_2]}$.
\end{enumerate}
\end{defi}
%
%
\begin{rem}
\label{R:atomic}Given variables~$a$ and~$b$,
we have (cf. Statement~\eqref{eq:atomic})
\begin{equation*}
a\, M_1 \dotsb M_n \,\eqbe\, b\, N_1 \dotsb N_m
\quad\implies\quad
a=b\ \htam{ and } M_i = N_i
\end{equation*}
for all terms~$M_i$
and $N_i$.
In this respect
the terms~$\varrho^\Xi_a$s
of an \emph{atomic reduction}~$\varrho$
behave similar to \emph{atomic terms} ($=$variables).
Hence the name.
\end{rem}
%
%
\begin{rem}
\label{R:atomic-red}
Given context--types
$\ct{\Gamma_1}{\Delta_1}$
and
$\ct{\Gamma_2}{\Delta_2}$
we have
\begin{equation*}
\ct{\Gamma_1}{[\Delta_1]}\leqa \ct{\Gamma_2}{[\Delta_2]}
\quad\iff\quad
\Gamma_1,\Delta_1 \leqa \Gamma_2,\Delta_2
\end{equation*}
by Definition~\ref{D:ar}\iref{D:ar-red}.
Cf. Definition~\ref{D:s}\iref{D:s-red}.
\end{rem}
%
%
\begin{prop}
\label{P:atomic-strong}
For context--types~$\ct{\Gamma_1}{[\Delta_1]}$ and~$\ct{\Gamma_2}{[\Delta_2]}$
we have
\begin{equation*} 
\ct{\Gamma_1}{[\Delta_1]}\,\leqa\, \ct{\Gamma_2}{[\Delta_2]}
\quad\implies\quad
\ct{\Gamma_1}{[\Delta_1]}\,\leqs\, \ct{\Gamma_2}{[\Delta_2]}\htam.
\end{equation*}
\end{prop}
\proof 
Assume that 
$\ct{\Gamma_1}{[\Delta_1]}\,\leqa\, \ct{\Gamma_2}{[\Delta_2]}$.
That is,
there is some atomic reduction~$\varrho$
from~$\ct{\Gamma_1}{[\Delta_1]}$ to~$\ct{\Gamma_2}{[\Delta_2]}$.
By Theorem~\ref{T:atomic-strong},
$\varrho$ is also a strong reduction.
Hence 
$\ct{\Gamma_1}{[\Delta_1]}\,\leqs\, \ct{\Gamma_2}{[\Delta_2]}$
by Proposition~\ref{P:strongsubs}
and Definition~\ref{D:strongred}.\qed

%
%
%
Below we have collected the calculation rules for~$\leqa$
which we use later on.
The reader can chose to skip them at first
and proceed to Remark~\ref{R:ar-transitive}.
\begin{lem}
\label{L:ar-glp}
Let~$\Gamma$, $\Delta$ and~$\Theta$ be contexts.
\begin{enumerate}[label=(\roman*),ref=\roman*]
\item\label{L:ar-glp-i}
If $\{\Gamma\}\subseteq \{\Delta\}$ then $\Gamma\leqa \Delta$.
\item\label{L:ar-glp-ii}
If $\Gamma\leqa \Delta$ then $\Theta,\Gamma \leqa \Theta,\Delta$.
\end{enumerate}
\end{lem}
\proof 
\noindent\iref{L:ar-glp-i}.\ 
Assume that~$\{\Gamma\}\subseteq\{\Delta\}$.
To prove~$\Gamma\leqa\Delta$,
we need to find an atomic reduction
from~$\Gamma$ to~$\Delta$
(see Definition~\ref{D:ar}\iref{D:ar-reduces}).
Let~$\varrho$ be the substitution from~$\Gamma$
to~$\Delta$ given by~$\varrho_c = c$
for all~$c^C\in\{\Gamma\}$
(see Def.~\ref{D:ct}\iref{D:ct-sub-ctx}).
To prove that~$\varrho$
is an atomic reduction,
we need to show that given a context~$\Xi$
and $a^A,b^B\in\{\Xi,\Gamma\}$
with $A\equiv[A_1,\dotsc,A_n]$
and $B\equiv[B_1,\dotsc,B_m]$,
\begin{equation*}
\varrho^\Xi_a\, M_1 \dotsb M_n \,\eqbe\, \varrho^\Xi_b\,N_1 \dotsb N_m
\quad\implies\quad a=b\ \htam{ and }\  M_i = N_i
\end{equation*}
for all~$M_i\in\ts[\Xi,\Delta]{A_i}$
and $N_i\in\ts[\Xi,\Delta]{B_i}$.
Since all~$\varrho_a^\Xi$ are distinct variables,
this follows immediately from Remark~\ref{R:atomic}.

\noindent\iref{L:ar-glp-ii}.\ 
Assume that~$\Gamma\leqa\Delta$,
that is, that there is some atomic reduction~$\varrho$
from~$\Gamma$ to~$\Delta$ (see Definition~\ref{D:ar}\iref{D:ar-reduces}).
To prove that~$\Theta,\Gamma\leqa\Theta,\Delta$,
we show that~$\varrho^\Theta$
is an atomic reduction from~$\Theta,\Gamma$ to~$\Theta,\Delta$.
For this we must prove that 
\begin{equation}
\label{eq:L:ar-glp-ii}
(\varrho^{\Theta})^{\Xi'}_a\, M_1 \dotsb M_n \,\eqbe\, 
(\varrho^{\Theta})^{\Xi'}_b\,N_1 \dotsb N_m
  \quad\implies\quad a=b\ \htam{ and }\  M_i = N_i
\end{equation}
for every context~$\Xi'$ and appropriate $a$, $b$, $M_i$ and~$N_i$
(see Definition~\ref{D:ar}\iref{D:ar-red}).
Since we have that~$(\varrho^{\Theta})^{\Xi'} = \varrho^{\Xi',\Theta}$
(see Lemma~\ref{L:xi-func}),
Statement~\eqref{eq:L:ar-glp-ii} follows
immediately from the fact that~$\varrho$ 
is an atomic reduction. (Indeed, pick $\Xi=\Xi',\Theta$).\qed

%
%
\begin{cor} \label{L:sr-glt}
Given types $C_1,\ldots,C_k$ and a permutation $\varphi$
of $\set{1,\ldots,k}$, we have
\begin{equation*}
[C_1,\ldots,C_k]\ \leqa\ [C_{\varphi(1)},\ldots,C_{\varphi(k)}].
\end{equation*}
\end{cor}
\proof  
Write $[C_1,\ldots,C_k]=[\Gamma]$
and $[C_{\varphi(1)},\dotsc,C_{\varphi(k)}] =[\varphi\cdot\Gamma]$.
We must prove that $\Gamma\leqa \varphi\cdot\Gamma$
(see Remark~\ref{R:atomic-red}).
This follows immediately from Lemma~\ref{L:ar-glp}\iref{L:ar-glp-i}
since~$\{\Gamma\}=\{\varphi\cdot\Gamma\}$.\qed

%
%
\begin{lem}
\label{L:atomic}
A substitution~$\varrho$ from~$\Gamma$ to~$\Delta$
is an atomic reduction provided that
\begin{enumerate}[label=(\roman*),ref=\roman*]
\item\label{L:atomic-i}
If $a^A,b^B\in\{\Gamma\}$
with $A\equiv[A_1,\dotsc,A_n]$
and $B\equiv[B_1,\dotsc,B_m]$, then
\begin{equation*}
\varrho_a\,M_1\dotsb M_n \,\eqbe\, \varrho_b \, N_1\dotsb N_m
\quad\implies\quad a=b\ \text{ and }\ M_i = N_i
\end{equation*}
for all $M_i\in\ts[\Xi,\Delta]{A_i}$
and $N_i\in\ts[\Xi,\Delta]{B_i}$
and every context~$\Xi$.
\item\label{L:atomic-ii}
If $a^A\in\{\Gamma\}$, $d^D\in\{\Xi\}$
with $A\equiv[A_1,\dotsc,A_n]$, $D\equiv[D_1,\dotsc,D_\ell]$,
then
\begin{equation*}
\varrho_a \, M_1\dotsb M_n \, \neqbe \, d\, N_1 \dotsb N_\ell
\end{equation*}
for all $M_i\in\ts[\Xi,\Delta]{A_i}$
and $N_i\in\ts[\Xi,\Delta]{D_i}$.
\end{enumerate}
\end{lem}
\proof 
Let $a,b\in\{\Gamma,\Xi\}$
with $A\equiv[A_1,\dotsc,A_n]$
and $B\equiv[B_1,\dotsc,B_m]$
and 
\begin{equation*}
\varrho_a^\Xi\,M_1\dotsb M_n \,\eqbe\, \varrho_b^\Xi \, N_1\dotsb N_m
\end{equation*}
for some~$M_i\in\ts[\Xi,\Delta]{A_i}$
and~$N_i\in\ts[\Xi,\Delta]{B_i}$
and some context~$\Xi$.
We need to prove~$a=b$ and $M_i = N_i$
(see Definition~\ref{D:ar}\iref{D:ar-ctx-red}).
We distinguish four cases.
\begin{enumerate}[label=(\roman*),ref=\roman*]
\item
If $a,b\in\{\Gamma\}$
then $\varrho_a^\Xi = \varrho_a$, so
 $a=b$ and~$M_i = N_i$
by Assumption~\iref{L:atomic-i}.
\item
If $a,b\in\{\Xi\}$
then $\varrho^\Xi_a = a$,
so $a=b$ and $M_i = N_i$
by Remark~\ref{R:atomic}.
\item
The situation that $a\in \{\Gamma\}$ and $b\in \{\Xi\}$
does not occur,
since 
by Assumption~\iref{L:atomic-ii}
we have 
$\varrho_a^\Xi\,M_1\dotsb M_n \,\neqbe\,
 b\, N_1\dotsb N_m\equiv\varrho_b^\Xi \, N_1\dotsb N_m$.
\item
Similarly,
the situation that
$a \in \{\Xi\}$ and $b\in\{\Gamma\}$
does not happen.\qed
\end{enumerate}

\begin{lem} \label{L:sr-flt}
Given types $C_1,\dotsc,C_k$
and a permutation~$\varphi$ of~$\{1,\dotsc,k\}$,
we have
\begin{equation*}
[[C_1,\dotsc,C_k]] \,\leqa\, [[C_{\varphi(1)},\dotsc,C_{\varphi(k)}]].
\end{equation*}
\end{lem}
\proof 
Writing $[C_1,\dotsc,C_k]=[\Gamma]$
and $[C_{\varphi(1)},\dotsc,C_{\varphi(k)}]$,
we must prove that $[[\Gamma]]\leqa [[\varphi\cdot\Gamma]]$.
By Remark~\ref{R:atomic-red}
we need to find an atomic reduction
from $F^{[\Gamma]}$ to~$G^{[\varphi\cdot\Gamma]}$.
We show that the substitution~$\varrho$ from~$F$ to~$G$ given by
\begin{equation}
\label{eq:sr-flt-rhoF}
\varrho_F\ \eqdf\ \lm{\Gamma} G (\varphi\cdot \Gamma)
\end{equation}
is an atomic reduction.
For this we use Lemma~\ref{L:atomic}. Let~$\Xi$ be a context.

\begin{enumerate}[label=(\roman*),ref=\roman*]
\item[\iref{L:atomic-i}]
Suppose that $\varrho_F M_1 \dotsb M_k \eqbe \varrho_F N_1 \dotsb N_k$
for some $M_i,N_i\in\ts[\Xi,G]{C_i}$.
We need to prove that $M_i=N_i$.
Indeed,
$\varrho_F M_1 \dotsb M_k \eqbe \varrho_F N_1 \dotsb N_k$
yields
\begin{equation*}
G M_{\varphi(1)} \dotsb M_{\varphi(k)} =
G N_{\varphi(1)} \dotsb N_{\varphi(k)}.
\end{equation*}
Hence $M_{\varphi(i)}=N_{\varphi(i)}$ and thus $M_i=N_i$.

\item[\iref{L:atomic-ii}]
Let $d^D\in\{\Xi\}$
with $D\equiv[D_1,\dotsc,D_\ell]$
and $\varrho_F M_1 \dotsb M_k \eqbe \varrho_d^\Xi N_1 \dotsb N_\ell$
for some $M_i\in\ts[\Xi,G]{C_i}$ and $N_i\in\ts[\Xi,G]{D_i}$.
We need to reach a contradiction.
Indeed, we get 
$G M_{\varphi(1)} \dotsb M_{\varphi(k)} =
b N_{1} \dotsb N_{k}$,
so $G=b$, quod non.\qed
\end{enumerate}

%
%
%
\begin{rem}
Given a type~$C\equiv[C_1,\dotsc,C_k]$,
the order $C_1,\dotsc,C_k$
of the components
is largely immaterial.
Witnesses of this principle include Corollary~\ref{L:sr-glt},
Lemma~\ref{L:sr-flt}
and Definition~\ref{D:ct}\iref{D:ct-sub-ctx}.
We will often use this principle implicitly. 
For instance, we will use
Lemma~\ref{L:sr-glp} to argue that
$\Gamma\leqs\Delta\implies \Gamma,\Theta \leqs \Delta,\Theta$.
(Of course, this is licit by Corollary~\ref{L:sr-glt}.)
\end{rem}
%
%
%
%
%
\begin{lem}
\label{L:sr-ci}
Let $A$, $B$ and $C_1,\dotsc,C_k$ be types. Then
\begin{equation*}
A\leqs B
\quad\implies\quad
[[A,C_1,\dotsc,C_k]] \leqa [[B,C_1,\dotsc,C_k]].
\end{equation*}
\end{lem}
\proof 
Assume that $A\leqs B$ to find an
atomic reduction from $F^{[A,C_1,\dotsc,C_k]}$ and $G^{[B,C_1,\dotsc,C_k]}$
(see Remark~\ref{R:atomic-red}).
Pick a strong reduction
$\sigma\colon A\leqs B$
with reducing term~$S\colon A\ra B$.
Define a substitution $\varrho$ from~$F$ to~$G$ by
\begin{equation*}
\varrho_F \eqdf \lm{a^{A} \Gamma}
    G (S a) \Gamma \htam{,}
\end{equation*}
where $\Gamma \eqdf c_1^{C_1},\dotsc,c_k^{C_k}$.
We prove that~$\varrho$ is an atomic reduction
by Lemma~\ref{L:atomic}.

\begin{enumerate}[label=(\roman*),ref=\roman*]
\item[\iref{L:atomic-i}]
Let $\Xi$ be a context and suppose that
\begin{equation}
\label{eq:ertyuio}
\varrho_F \, M\, M_1\dotsb M_k
\,\eqbe\,
\varrho_F \, N\, N_1\dotsb N_k
\end{equation}
for some~$M,N\in\ts[\Xi,G]{A}$
and $M_i,N_i\in\ts[\Xi,G]{C_i}$
in order to prove~$M_i=N_i$
and $M=N$.
Equation~\eqref{eq:ertyuio} yields
$G(SM)M_1\dotsb M_k \eqbe G(SN)N_1\dotsb N_k$,
so $M_i = N_i$ and $SM \eqbe SN$.
Hence $M=N$ too, since~$\sigma$ is a strong reduction.

\item[\iref{L:atomic-ii}]
Let~$\Xi$ be a context and
$d^D\in\{\Xi\}$ 
with $D=[D_1,\dotsc,D_\ell]$.
Suppose that
\begin{equation*}
\varrho_F \, M\, M_1\dotsb M_k
\,\eqbe\,
d \, N_1 \dotsb N_\ell
\end{equation*}
for some $M\in\ts[\Xi,G]{A}$,
$M_i\in\ts[\Xi,G]{C_i}$
and $N_i\in\ts[\Xi,G]{D_i}$
to reach a contradiction.
We get
$G(SM) M_1\dotsb M_k \eqbe dN_1\dotsb N_\ell$.
But then $G = d$, which is absurd.\qed
\end{enumerate}

%
%
\begin{lem}
\label{L:sr-c}
$A \leqa {\lbb A \rbb}$ for every type~$A$.
\end{lem}
\proof 
Write $A=[\Gamma]$
for some context~$\Gamma$.
It suffices to 
find an atomic reduction~$\varrho$
from~$\Gamma$ to~$F^{\lbb \Gamma \rbb}$
(see Remark~\ref{R:atomic-red}).
Let~$\varrho$ be the substitution
from $\Gamma$ to $F^{\lbb \Gamma \rbb}$
given by
\begin{alignat*}{2}
\varrho_{d} & \eqdf \lm{\Delta} F \, \lm{\Gamma} d \, \Delta
 &\quad& \htam{ for each } d^{[\Delta]} \htam{ from } \{\Gamma\}
\htam{.}
\end{alignat*}
We prove that~$\varrho$ is atomic
using Lemma~\ref{L:atomic}.
Let $\Xi$ be some context.

\begin{enumerate}[label=(\roman*),ref=\roman*]
\item[\iref{L:atomic-i}]
Let $d^D,e^E\in \{\Gamma\}$
with $D\equiv [D_1,\dotsc,D_k]$
and $E\equiv [E_1,\dotsc,E_\ell]$.
Assume  that
\begin{equation}
\label{eq:sr-c-1}
\varrho_d\, M_1\dotsb M_k \,\eqbe\, \varrho_e \,N_1\dotsb N_\ell
\end{equation}
for some 
$M_i\in \ts[\Xi,F]{D_i}$
and
$N_i\in\ts[\Xi,F]{E_i}$,
to prove $d=e$ and $M_i=N_i$.
It is easy to see that Equation~\eqref{eq:sr-c-1} implies that
\begin{equation*}
F \lm{\Gamma} d\, M_1 \dotsb M_k 
\,=\,
F \lm{\Gamma} e\, N_1 \dotsb N_\ell \htam.
\end{equation*}
Hence $d\,M_1\dotsb M_k = e\,M_1\dotsb M_\ell$
and thus $d=e$ and $M_i = N_i$.

\item[\iref{L:atomic-ii}]
Given $d^D\in\{\Gamma\}$, $b^B\in\{\Xi\}$
with $D\equiv[D_1,\dotsc,D_k]$, $B\equiv[B_1,\dotsc,B_n]$,
assume
that
there are $M_i\in\ts[\Xi,F]{D_i}$ and
 $N_i\in\ts[\Xi,F]{B_i}$ such that
\begin{equation}
\label{eq:sr-c-2}
\varrho_d \, M_1\dotsb M_k \, \eqbe\, b\, N_1\dotsb N_m
\end{equation}
in order to obtain a contradiction.
This is easy; Equation~\eqref{eq:sr-c-2} implies
\begin{equation*}
F\lm{\Gamma} d \,M_1\dotsb M_k \,=\, b \, N_1\dotsb N_m\htam,
\end{equation*}
and thus $F=b$, qoud non.\qed
\end{enumerate}

%
%
\begin{lem}
\label{L:sr-sp}
Given types~$A_1,\dotsc,A_n$ we have
\begin{equation*}
[\,[A_1,\dotsc,A_n]\,]\ \leqa\  [\,[A_1],\dotsc,[A_n],\,[0^n]\,].
\end{equation*}
\end{lem}
\proof 
By Remark~\ref{R:atomic-red}
it suffices to show that the substitution
from the context~$\smash{F^{[A_1,\dotsc,A_n]}}$
to the context $\Theta\eqdf\smash{ F_1^{[A_1]},\dotsc,F_n^{[A_n]},p^{[0^n]}}$
given by
\begin{equation*}
\varrho_F\ \eqdf\ \lm{m_1^{A_1}\dotsb m_n^{A_n}}\ 
p\, (F_1 m_1)\,\dotsb\,(F_n m_n)
\end{equation*}
is an atomic reduction.
To prove this,
we use Lemma~\ref{L:atomic}.
\begin{enumerate}[label=(\roman*),ref=\roman*]
\item[\iref{L:atomic-i}]
Let~$\Xi$ be a context
and suppose that
\begin{equation*}
\varrho_F\, M_1\dotsb M_n \ \eqbe\  \varrho_F\, N_1\dotsb N_n
\end{equation*}
for some $M_i,N_i\in\ts[\Xi,\Theta]{[A_1,\dotsc,A_n]}$
in order to show $M_i = N_i$.
We have
\begin{equation*}
p\, (F_1 M_1)\,\dotsb\,(F_n M_n)
\ \eqbe\ 
p\, (F_1 N_1)\,\dotsb\,(F_n N_n).
\end{equation*}
So we get $F_i M_i \eqbe F_i N_i$.
Thus $M_i = N_i$.

\item[\iref{L:atomic-ii}]
As easy as before.\qed
\end{enumerate}

%
%
\begin{lem}
\label{L:sr-ut}
Given a set of types  $\mathcal{A} \equiv \{A_1,\dotsc,A_n\}$ 
we have
\begin{equation*}
[\,\lbb C_1,\dotsc,C_k \rbb \,] \ \leqa \ [\,\lbb A_1 \rbb,\,\dotsc,\, \lbb A_n \rbb\, ]
\end{equation*}
for all $C_1,\dotsc,C_k \in \mathcal{A}$.
\end{lem}
\proof 
Writing $\Gamma\eqdf c_1^{C_1},\dotsc,c_k^{C_k}$
it suffices to prove that the substitution~$\varrho$ 
from the context~$\smash{F^{\lbb\Gamma\rbb}}$ 
to~$\Theta\eqdf\smash{F_{A_1}^{\lbb A_1 \rbb},\dotsc,F_{A_n}^{\lbb A_n \rbb}}$
given by the assignment
\begin{equation*}
\varrho_{F} \ \eqdf\ 
\lm{m^{[\Gamma]}}\  F_{C_1}\lm{c_1^{C_1}} \ \dotsb\  F_{C_k}\lm{c_k^{C_k}}
\ m\, \Gamma
\end{equation*}
is an atomic reduction
(see Remark~\ref{R:atomic-red}).
For this we use Lemma~\ref{L:atomic}.

\begin{enumerate}[label=(\roman*),ref=\roman*]
\item[\iref{L:atomic-i}]
Let~$\Xi$ be a context
and suppose that
\begin{equation*}
\varrho_F M \,\eqbe\, \varrho_F N
\end{equation*}
for some $M,N\in\ts[\Xi,\Theta]{[\Gamma]}$
in order to prove $M=N$.
We get
\begin{equation*}
F_{C_1}\lm{c_1^{C_1}} \ \dotsb\  F_{C_k}\lm{c_k^{C_k}}
\ M\, \Gamma
\ \eqbe\ 
F_{C_1}\lm{c_1^{C_1}} \ \dotsb\  F_{C_k}\lm{c_k^{C_k}}
\ N\, \Gamma.
\end{equation*}
Hence $M\Gamma\eqbe N\Gamma$, so $M=N$.

\item[\iref{L:atomic-ii}]
Again trivial.\qed
\end{enumerate}

\begin{cor}
\label{C:sr-ut}
For every type~$A$ and $k\geq 1$, we have
$[\lbb A^k \rbb ] \leqa [\lbb A \rbb]$.
\end{cor}
\proof 
Apply Lemma~\ref{L:sr-ut} with~$\mathcal{A}\eqdf \{A\}$ 
and $C_1,\dotsc,C_k \eqdf A,\dotsc, A$.\qed

%
%
\begin{lem}
\label{L:sr-rc}
Let~$A$ be a type. Then $[\lbb A \rbb ] \leqa [3,A]$.
\end{lem}
\proof 
Let~$\varrho$ be the substitution from~$\smash{F^{\lbb A \rbb}}$
to~$\Theta\eqdf \Phi^3,a^A$ given by
\begin{equation*}
\varrho_F \,\eqdf\, \lm{m^{[A]}} \Phi\lm{f^1} m(fa)\htam.
\end{equation*}
We prove that~$\varrho$ is an atomic reduction
(and thus $[\lbb A \rbb ] \leqa [A,3]$)
using Lemma~\ref{L:atomic}.
\begin{enumerate}[label=(\roman*),ref=\roman*]
\item[\iref{L:atomic-i}]
Let~$\Xi$ be a context and suppose that
\begin{equation*}
\varrho_F  M \,\eqbe\, \varrho_F  N
\end{equation*}
for some $M,N\in\ts[\Xi,\Theta]{[A]}$
in order to prove $M=N$.
We have
\begin{equation*}
\Phi\lm{f^1} M(fa)
\eqbe
\Phi\lm{f^1} N(fa)
\end{equation*}
and thus $M(fa)\eqbe N(fa)$.
Since $f$ does not occur in~$M$ and~$N$ we have
\begin{equation*}
M(R a)\ \eqbe N(Ra)
\end{equation*}
for every term~$R$ of type~$1$.
If we pick $R\eqdf \lm{x^0} a_1$ for fresh $a_1^0$, we get
\begin{equation*}
Ma_1 \eqbe Na_1.
\end{equation*}
So we see that $M=N$.

\item[\iref{L:atomic-ii}] Easy.\qed
\end{enumerate}

%
%
\noindent For the proof of Lemma~\ref{L:sr-rc2}
we need the following fact concerning terms.
\begin{lem}
\label{L:sr-rc2h}
Let~$E$ be a type and let~$c^E,d^E$ be variables.
Then we have
\begin{equation}
\label{eq:sr-rc2-1}
\left.
\begin{aligned}
M[x\subs c] \,&=\, N[x\subs c] \\
M[x\subs d] \,&=\, N[x\subs d]
\end{aligned}
\quad \right]
\qquad\implies\quad
M=N
\end{equation}
for all terms~$M,N$
(which might contain $c$ and~$d$).
\end{lem}
\proof 
Write $\Theta\eqdf x^E,c^E,d^E$.
Given a context--type $\ct{\Xi}C$
and $M\in\ts[\Xi,\Theta]{C}$
let $P(M)$ be  the property
that Statement~\eqref{eq:sr-rc2-1}
holds for~$M$ and any~$N\in\ts[\Xi,\Theta]{C}$.
With induction we prove that~$P(M)$
holds for every~$M$. This is sufficient.
\begin{enumerate}[label=(\roman*),ref=\roman*]
\item
Suppose that $M\equiv a\, M_1\dotsb M_n$
for some $a^A\in\{\Xi,\Theta\}$
with $A\equiv[A_1,\dotsc,A_n]$
and $M_i\in\ts[\Xi,\Theta]{A_i}$
with~$P(M_i)$
and let~$N\in\ts[\Xi,\Theta]{0}$
be such that
\begin{equation}
\label{eq:sr-rc2-2}
M[x\subs c]=N[x\subs c]
\qquad\text{and}\qquad
M[x\subs d]=N[x\subs d].
\end{equation}
Write $N\equiv b\, N_1\dotsb N_{m}$
where $b^B\in\{\Xi,\Theta\}$
with $B\equiv[B_1,\dotsc,B_m]$.
We need to prove that~$M=N$.
Note that Statement~\eqref{eq:sr-rc2-2}
implies that
\begin{equation}
\label{eq:sr-rc2-3}
a[x\subs c] \,=\, b [x\subs c]
\qquad\text{and}\qquad
a[x\subs d] \,=\, b[x\subs d].
\end{equation}
By examining
the different cases
for~$a$
(viz., $a=c$,\ $a=d$,\ $a=x$ and $a\in\{\Xi\}$)
and similarly for~$b$,
one easily sees that Statement~\eqref{eq:sr-rc2-3}
implies~$a=b$.

Statement~\eqref{eq:sr-rc2-2} also implies 
$M_i[x\subs c]\,=\,N_i[x\subs c]$ and 
$M_i[x\subs d]\,=\,N_i[x\subs d]$.

Consequently, $M_i = N_i$ as $P(M_i)$ by assumption.
Hence $M=N$.

\item Suppose that $M\equiv \lm{\Delta} M'$
for some~$M'\in\ts[\Delta,\Xi,\Theta]{0}$
with $P(M')$.
Assume that
\begin{equation}
\label{eq:sr-rc2-4}
M[x\subs c]=N[x\subs c]
\qquad\text{and}\qquad
M[x\subs d]=N[x\subs d]
\end{equation}
for some~$N\in\ts[\Xi,\Theta]{[\Delta]}$
in order to show that~$M=N$.
Writing $N\equiv \lm{\Delta}N'$
with $N'\in\ts[\Delta,\Xi,\Theta]0$,
we see that
Statement~\eqref{eq:sr-rc2-4} implies
that
\begin{equation*}
\lm{\Delta} M'[x\subs c] \,=\, \lm{\Delta} N'[x\subs c].
\end{equation*}
Hence $M'[x\subs c] = N'[x\subs c]$.
Similarly, we get
 $M'[x\subs d] = N'[x\subs d]$.

Then $P(M')$ implies $M'=N'$, so that $M=N$.\qed
\end{enumerate}

\begin{lem}
\label{L:sr-rc2}
For any type~$A$ we have
$[\lbb A \rbb ] \leqa [[0,0],A,A]$.
\end{lem}
\proof 
We need to find an atomic reduction
from~$\smash{F^{\lbb A \rbb }}$
to $\Theta\eqdf b^{[0,0]},c^A,d^A$
(see Remark~\ref{R:atomic-red}).
Let~$\varrho$
be the substitution from~$F$ to~$\Theta$ given by
\begin{equation*}
\varrho_F\,\eqdf\,
\lm{m^{[A]}} b\,(mc)(md).
\end{equation*}
We prove that~$\varrho$
is an atomic reduction using Lemma~\ref{L:atomic}.
\begin{enumerate}[label=(\roman*),ref=\roman*]
\item[\iref{L:atomic-i}]
Let $\Xi$ be a context 
and suppose that
\begin{equation*}
\varrho_F M\,\eqbe \varrho_F N
\end{equation*}
for some $M,N\in\ts[\Xi,\Theta]{[A]}$
in order to prove~$M=N$.
By reduction we get
\begin{equation*}
b\, (Mc)(Md)\,\eqbe\, b\,(Nc)(Nd).
\end{equation*}
Hence $Mc\eqbe Nc$, $Md\eqbe Nd$.
Writing $M\equiv \lm{x^A} M'$,
 $N\equiv\lm{x^A} N'$,
we get
\begin{equation*}
N'[x\subs c]\,=\,M'[x\subs c]
\qquad\text{and}\qquad
N'[x\subs d]\,=\,M'[x\subs d].
\end{equation*}
Hence $M'=N'$ by Lemma~\ref{L:sr-rc2h},
and thus $M=N$.

\item[\iref{L:atomic-ii}]
Simple.\qed
\end{enumerate}

%
%
\begin{lem}
\label{L:sr-dder}
Let $\Gamma$ be a context and
$F^A \in\{\Gamma\}$ with $A\equiv[[\Gamma_1],\dotsc,[\Gamma_n]]$.
Then 
\begin{equation*}
\Theta\leqa \Gamma,\Gamma_k\quad\implies\quad\Theta\leqa \Gamma
\end{equation*}
for every context~$\Theta$ and $k\in\{1,\dotsc,n\}$
such that
\begin{equation}
\label{eq:sr-dder}
[\Gamma_i,\Delta,\Gamma]\text{\  is inhabited}
\qquad\qquad\text{for all\ } \smash{t^{[\Delta]}}\in\{\Theta\},\ i\neq k.
\end{equation}
\end{lem}
\proof 
Assume that~$\Theta\leqa\Gamma,\Gamma_k$
for some~$\Theta$ and~$k$.
By Definition~\ref{D:ar},
there is an atomic reduction~$\varrho$
from~$\Theta$ to~$\Gamma,\Gamma_k$.
In order to prove that~$\Theta\leqa \Gamma$,
we need to find an atomic reduction from~$\Theta$ to~$\Gamma$.
Pick terms~$H^t_i \in \ts[\Gamma_i,\Delta,\Gamma]0$
for every~$i\neq k$ and $\smash{t^{[\Delta]}}\in\{\Theta\}$;
this is possible by Statement~\eqref{eq:sr-dder}.
Now, let~$\sigma$ be the substitution from~$\Theta$ to~$\Gamma$
given by
\begin{equation*}
\sigma_t\,\eqdf\,
\lm{\Delta} F\, M^t_1\dotsb M^t_n;
\qquad M^t_i\,\eqdf\,
\begin{cases}
\ \lm{\Gamma_k} \varrho_t \Delta & \text{ if } i=k \\
\ \lm{\Gamma_i} H^t_i & \text{ otherwise}
\end{cases}
\end{equation*}
for every $\smash{t^{[\Delta]}}\in\{\Theta\}$.
We use Lemma~\ref{L:atomic} to prove that~$\sigma$ is an atomic reduction.

\begin{enumerate}[label=(\roman*),ref=\roman*]
\item[\iref{L:atomic-i}]
Let~$s^S,t^T\in\{\Theta\}$ with $S\equiv [S_1,\dotsc, S_k]$
and $T\equiv[T_1,\dotsc, T_\ell]$.
Suppose
\begin{equation*}
\sigma_s\, U_1\dotsb U_k \,\eqbe\, \sigma_t\, V_1\dotsb V_\ell
\end{equation*}
for certain $U_i\in\ts[\Xi,\Gamma]{S_i}$, $V_i\in\ts[\Xi,\Gamma]{T_i}$
and some~$\Xi$,
to prove $U_i=V_i$.
Then
\begin{equation*}
F\, M^s_1[\Delta\subs\svec{U}]\dotsb M^s_n[\Delta\subs\svec{U}]
\ \eqbe\  
F\, M^t_1[\Delta\subs\svec{V}]\dotsb M^t_n[\Delta\subs\svec{V}].
\end{equation*}
Hence $M^s_i[\Delta\subs\svec{U}]\eqbe M^t_i[\Delta\subs\svec{V}]$.
For~$i=k$, we get
\begin{equation*}
\lm{\Gamma_k}\varrho_s\, U_1\dotsb U_k
\ \eqbe\ 
\lm{\Gamma_k}\varrho_t\, V_1\dotsb V_\ell.
\end{equation*}
Thus $\varrho_s\,U_1\dotsb U_k\eqbe \varrho_t\,V_1\dotsb V_\ell$.
Hence $s=t$ and $U_i=V_i$.

\item[\iref{L:atomic-ii}]
Trivial.\qed
\end{enumerate}

\begin{lem}
\label{L:sr-der}
Let $\Gamma$ be a context
and~$\Delta$ a derivative of~$\Gamma$
(see Definition~\ref{D:der}). Then
\begin{equation*}
\Theta\leqa \Delta 
\quad\implies\quad
\Theta\leqa\Gamma
\end{equation*}
for every context~$\Theta$
such that
\begin{equation}
\label{eq:sr-der}
[\Gamma,\Xi]\text{\  is inhabited}
\qquad\qquad\text{for all }t^{[\Xi]}\in\{\Theta\}.
\end{equation}
\end{lem}
\proof 
Let $\Gamma$ and $\Theta$ with $\Theta\leqa \Gamma$ be given
and suppose Statement~\eqref{eq:sr-der} holds.
We prove that
$\Theta\leqa \Delta$
for every derivative $\Delta$ of~$\Gamma$ with induction on~$\Delta$.

Let~$\Delta$ be a derivative of the context~$\Gamma$
and let~$\Delta'$ be a direct derivative of~$\Delta$.
Assume that $\Theta\leqa \Delta$.
We need to prove that~$\Theta\leqa \Delta'$.

By Definition~\ref{D:der},
$\Delta'\equiv\Delta,\Delta_k$ for some
 $F^A \in\{\Delta\}$
with $A\equiv[[\Delta_1],\dotsc,[\Delta_n]]$.
So we apply Lemma~\ref{L:sr-dder}
to prove $\Theta\leqa \Delta'$.
We must show that $[\Delta_i,\Xi,\Delta]$
is inhabited for every~$t^{[\Xi]}\in\{\Theta\}$ and $i\neq k$.

Since~$\Delta$ is a derivative of~$\Gamma$,
we have~$\{\Gamma\}\subseteq \{\Delta\}$.
Hence $\{\Gamma,\Xi\}\subseteq\{\Delta_i,\Xi,\Delta\}$.
So to prove that $[\Delta_i,\Xi,\Delta]$ is inhabited,
it suffices to show $[\Gamma,\Xi]$ is inhabited.
This is Statement~\eqref{eq:sr-der}.\qed

%
%
\begin{rem}
\label{R:ar-transitive}
Surprisingly,
it is not clear whether the relation~$\leqa$ is transitive.
\end{rem}
%
%
%
%
%
\subsection{Atomic types}
\label{SS:atomic-types}
We are interested in types~$A$ with the property
\begin{equation}
\label{eq:summing}
[\Gamma_1]\leqs A \quad\htam{ and }\quad [\Gamma_2]\leqs A
\quad\implies\quad [\Gamma_1,\Gamma_2]\leqs A\htam,
\end{equation}
as this property makes it easier
to find reductions to~$A$.
For instance,
to prove~$[3,0,0]\leqs A$,
it suffices to show that both $[3]\leqs A$ and $[0]\leqs A$.
In this subsection we give a criterion (namely $[1,1]\leqa A$)
for a type to satisfy Statement~\eqref{eq:summing}.

\label{SS:atomic}
\begin{defi}\label{D:a}
A context--type $\ct{\Gamma}{A}$
is \keyword{atomic} if
$[1,1]\leqa \ct{\Gamma}A$.
\end{defi}
\begin{rem}
\label{R:atomic-type}
A context--type $\ct{\Gamma}{[\Delta]}$
is atomic iff $\Gamma,\Delta$ is atomic by Remark~\ref{R:atomic-red}.\\
In particular,
a type~$[\Delta]$ is atomic iff the context~$\Delta$ is atomic.
\end{rem}
%
%
\begin{lem}
\label{L:atomic-pair}
A context~$\Theta$ is atomic iff there are terms $X_1,X_2\in\ts[\Theta]{1}$
with:
\begin{enumerate}[label=(\roman*),ref=\roman*]
\item\label{L:atomic-pair-i}
For every context~$\Xi$
and for all $M,N\in\ts[\Xi,\Theta]{0}$,
\begin{equation*}
X_i M \eqbe X_j N\quad\implies\quad  i=j\ \text{ and }\ M=N\htam.
\end{equation*}
\item\label{L:atomic-pair-ii}
For every context~$\Xi$
and all $d^D\in\{\Xi\}$
with $D\equiv[D_1,\dotsc,D_\ell]$,
\begin{equation*}
X_i M \ \neqbe\  d \, N_1 \dotsb N_\ell,
\end{equation*}
where $M\in\ts[\Xi,\Theta]{0}$ and $N_i\in\ts[\Xi,\Theta]{D_i}$.
\end{enumerate}
\end{lem}
\proof 
Simply expand Definition~\ref{D:ct}\iref{D:ct-sub-ctx}
in Lemma~\ref{L:atomic}.\qed

\begin{defi}
\label{D:atomic-pair}
Let~$\Theta$ be a context.
A pair of terms $X_1,X_2\in\ts[\Theta]{1}$
which satisfies conditions~\iref{L:atomic-pair-i} and~\iref{L:atomic-pair-ii}
of Lemma~\ref{L:atomic-pair}
will be called an \keyword{atomic pair}.
\end{defi}
%
%
\noindent Before we give some examples of atomic types,
we prove (as promised)
that an atomic type satisfies Statement~\eqref{eq:summing}.
The result is recorded in Corollary~\ref{C:A-summing}.
%
\begin{lem}
\label{sr-summing-lem}
Let $\Theta=t_1^{T_1},\dotsc,t_n^{T_n}$
be an atomic context.
Then $\Theta_1,\Theta_2 \leqa \Theta$,
where~$\Theta_i$ 
are clones of~$\Theta$,
defined by $\Theta_i\eqdf t_{i1}^{T_1},\dotsc t_{in}^{T_n}$.
\end{lem}
%
\proof 
Since $\Theta$ is atomic,
$[1,1]\leqa \Theta$ (see Definition~\ref{D:a}).
So there is an atomic reduction~$\varrho$
from the context $f^1_1,f^1_2$ to~$\Theta$
(see Remark~\ref{R:atomic-red}).
We need to find an atomic reduction~$\sigma$ 
from~$\Theta_1,\Theta_2\equiv t_{11},\dotsc,t_{1n},t_{21},\dotsc,t_{2n}$ 
to~$\Theta\equiv t_1,\dotsc,t_n$.
We do this by replacing $t_{ij}$ by $\varrho_{f_i} t_j$.
More formally,
write $T_i \equiv [\Gamma_i]$ and
define the substitution~$\sigma$ 
from $\Theta_1,\Theta_2$ to~$\Theta$ by
\begin{alignat*}{2}
\sigma_{t_{ij}} &\,\eqdf\, \lm{\Gamma_j}
\varrho_{f_i} t_j\, \Gamma_j&
\qquad& \htam{ for all }j\in\{1,\dotsc,n\},\ i\in\{1,2\}.
\end{alignat*}
We use Lemma~\ref{L:atomic} to
prove that $\sigma$ is an atomic reduction.
Let $\Xi$ be a context.
\begin{enumerate}[label=(\roman*),ref=\roman*]
\item[\iref{L:atomic-i}]
Let $t_{ij},t_{k\ell}\in \{\Theta_1,\Theta_2\}$ be given.
Suppose that 
\begin{equation*}
\sigma_{t_{ij}} \,\svec{M} \,\eqbe\, \sigma_{t_{k\ell}} \,\svec{N}
\end{equation*}
for some tuples $\svec{M}$ and $\svec{N}$
with free variables from~$\Xi,\Theta$
which fit in~$\Gamma_j$ and $\Gamma_\ell$, respectively
(see Definition~\ref{D:ctx}\iref{D:ctx-fits}).
We need to prove that 
$\svec{M}=\svec{N}$,
$i=k$ and  $j=\ell$.
If we expand the definition of $\sigma$, we get 
\begin{equation*}
\varrho_{f_i}\,t_j \svec{M} \,\eqbe\, \varrho_{f_k}\, t_\ell\svec{N}\htam.
\end{equation*}
Since $\varrho$ is an atomic reduction,
this implies $f_i = f_k$ (so $i=k$) and $t_j \svec{M} = t_\ell \svec{N}$.
The latter implies $t_j = t_\ell$ (so $j=\ell$) and $\svec{M}=\svec{N}$.
\item[\iref{L:atomic-ii}]
Let $t_{ij}\in\{\Theta_1,\Theta_2\}$
and $b^B\in\{\Theta\}$ with $B\equiv[B_1,\dotsc,B_m]$.
Assume
\begin{equation}
\label{eq:ewqewq}
\sigma_{t_{ij}} \,\svec{M} \, \eqbe \, b\,\svec{N} 
\end{equation}
for some tuples~$\svec{M}$ and $\svec{N}$ with free variables
from $\Xi,\Theta$ which fit on $\Gamma_j$ and~$\Delta$, respectively.
Equation~\eqref{eq:ewqewq} 
implies $b\svec{N} \eqbe \varrho_{f_i} t_j \svec{M}$.
On the other hand
we have 
$b\svec{N} \neqbe \varrho_{f_i} (t_j \svec{M})$
 as~$\varrho$ is an atomic reduction.
A contradiction.\qed
\end{enumerate}

%
\begin{prop}
\label{sr-summing}
\label{L:sr-summing}
Let $\Gamma$, $\Delta$ and $\Theta$
be contexts
and suppose~$\Theta$ is atomic.
Then
\begin{alignat}{6}
\label{eq:summing-ctx}
\Gamma&\leqs\Theta&  \ \text{ and }\  \Delta&\leqs\Theta&
\quad\implies\quad
\Gamma,\Delta&\leqs\Theta&\htam. 
\end{alignat}
\end{prop}
%
\proof 
Assume that $\Gamma\leqs \Theta$
and $\Delta\leqs \Theta$;
we must prove that $\Gamma,\Delta\leqs \Theta$.
Let $\Theta_1$ and $\Theta_2$ be clones of $\Theta$.
By Lemma~\ref{sr-summing-lem}
and Proposition~\ref{P:atomic-strong}
we see that we have
$\Theta_1,\Theta_2 \leqs \Theta$.
Further,
$\Gamma\leqs\Theta$ \& $\Delta\leqs\Theta$
implies $\Gamma\leqs\Theta_1$ \& $\Delta\leqs\Theta_2$.
So we see that
\begin{equation*}
\Gamma,\Delta
\,\leqs\,
\Theta_1,\Delta
\,\leqs\,
\Theta_1,\Theta_2
\,\leqs\,
\Theta
\end{equation*}
by Lemma~\ref{L:sr-glp}
and transitivity of~$\leqs$.\qed

\begin{cor}
\label{C:A-summing}
An atomic type~$A$ satisfies Statement~\eqref{eq:summing}.
\end{cor}
\proof 
Let $A\equiv[\Theta]$ be an atomic type.
Then~$\Theta$ is an atomic context (see Remark~\ref{R:atomic-type}).
Hence~$\Theta$ satisfies Statement~\eqref{eq:summing-ctx}
by Proposition~\ref{L:sr-summing}.
But then the type $A=[\Theta]$ satisfies Statement~\eqref{eq:summing}
because of Definition~\ref{D:s}\iref{D:s-red}.\qed

\noindent Atomic types are quite common;
in fact, 
we will spend the remainder of this section
showing that a type~$A$ is atomic
if it is from
$\HH{\omega+2}$ (see Corollary~\ref{C:a-w2}), 
$\HH{\omega+3}$ (Corollary~\ref{C:a-w3})
or $\HH{\omega+4}$ (Corollary~\ref{C:a-w4}).
%
%
\begin{lem}
\label{L:a-e}
Let~$\Gamma$, $\Delta$ be contexts
with $\{\Gamma\}\subseteq \{\Delta\}$.
We have
\begin{enumerate}[label=(\roman*),ref=\roman*]
\item
\label{L:a-e-i}
$\Gamma\text{ is atomic}\implies\Delta\text{ is atomic}$,
\item 
\label{L:a-e-ii}
$\Theta \leqa \Gamma\implies\Theta\leqa \Delta$
for every context~$\Theta$.
\end{enumerate}
\end{lem}
\proof 
By expanding Definition~\ref{D:a} and using Remark~\ref{R:atomic-red},
one easily sees that part~\iref{L:a-e-i} is a special case of~\iref{L:a-e-ii}.
Let us prove part~\iref{L:a-e-ii}.

Let~$\Theta$ be a context
with $\Theta\leqa\Gamma$.
We need to prove~$\Theta\leqa\Gamma$.
That is,
we need to find an atomic reduction~$\varrho$
from~$\Theta$ to~$\Delta$.
We know there is a substitution~$\varrho$
from~$\Theta$ to~$\Gamma$
which is an atomic reduction.
Since~$\{\Gamma\}\subseteq\{\Delta\}$,
the map~$\varrho$
can be considered a substitution from~$\Theta$ to~$\Delta$
(see Definition~\ref{D:ct}\iref{D:ct-sub-ctx}).
We prove that~$\varrho$ is an atomic reduction from~$\Theta$ to~$\Delta$.

Let $a^A,b^B\in\{\Xi,\Theta\}$
with $A\equiv[A_1,\dotsc,A_n]$
and $B\equiv[B_1,\dotsc,B_m]$
be given
where~$\Xi$ is some context.
We need to show that
\begin{equation}
\label{eq:a-e_1}
\varrho_a\,M_1\dotsb M_n \,\eqbe\, \varrho_b\,N_1\dotsb N_m
\quad\implies\quad
a=b\quad\htam{ and }\quad M_i = N_i
\end{equation}
for all~$M_i\in\ts[\Xi,\Delta]{A_i}$
and~$N_i\in\ts[\Xi,\Delta]{B_i}$.

Let us shorten
``Statement~\eqref{eq:a-e_1} holds
for~$M_i\in\ts[\Xi_0]{A_i}$
and~$N_i\in\ts[\Xi_0]{B_i}$''
to ``\eqref{eq:a-e_1} holds for~$\Xi_0$''.
We need to prove that \eqref{eq:a-e_1} holds for~$\Xi,\Delta$.

Recall that $\{\Gamma\}\subseteq\{\Delta\}$.
Pick a context~$\Gamma^c$ such that~$\{\Gamma^c,\Gamma\} = \{\Delta\}$.
Then $\{\Xi,\Delta\}= \{\Xi,\Gamma^c,\Gamma\}$.
Thus $\ts[\Xi,\Delta]{C} = \ts[\Xi,\Gamma^c,\Gamma]{C}$
for all types~$C$.
Hence 
to prove \eqref{eq:a-e_1} holds for~$\Xi,\Delta$,
it suffices to show that \eqref{eq:a-e_1} 
holds for~$\Xi,\Gamma^c,\Gamma$.

Thus,
writing $\Xi'\eqdf \Xi,\Gamma^c$,
we need to prove that \eqref{eq:a-e_1} holds for~$\Xi',\Gamma$.
Since we have $a,b\in\{\Xi,\Theta\}\subseteq\{\Xi',\Theta\}$,
this follows immediately from the fact that
$\varrho$ is an atomic reduction from~$\Theta$ to~$\Gamma$.\qed

%
%
\begin{lem}
\label{L:rap-wide}
Let~$A_1$ and~$A_2$ be types.
Then $[[A_1],[A_2]]$ is atomic.
\end{lem}
\proof 
By Remark~\ref{R:atomic-type}
we must to show  
that~$\Theta\eqdf\smash{F_1^{[A_1]},F_2^{[A_2]}}$ is atomic.
To this end,
we apply Lemma~\ref{L:atomic-pair}.
Writing $A_i=[\Gamma_i]$, define
\begin{align*}
X_i &\eqdf \lm{z^0} F_i\lm{\Gamma_i} z\htam.
\end{align*}
We prove that~$X_1,X_2$ is an atomic pair (see Definition~\ref{D:atomic-pair}),
i.e.,
that the terms $X_1,X_2$ satisfy conditions~\iref{L:atomic-pair-i}
and~\iref{L:atomic-pair-ii} of Lemma~\ref{L:atomic-pair}.
Let~$\Xi$ be a context.

\begin{enumerate}[label=(\roman*),ref=\roman*]
\item[\iref{L:atomic-pair-i}]
Assume $X_i M \eqbe X_j N$ for some $M,N\in\ts[\Xi,\Theta]{0}$
in order to show
that $M=N$ and $i=j$.
By reduction, we get an equality between lnfs,
\begin{equation*}
	F_i \lm{\Gamma_i} M \,=\, F_j \lm{\Gamma_j} N\htam{.}
\end{equation*}
Hence  $M=N$ and $F_i=F_j$.
The latter implies $i=j$ as $F_1\neq F_2$.

\item[\iref{L:atomic-pair-ii}]
Trivial.  Indeed,
if $X_i M \eqbe d\svec{N}$
for appropriate $M$, $d$ and $\svec{N}$,
then 
\begin{equation*}
F_i \lm{\Gamma_i} M \,=\, d\svec{N},
\end{equation*}
so $F_i = d$, which is absurd.\qed
\end{enumerate}

\begin{cor}
\label{C:a-w2}
If~$A\in\HH{\omega+2}$ then~$A$ is atomic.
\end{cor}
\proof 
Writing~$A=[\Delta]$,
we need to prove that~$\Delta$ is atomic
(see Remark~\ref{R:atomic-type}).
We claim there is a context~$\Gamma\equiv \smash{f_1^{[B_1]},f_2^{[B_2]}}$
such that $\{\Gamma\}\subseteq\{\Delta\}$.
Then 
since $[[B_1],[B_2]]$ (and thus~$\Gamma$) is atomic by Lemma~\ref{L:rap-wide},
we know that~$\Delta$ (and thus~$A$) 
is atomic by Lemma~\ref{L:a-e}\iref{L:a-e-i}.

To ground the claim,
it suffices to find two components of~$A$ 
of the form~$[B]$.
Since $A\in\HH{\omega+2}$,
we know that~$A$ is small and has at least two components~$C_1,C_2$
with~$\rk(C_i)\geq1$  (see Theorem~\ref{T:Hierarchy}).
Since $\rk(C_i)\geq 1$,
the type~$C_i$ must have at least one component.
Also~$C_i$ has at most one component
since $C_i$ is not fat as~$A$ is small
(see Definition~\ref{D:e}\iref{D:e-small}).
So we see that~$C_i\equiv[B_i]$ for some type~$B_i$.\qed

%
%
\noindent To prove that all types $A\in\HH{\omega+4}$ are atomic,
we need two lemmas.
\begin{lem}
\label{L:large-pairing}
If~$A$ is a large type (see Definition~\ref{D:e}\iref{D:e-large}),
 then $[[0,0]]\leqa A$.
\end{lem}
\proof 
Write $A\equiv[\Gamma]$.
It suffices to prove that $b^{[0,0]}\leqa \Gamma$
(see Remark~\ref{R:atomic-red}).

One can verify that
since
$A$ is large 
there is derivative 
$\Delta$ of~$\Gamma$
and ~$p^P\in\{\Delta\}$
such that $P$ is fat
(see Definition~\ref{D:der}).
Further,
note that  $[\Gamma,0,0]$ is inhabited.
Hence to prove~$b\leqa \Gamma$,
it suffices
to show that $b\leqa \Delta$
by Lemma~\ref{L:sr-der}.

Since~$P$ is fat 
$P\equiv[[\Gamma_1],\dotsb,[\Gamma_k]]$
with~$k\geq 1$ (see Definition~\ref{D:e}\iref{D:e-fat}).
Define
\begin{equation*}
\varrho_b\ \eqdf\  \lm{x^0y^0} 
    p \,(\lm{\Gamma_1} x) \ (\lm{\Gamma_2} y) \dotsb (\lm{\Gamma_k} y)\htam.
\end{equation*}
Then $\varrho_b\in\ts[\Delta]{[0,0]}$
yields a substitution~$\varrho$ from~$b$ to~$\Delta$.
We prove that~$\varrho$ is an atomic reduction
(and thus $b\leqa \Delta$) using Lemma~\ref{L:atomic}.
\begin{enumerate}[label=(\roman*),ref=\roman*]
\item[\iref{L:atomic-i}]
Given a context $\Xi$  and
$M_i\in\ts[\Xi,\Delta]{0}$ and
$N_i\in\ts[\Xi,\Delta]{0}$ with
\begin{equation*}
\varrho_b\, M_1 M_2 \,\eqbe\, \varrho_b\,N_1 N_2
\end{equation*}
we need to prove that $M_i = N_i$.
By reduction we get
\begin{equation*}
    p \,(\lm{\Gamma_1} M_1) \ 
             (\lm{\Gamma_2} M_2) \dotsb (\lm{\Gamma_k} M_2)
  \,=\,
    p \,(\lm{\Gamma_1} N_1) \ (\lm{\Gamma_2} N_2) \dotsb (\lm{\Gamma_k} N_2).
\end{equation*}
Hence $M_1=N_1$ and $M_2=N_2$.

\item[\iref{L:atomic-ii}]
Simple as before.\qed
\end{enumerate}

\begin{lem}
\label{L:rap-fat}
Let~$A$ be a type such that
 $[[0,0]]\leqa A$.
Then $A$ is atomic.
\end{lem}
\proof 
Write~$A\equiv[\Theta]$.
We need to prove that~$\Theta$ is atomic
(see Remark~\ref{R:atomic-type}).
We will define a pair~$X_1,X_2\in\ts[\Theta]{1}$
and show it is atomic (see Definition~\ref{D:atomic-pair}).

Since $[[0,0]]\leq A$,
there is an atomic reduction~$\varrho$
from~$b^{[0,0]}$ to~$\Theta$.
Define 
\begin{equation*}
\mathsf{s} \,\eqdf\, \lm{x^0} \varrho_b\,xx.
\end{equation*}
Let~$\Xi$ be a context.
Given $M,N\in\ts[\Xi,\Theta]0$,
we have 
\begin{equation}
\label{eq:rap-fat-1}
\mathsf{s} M\,\eqbe\, \mathsf{s} N\quad \implies\quad M=N.
\end{equation}
Moreover,
we claim that $\mathsf{s} M \neqbe M$ for all~$M\in\ts[\Xi,\Theta]0$.

To prove the claim,
write $\mathsf{s}\equiv\lm{x^0} S$
for some $S\in\ts[x,\Theta]{0}$.
Note that either~$x$ occurs in~$S$ or not,
and if~$x$ does not occur in~$S$
then $\mathsf{s}M\eqbe\mathsf{s}N$ for all~$N,M$,
which contradicts Statement~\eqref{eq:rap-fat-1}.
Hence $x$ occurs in~$S$.

Now, let~$M\in\ts[\Xi,\Theta]{0}$ be given;
we prove $\mathsf{s} M\neqbe M$.
Recall that we consider all terms to be in long normal form.
In particular, $S$ is in lnf.
Note that 
if we replace~$x$ in~$S$ with~$M$,
the resulting term is immediately
in long normal form---no reduction is needed.
Hence if $S\neq x$, we see that
$M$ is a strict subterm of~$S[x\subs M]= M$,
which is absurd.
So $S\equiv x$ and thus $\mathsf{s}= \lm{x^0} x$.

This is also absurd.
Indeed, we get $\varrho_b\, d d\eqb \mathsf{s}\, d \eqbe d$
for any fresh variable~$d^0$,
which contradicts that~$\varrho$ is an atomic reduction.

Now that we know $\mathsf{s}M\neqbe M$
for all~$M\in\ts[\Xi,\Theta]{0}$,
cunningly define
\begin{align*}
X_1 &\eqdf \lm{x^0} \mathsf{b}\,xx\htam;&
X_2 &\eqdf \lm{x^0} \mathsf{b}\,x(\mathsf{s}x)\htam.
\end{align*}
Then $X_i\in\ts[\Theta]1$.
We prove $X_1,X_2$
satisfies \iref{L:atomic-pair-i}
and~\iref{L:atomic-pair-ii}
of Lemma~\ref{L:atomic-pair}.

\begin{enumerate}[label=(\roman*),ref=\roman*]
\item[\iref{L:atomic-pair-i}]
Given~$M,N\in\ts[\Xi,\Theta]0$,
assume $X_i M \eqbe X_j N$
to show $i=j\ \&\ M=N$.
We distinguish three cases.
\begin{enumerate}[label=(\roman*),ref=\roman*]
\item
$X_1 M \eqbe X_1 N$.\ 
Then $\mathsf{b}\,M M \eqbe \mathsf{b}\,NN$,
so $M = N$.
\item
$X_2 M \eqbe X_2 N$.\ 
Then $\mathsf{b}\,M (\mathsf{s} M)\eqbe \mathsf{b}\,N(\mathsf{s}N)$,
so $M = N$.
\item
$X_1 M \eqbe X_2 N$.\ 
Then $\mathsf{b}\,M M\eqbe \mathsf{b}\, N(\mathsf{s} N)$.
So we have both $M = N$ and $M \eqbe \mathsf{s} N$.
Consequently, $N \eqbe \mathsf{s} N$, which is absurd.
\end{enumerate}
\item[\iref{L:atomic-pair-ii}]
As simple as before.\qed
\end{enumerate}

%
%
\begin{cor}
\label{C:rap-large}
Each large type $A$ is atomic.
\end{cor}
\proof 
Combine Lemma~\ref{L:rap-fat} and Lemma~\ref{L:large-pairing}.
\qed

\begin{cor}
\label{C:a-w4}
If~$A\in\HH{\omega+4}$ then~$A$ is atomic.
\end{cor}
\proof 
Since~$A$ is large by definition
of~$\HH{\omega+4}$,
$A$ is atomic by Corollary~\ref{C:rap-large}.\qed

%
%
%
%
\begin{lem}
\label{L:a-der}
A context~$\Gamma$ is atomic
if one of its derivatives~$\Delta$ is atomic.
\end{lem}
\proof 
Follows from Lemma~\ref{L:sr-der}
as the type $[\Gamma,0]$ is inhabited.\qed

%
%
\begin{lem}
\label{L:rap-tall}
Let~$A$ be a small type with~$\rk(A)\geq 4$.
Then~$A$ is atomic.
\end{lem}
\proof 
There is a component~$B$ of~$A$
such that~$\rk(B)\geq 3$
(see Definition~\ref{D:e}\iref{D:e-rank}).
In other words,
writing~$A=[\Gamma]$,
there is an~$b^B\in\{\Gamma\}$ such that~$\rk(B)\geq 3$.
Similarly,
if we write $B\equiv\lbb\Theta\rbb$ for some~$\Theta$
(recall that~$A$ is small),
then there must be a~$c^C\in \{\Theta\}$ with~$\rk(C)\geq 1$.
So~$C\equiv[D]$ for some~$D$.

Note that $\Gamma,\Theta$
is a direct derivative of~$\Gamma$,
so to prove~$A$ is atomic,
it suffices to show that~$\Gamma,\Theta$ is atomic
(by Lemma~\ref{L:a-der}).
To this end, 
consider the context~$\Xi\eqdf b^B,c^C$.
We have $\{\Xi\}\subseteq \{\Gamma,\Theta\}$
and $[\Xi] = [\lbb\Theta\rbb,[D]]$,
so~$\Xi$ is  atomic by Lemma~\ref{L:rap-wide}
and hence $\Gamma,\Theta$ is atomic by Lemma~\ref{L:a-e}\iref{L:a-e-i}.\qed

\begin{cor}
\label{C:a-w3}
If~$A\in\HH{\omega+3}$ then~$A$ is atomic.
\end{cor}
\proof 
Since~$A$ is small and~$\rk(A)\geq 4$ by definition,
$A$ is atomic by Lemma~\ref{L:rap-tall}.\qed

%% file: part23.tex
\section{\texorpdfstring%
	{Order type of $\leqh$}
	{Order type of <=h}}
\label{S:part23}
The order type of
the reducibility relation~$\leqh$ 
(see Definition~\ref{defn:reductions}\iref{defn:reductions-h}) is~$\omega+5$.
At least,
this is what is shown in Subsection~\ref{SS-sketch-of-the-proof}
using statements promised to be proven later on.
In this section,
we deliver on these promises;
they are  $H_\alpha \leqh \HH\alpha$ (Subsection~\ref{S:part2}),
$\HH\alpha\leqh H_\alpha$ (Subsection~\ref{S:part3}),
and $\alpha\leq\beta\implies H_\alpha \leqh H_\beta$ 
(Subsection~\ref{SS:HaleqHb}).

We refer the reader to Theorem~\ref{T:Hierarchy}
for the definition of~$H_\alpha$ and~$\HH\alpha$.
\subsection{\texorpdfstring%
	{$H_\alpha \leqh \HH\alpha$}
	{Ha reduces to HHa}}
	\label{S:part2}
Let us begin with a harvest.
We use the theory of strong reductions and atomic types
to easily prove that  $H_\alpha\leqs A$
for all~$A\in\HH\alpha$ and~$\alpha\in\omega+5$.
Loosely stated,
we do this by recognizing
the tree of $H_\alpha$ as part of the tree of~$A$
(see Subsection~\ref{S:structure}).

Recall that an atomic reduction is also a strong reduction,
so for example $H_\alpha \leqa A$
implies $H_\alpha \leqs A$
(see Proposition~\ref{P:atomic-strong}).
We use this fact without further mention.
%
%
\begin{lem}
$H_0 \leqs A$ for all~$A\in\HH{0}$.
\end{lem}
\proof 
We need to prove that~$0\leqs A$
whenever~$A$ is uninhabited.
We will prove~$0\leqs A$ for all types~$A$.
Writing~$A\equiv[\Gamma]$,
we need to prove~$[\ectx] \leqs [\Gamma]$.
So it suffices to show that~$\ectx \leqs \Gamma$
(see Definition~\ref{D:s}\iref{D:s-red}).
This follows immediately from Lemma~\ref{L:ar-glp}\iref{L:ar-glp-i}.\qed

%
%
\begin{lem}
$H_n\leqs A$ for all $A\in\HH{n}$ where~$n\geq 1$.
\end{lem}
\proof 
Trivial, since $\HH{n}=\{H_n\}$ for each $n\in\mathbb{N}$.\qed

%
%
\begin{lem}
\label{L:Hw_leqs_HHw}
$H_\omega \leqs A$ for each $A\in\HH{\omega}$.
\end{lem}
\proof 
We need to prove that~$[1,0]\leqs A$.
Recall that since~$A\in\HH{\omega}$, 
we have $A$ is small, $\rk(A)=2$ and $A$ has
exactly one component of rank~$1$.
So precisely one of the components of~$A$ is~$1$;
the remaining components are~$0$.
By a permutation of the components we get $A\iss [1,0^k]$ for some~$k$
(see Corollary~\ref{L:sr-glt}).
Hence it suffices to prove that~$[1,0]\leqs[1,0^k]$.
This follows from Lemma~\ref{L:ar-glp}\iref{L:ar-glp-i}.\qed

%
%
\begin{lem}
$H_{\omega+1}\leqs A$ for every $A\in\HH{\omega+1}$.
\end{lem}
\proof 
We need to prove that~$[2]\leqs A$.
Note that~$A$ is small, $\rk(A)=3$
and~$A$ has exactly one component of rank~$\geq 1$.
So one of the components of~$A$ is of the form~$\lbb 0^{\ell} \rbb$
where~$\ell\geq 1$ and
the remaining components are~$0$.
Hence 
$A\iss [\lbb 0^{\ell+1} \rbb, 0^k]$ by Corollary~\ref{L:sr-glt}.
So it suffices to prove that~$[2]\leqs[\lbb 0^{\ell} \rbb, 0^k]$.

Since $[0]\leqs [0^{\ell}]$ by Lemma~\ref{L:ar-glp}\iref{L:ar-glp-i},
we have
\begin{equation*}
[2]\equiv[\lbb 0 \rbb ] 
\,\leqs\, [\lbb 0^\ell \rbb]
\,\leqs\, [\lbb 0^\ell \rbb, 0^k]
\end{equation*}
by Lemma~\ref{L:sr-ci} and Lemma~\ref{L:ar-glp}\iref{L:ar-glp-i},
respectively.\qed

%
%
\begin{lem}
$H_{\omega+2}\leqs A$ for every $A\in\HH{\omega+2}$.
\end{lem}
\proof 
Note that $A$ is small and has at least two components of rank $\geq 1$,
so
after a permutation of~$A$s components
we get $A \iss B\eqdf[\lbb\Delta_1\rbb,\lbb\Delta_2\rbb,\Gamma]$
for some contexts $\Delta_1$, $\Delta_2$ and~$\Gamma$.
We need to show that $[1,1,0]\leqs B$.
Since~$B$ is atomic by Corollary~\ref{C:a-w2},
it suffices to prove $[0]\leqs B$ and~$[1]\leqs B$
(see  Proposition~\ref{L:sr-summing}).

We have~$[0]\leqs A \iss B$ since~$A$ is inhabited.
Concerning $[1]\leqs B$,
note that $0\leqs [\Delta_1]$ by Lemma~\ref{L:ar-glp}\iref{L:ar-glp-i}
and so $[1]\equiv[[0]]\leqs [\lbb \Delta_1 \rbb ] \leqs B$
by Lemma~\ref{L:sr-ci} and Lemma~\ref{L:ar-glp}\iref{L:ar-glp-i}.\qed

%
%
\begin{lem}
$H_{\omega+3}\leqs A$ for every $A\in\HH{\omega+3}$.
\end{lem}
\proof 
We need to prove that~$[3,0]\leqs A$.
By Proposition~\ref{L:sr-summing}
it suffices to show that $[3]\leqs A$ and $[0]\leqs A$
since~$A$ is atomic by Corollary~\ref{C:a-w3}.

As $A$ is inhabited, $[0]\leqs A$ is trivial.

Concerning $[3]\leqs A$.
Since~$A$ is small and $\rk(A)\geq 4$,
there is a component~$[A_1]$ of~$A$ with~$\rk(A_1)\geq 2$.
Then $[[A_1]]\leqs A$ by Lemma~\ref{L:ar-glp}\iref{L:ar-glp-i},
so it suffices to show $[[[1]]]\equiv [3]\leqs [[A_1]]$.
By Lemma~\ref{L:sr-ci} it is enough to prove that $[1]\leqs A_1$.

By similar reasoning for~$A_1$, 
we are left with the problem to prove
$0\leqs A_2$ where~$[A_2]$ is some component of~$A_1$.
Lemma~\ref{L:ar-glp}\iref{L:ar-glp-i} gives the solution.\qed

%
%
\begin{lem}
$H_{\omega+4}\leqs A$ for every $A\in\HH{\omega+4}$.
\end{lem}
\proof 
We need to prove that~$[[0,0],0]\leqs A$.
Since~$A$ is atomic by
Corollary~\ref{C:a-w4}
it suffices to show by Proposition~\ref{L:sr-summing}
that~$[0]\leqs A$ and~$[[0,0]]\leqs A$.
The former inequality is trivial since~$A$ is inhabited.
The latter is Lemma~\ref{L:large-pairing}.\qed

\subsection{\texorpdfstring%
	{$\HH\alpha \leqh H_\alpha$}
	{HHa reduces to Ha}}
	\label{S:part3}
In this subsection
we prove that $A\leqh H_\alpha$ for all~$\alpha\in \omega+5$
and~$A\in\HH\alpha$.
(In fact,
we show that $A\leqs H_\alpha$ for all~$\alpha\neq 0$.)
This is more difficult than proving $H_\alpha \leqs A$
(which involved only `chopping'),
as it requires
the \scare{encoding} of the inhabitants
of~$A$ using the simpler inhabitants of~$H_\alpha$. 

\subsubsection{Ad  \texorpdfstring{$0,\ldots,\omega$ and $\omega+1$}
			       {0, ..., w, w+1}}

\begin{lem} \label{L:part3-bit0}
$A\leqh H_0$ for all $A\in\HH{0}$.
\end{lem}
\proof 
We need to prove that
$A\leqh 0$.
Since~$A$ is uninhabited (by definition of~$\HH0$),
all the components of~$A$ are inhabited by Theorem~\ref{T:inhabitation}.
Write $A=[\Gamma]$
and pick for each~$b^B\in\{\Gamma\}$
an inhabitant~$N_b$ of~$B$.
Then $\varrho_{b}\eqdf N_b$
yields a substitution~$\varrho$ from~$A$ to~$0$.
For a rather dull reason
the map~$\hat\varrho\colon \ts{A}\ra\ts{0}$ is injective:
$\ts{A}$ is empty.
Hence $A\leqh 0$.\qed

\begin{lem} \label{L:part3-bitsw}
$A \leqs H_k$ for all~$A\in\HH{k}$ where $k>0$.
\end{lem}
\proof 
Trivial, since~$\HH{k} = \{ H_k\}$.\qed

\begin{lem} \label{L:part3-bitw}
$A \leqs H_\omega$ for all~$A\in\HH{\omega}$.
\end{lem}
\proof 
Again we have $A\iss [1,0^k]$ for some~$k>0$
(see the proof of Lemma~\ref{L:Hw_leqs_HHw}).
So we need to prove that~$[1,0^k]\leqs[1,0]$.
By Definition~\ref{D:s}\iref{D:s-red},
it suffices to show that
\begin{equation*}
f^1,d_1^0,\dotsc,d^0_k\,\leqs\, f^1,c^0.
\end{equation*}
The terms
\begin{gather*}
\varrho_f  \eqdf f^{(k)}, \qquad
\varrho_{d_{i+1}} \eqdf f^{(i)}c \htam.
\end{gather*}
constitute a substitution from~$f^1,d^0_1,\dotsc,d_k^0$
to~$f^1,c^0$.
It suffices to prove that~$\varrho$
is an atomic reduction
(see Proposition~\ref{P:atomic-strong}).
Let~$\Xi$ be a context.
Note that for all 
$M,N\in\ts[\Xi,f,c]{0}$,
\begin{equation*}
\label{eq:part3-bitw}
\begin{alignedat}{6}
\varrho_{d_i} & \eqbe && \varrho_{d_j} &&\implies\quad  &i&=j, \\
\varrho_f M & \eqbe && \varrho_f N \quad&&\implies \quad &M&=N, \\
\varrho_f M &\neqbe\  && \varrho_{d_i}.
\end{alignedat}
\end{equation*}
Hence condition~\iref{L:atomic-i} of Lemma~\ref{L:atomic} is met.
Since the other condition can be easily verified,
Lemma~\ref{L:atomic} implies that~$\varrho$ is an atomic reduction.\qed

\noindent
Before we proceed to~``ad $\omega+1$'',
we need a lemma.
\begin{lem} \label{L:part3-0k-leqs-2}
$[\lbb 0^k \rbb ] \leqs [\lbb0\rbb]\equiv [2]$
for all~$k>0$.
\end{lem}
\proof 
Apply Corollary~\ref{C:sr-ut} with~$A\eqdf 0$.\qed

\begin{lem}
\label{L:sametrick}
Given $A\in\HH{\omega+1}$, we have $A \leqs H_{\omega+1}$.
\end{lem}
\proof 
Let $A\in\HH{\omega+1}$ be given.
We need to prove that~$A\leqs [2]$.
One can easily verify that $A\iss [\lbb0^k\rbb,0^l]$
for some $k\geq 1$ and $l\geq0$. 
By Lemma~\ref{L:part3-0k-leqs-2},
we have $[\lbb0^k\rbb,0^l] \leqs [2,0^k]$.
So it suffices to show that $[2,0^k]\leqs [2]$.
To this end 
note that the terms
\begin{align*}
\varrho_F & \eqdf 
   \lm{f^1} F \lm{z^0} F \lm{x_1^0} \cdots F \lm{x_{l}^0}
				fz \htam; \\
\varrho_{c_i} & 
   \eqdf F \lm{x_1^0} \cdots F \lm{x_i^0} x_1 \htam{.}
\end{align*}
give a strong reduction from $[F^2,c_1^0,\dotsc,c_k^0]$
to~$[2]$
(cf. Lemma~\ref{L:part3-bitw}).
\qed

\subsubsection{Ad \texorpdfstring{$\omega+2$}{w+2}}
We need to prove that~$A\leqs [1,1,0]$
for all~$A\in\HH{\omega+2}$.
With atomicity,
we easily reduce the problem to showing that $[2]\leqs[1,1,0]$
(see Lemma~\ref{L:2-leq-110}).
Interestingly,
we have $[2]\nleqa [1,1,0]$ (see Lemma~\ref{L:2-nleqa-110}).
Consequently,
the proof of $[2]\leqs [1,1,0]$
has quite a unique flavor (see Proposition~\ref{sr-2-to-1-1-0}).

\begin{lem}
\label{L:2-leq-110}
Suppose that $[2]\leqs[1,1,0]$.
Then $A \leqs [1,1,0]$
for all~$A\in\HH{\omega+2}$.
\end{lem}

\proof 
Let $A\in\HH{\omega+2}$ be given.
Then $A$ is small and~$\rk(A)\leq 3$,
so 
\begin{equation*}
A\ \iss\ 
 [\lbb0^{k_1}\rbb, \ldots, \lbb0^{k_n}\rbb, 1^l, 0^m] 
\end{equation*}
for some~$k_i,n,m,\ell$ 
by Corollary~\ref{L:sr-glt}.
We need to prove that $A\leqs[1,1,0]$.
By Lemma~\ref{L:part3-0k-leqs-2},
we have $[\lbb \smash{0^{k_i}} \rbb ] \leqs [2]$,
and so $A\leqs [2^n, 1^\ell, 0^m]$
by Lemma~\ref{L:sr-glp}.
Hence it suffices to show that $[2^n,1^\ell,0^m]\leqs[1,1,0]$
by transitivity.
Since~$[1,1,0]$ is atomic
by Lemma~\ref{L:rap-wide},
we  apply Proposition~\ref{sr-summing}.
It remains to be shown that
\begin{equation*}
[2]\leqs[1,1,0];\qquad[1]\leqs [1,1,0];\qquad[0]\leqs [1,1,0].
\end{equation*}
The first statement is valid by assumption,
the latter two by Lemma~\ref{L:ar-glp}\iref{L:ar-glp-i}.\qed

\begin{lem}
\label{L:2-nleqa-110}
We have $[2]\nleqa [1,1,0]$.
\end{lem}
\proof 
Let~$\varrho$ be a substitution
from~$F^2$ to~$\Theta\eqdf f^1,g^1,c^0$.
In order to show that $[2]\nleqa [1,1,0]$,
we prove that~$\varrho$ is not atomic
by finding terms $M,N\in\ts[\Theta]{1}$
with $\varrho_F M \eqbe \varrho_F N$
while $M\neq N$ (see Definition~\ref{D:ar}).
Write
\begin{equation*}
\varrho_F \,\equiv\,
\lm{h^1} w_0 \,h w_1 \dotsb h w_n c
\end{equation*}
where $w_i \in \ts[\Theta]{1}$
and define
$M\eqdf \lm{x^0}x$
and $N\eqdf \lm{x^0} w_1\dotsb w_n c$.
Then 
\begin{alignat*}{3}
\varrho_F M 
\,&\eqb\,
w_0\, M w_1  \dotsb M w_n c\\
\,&\eqb\,
w_0\, w_1 \dotsb w_n c \\
\,&\eqb\,
w_0\, N K &&\qquad \text{for all }K\in\ts[\Theta]0\\
\,&\eqb\, 
w_0\, N (w_1 \dotsb N w_n c) \ \eqb\, \varrho_F N,
\end{alignat*}
while $N\neq M$.\qed

\begin{prop} \label{sr-2-to-1-1-0}
We have $[2]\leqs[1,1,0]$.
\end{prop}
\proof 
It suffices to prove that $F^2\leqs f^1,g^1,c^0$
(see Definition~\ref{D:s}\iref{D:s-red}).
For this
we need to find a substitution~$\varrho$
from~$F^2$ to~$\Theta\eqdf f^1,g^1,c^0$
such that the map 
\begin{equation*}
\varrho^\Xi\colon \ts[\Xi,F^2]{0}\ra \ts[\Xi,\Theta]0
\end{equation*}
is injective for every context~$\Xi$
(see Proposition~\ref{P:strongsubs}).
The assignment
\begin{equation*}
\varrho_F \,\eqdf\, \lm{h^1} fhghc
\end{equation*}
gives a substitution~$\varrho$ from~$F^2$ to~$\Theta$.
We prove that $\varrho^\Xi$ is injective
for given context~$\Xi$.

Let us examine $\hat\varrho^\Xi$,
informally.
Occurrences of~$F\lm{x^0}M$
are recursively replaced by~$fM[\,x{\subs}gM[x{\subs}c]\,]$.
E.g., consider the
inhabitant~$M=F\lm{x^0} F\lm{y^0} pxy$
of~$p^{[0,0]},F^2$:
\[ \myvcenter{\input{figures/tree_lxy_xy.tex}} \rightsquigarrow
\myvcenter{\input{figures/tree_lxy_xy-1.tex}} \rightsquigarrow
\myvcenter{\input{figures/tree_lxy_xy-2.tex}}.
\]

We are interested in the following ``subterms'' of the image:
\[
N= \kern-1.2em\myvcenter{\input{figures/tree_lxy_xy-3.tex}} \qquad
N_1= \kern-1.2em\myvcenter{\input{figures/tree_lxy_xy-4.tex}} \qquad
N_2= \kern-0.5em\myvcenter{\input{figures/tree_lxy_xy-5.tex}}.
\]
Replacing the maximal subterms of the form~$gK$ with distinct~$z_i$
yields~$N$.  This is almost the original term: the~$f$ are to be replaced
by~$F\lambda$ and the~$z_i$ need to be appropriately bound to them.

If we repeat the process on the aforementioned maximal subterms~$gK$ 
(which we replaced with~$z_i$ to get~$N$),
but instead simply remove the maximal subterms of~$K$ 
of the form~$gK'$, 
we get the terms~$N_i$.

An~$N_i$ is almost
a subterm of~$N$: lay~$N_i$ on top of~$N$ with~$c$ and~$z_i$ aligned.
There will always be a~$f$ under the top of $N_i$.  This is the~$f$ that
has to be bound to~$z_i$.

Thus the orignal term can be read back from the image. A rigorous
proof of the correctness of this method, requires nothing but tedious
bookkeeping and is therefore omitted.\qed

\subsubsection{Ad \texorpdfstring{$\omega+3$}{w+3}}
We need to show that $A\leqs [3,0]$
for all~$A\in\HH{\omega+3}$.
To this end,
we prove that $A\leqs [3,0]$
for every small type~$A$
(since every~$A\in\HH{\omega+3}$ is small).

\begin{lem}
\label{L:wp3-1}
Let $B_1,\dotsc,B_m$ be types.
We have
\begin{equation*}
[\,\lbb B_1,\dotsc,B_m\rbb\,]\ \leqs\  [B_1,\dotsc,B_m, \,3^m].
\end{equation*}
\end{lem}
\proof 
Combine Lemma~\ref{L:sr-rc}, Lemma~\ref{L:sr-ut}
and Corollary~\ref{L:sr-glt}.\qed

\begin{lem}
\label{L:small-leq-30}
Let~$A$ be a small type. Then $A\leqs [3,0]$.
\end{lem}
\proof 
One can easily verify that every component~$C$ of~$A$
is either~$0$ or of the form $C\equiv[B]$ where~$B$ is small.
So, if we repeatedly apply Lemma~\ref{L:wp3-1}
(with the help of Lemma~\ref{L:sr-glp} and Corollary~\ref{L:sr-glt}) 
we eventually see that, 
for some $k$, $\ell$ and~$m$,
\begin{equation*}
A\ \leqs\ [3^k, 1^\ell, 0^m].
\end{equation*}
We illustrate this with an example.
\begin{alignat*}{3}
[0,\lbb0,1,2\rbb ] \ 
& \leqs\  && [0,0,1,2,3^3] \ \equiv\  [0,0,1,\lbb 0\rbb, 3^3] \\
& \leqs && [0,0,1,0,3,3^3] \ \iss \ [3^4,1,0^2].
\end{alignat*}

So it remains to be shown that $[3^k,1^\ell,0^m]\leqs [3,0]$.
Since $[3,0]$ is atomic by Lemma~\ref{L:rap-tall},
it suffices
(see Proposition~\ref{sr-summing}) 
to prove that
\begin{equation*}
[0]\leqs [3,0]; \qquad
[3]\leqs[3,0]; \qquad 
[1]\leqs[3,0].
\end{equation*}
The first two statements follow 
immediately from Lemma~\ref{L:ar-glp}\iref{L:ar-glp-i}.
Concerning the last one, we have
 $[1]\leqs [\lbb 1 \rbb]\equiv[3]\leqs[3,0]$
by Lemma~\ref{L:sr-c}.\qed

\subsubsection{Ad  \texorpdfstring{$\omega+4$}{w+4}}
We need to show that $A\leq [[0,0],0]$
for all~$A\in\HH{\omega+4}$.
We prove more.
\begin{lem}\label{top}
Let~$A$ be a type.
Then  $A \leqs [[0,0], 0]$.
\end{lem}
\proof 
Note that~$[[0,0],0]$
is atomic by Lemma~\ref{L:rap-fat},
since we have 
\begin{equation}
\label{eq:top-1}
[[0,0]]\ \leqa\ [[0,0],0].
\end{equation}
Hence to prove that
$A\leqs [[0,0],0]$,
it suffices to show that~$[C]\leqs [[0,0],0]$
for every component~$C$ of~$A$
(see Proposition~\ref{sr-summing}).

Let~$C\equiv[C_1,\dotsc,C_k]$
be a component of~$A$. 
We prove that~$[C]\leqs[[0,0],0]$.
Using induction,
we may assume that
we already have~$C_i\leqs[[0,0],0]$ for all~$i$.

Since
$[[C_1,\dotsc,C_k]]\leqs [[C_1],\dotsc,[C_k],[0^k]]$
by Lemma~\ref{L:sr-sp},
it suffices to show that
\begin{equation*}
[[C_1],\dotsc,[C_k],[0^k]] \ \leqs\ [[0,0],0].
\end{equation*}
Since~$[[0,0],0]$ is atomic,
this reduces to $[[C_i]]\leqs[[0,0],0]$
and $[[0^k]]\leqs [[0,0],0]$.
For the latter inequality, note
that the substitution from~$\smash{p^{[0^k]}}$
to~$b^{[0,0]}$ given by
\begin{equation*}
\varrho_p \,\eqdf\, 
\lm{x_1\dotsb x_n} b x_1\, bx_2\, \dotsb bx_n x_n
\end{equation*}
is an atomic reduction
and hence $[[0^k]]\leqs [[0,0]]\leqs [[0,0],0]$.
Concerning the first inequality,
write $C_i \equiv [D_1,\dotsc,D_\ell]$
and note that
we have 
\begin{alignat*}{3}
[[C_i]] &\equiv&& [\lbb D_1,\dotsc,D_\ell\rbb ] \\
\ &\leqs \ &&[\lbb D_1, \rbb, \lbb D_2 \rbb, \dotsc, \lbb D_\ell \rbb ] 
   &&\text{by Lemma~\ref{L:sr-ut}} \\
\ &\leqs\ &&[\,D_1,D_1,[0,0],\, \lbb D_2 \rbb, \dotsc,  \lbb D_\ell\rbb ]
   &&\text{by Lemma~\ref{L:sr-rc2}} \\
& && \vdots \\
\ &\leqs\ && [\,D_1,\dotsc,D_\ell,\ D_1,\dotsc,D_\ell,\ [0,0]^\ell\,]
   \qquad&&\text{by Corollary~\ref{L:sr-glt}}.
\end{alignat*}
As we have
$C_i\equiv[D_1,\dotsc,D_\ell]\leqs [[0,0],0]$
and $[[0,0]]\leqs [[0,0],0]$,
we get 
\begin{equation*}
[\,D_1,\dotsc,D_\ell,\ D_1,\dotsc,D_\ell,\ [0,0]^\ell\,]
\ \leqs\ [[0,0],0]
\end{equation*}
by Proposition~\ref{sr-summing}.
Hence $[[C_i]]\leqs [[0,0],0]$.
So we are done.\qed

%
\subsection{\texorpdfstring%
	{$\alpha\leq\beta\implies H_\alpha \leqh H_\beta$}
	{a<=b implies H(a) <=h H(b)}}
	\label{SS:HaleqHb}
We prove that $\alpha\leq\beta$ implies $H_\alpha\leqsh H_\beta$
by showing  that
\begin{equation*}
[0^k] 
\ {\underset{(i)}{\leqsh}}\ [0^{k+1}]
\ {\underset{(ii)}{\leqsh}}\ [1,0]
\ {\underset{(iii)}{\leqsh}}\ [2]
\ {\underset{(iv)}{\leqsh}}\ [1,1,0]
\ {\underset{(v)}{\leqsh}}\ [3,0]
\ {\underset{(vi)}{\leqsh}}\ [[0,0],0]\htam.
\end{equation*}
\begin{enumerate}[label=(\roman*),ref=\roman*]
\item Follows directly from Lemma~\ref{L:ar-glp}\iref{L:ar-glp-i}.
\item Similar to Lemma~\ref{L:part3-bitw}, but easier.
\item
On the one hand, $[1,0]\leqsh[2,0]$ 
as $f^1\leqs F^2$ via 
$\varrho_f \eqdf \lm{x^0} F\lm{y^0} x$.
On the other hand,  $[2,0]\leqsh [2]$ by Lemma~\ref{L:sametrick}.
\item
This is Proposition~\ref{sr-2-to-1-1-0}.
\item
Follows from Lemma~\ref{L:small-leq-30} since $[1,1,0]$ is small.
\item
A consequence of Lemma~\ref{top}.
\end{enumerate}

%% file: figures/tree_lxy_xy.tex
\begin{tikzpicture}[level distance=0.7cm, sibling distance=1.0cm]
\node{$F\lm{x^0}$}
 child { node{$F\lm{y^0}$}
 child { node{$p$}
   child { node{$x$} }
   child { node{$y$} } } };
\end{tikzpicture}

%% file: figures/tree_lxy_xy-1.tex
\begin{tikzpicture}[level distance=0.7cm, sibling distance=1.0cm]
\node{$F\lm{x^0}$}
 child { node{$f$}
 child { node{$p$}
   child { node{$x$} }
   child { node{$g$}
     child { node{$p$}
       child { node{$x$} }
       child { node{$c$} } } }
   } };
\end{tikzpicture}

%% file: figures/tree_lxy_xy-2.tex
\begin{tikzpicture}[level distance=0.7cm, sibling distance=1.0cm]
\node{$f$}
 child { node{$f$}
 child[sibling distance=2.5cm] { node{$p$}
   child { node{$g$}
     child[sibling distance=1.0cm] { node{$f$}
       child { node{$p$}
         child { node{$c$} } 
	 child { node{$g$}
	   child { node[color=gray]{$p$}
	     child[color=gray] { node{$c$} }
	     child[color=gray] { node{$c$} } } }
   } } } 
   child { node{$g$}
     child[sibling distance=1.0cm] { node{$p$}
       child { node{$g$}
         child { node[color=gray]{$f$}
           child[color=gray] { node{$p$}
             child { node{$c$} } 
  	     child { node{$g$}
	       child { node{$p$}
	         child { node{$c$} }
	         child { node{$c$} } } }
	     } } }
         child { node{$c$} }
   } }
 } };
\end{tikzpicture}

%% file: figures/tree_lxy_xy-3.tex
\begin{tikzpicture}[level distance=0.7cm, sibling distance=1.0cm]
\node{$f$}
 child { node{$f$}
 child { node{$p$}
   child { node{$z_1$} }
   child { node{$z_2$} }
 } };
\end{tikzpicture}

%% file: figures/tree_lxy_xy-4.tex
\begin{tikzpicture}[level distance=0.7cm, sibling distance=1.0cm]
\coordinate child { node{$f$}
       child { node{$p$}
         child { node{$c$} } 
	 child { }
   } } ;
\end{tikzpicture}

%% file: figures/tree_lxy_xy-5.tex
\begin{tikzpicture}[level distance=0.7cm, sibling distance=1.0cm]
\coordinate
     child { node{$p$}
       child 
       child { node{$c$} }
    } ;
\end{tikzpicture}

%% file: part4.tex
\section{\texorpdfstring%
	{Order type of $\leqbe$ and $\leqhp$}
	{Order type of <=be and <=h+}}
	\label{S:part4}
We are in the home stretch now.
We have proven that the order type
of the reduction relation~$\leqh$ 
is as depicted in the diagram on page~\pageref{T:Hierarchy}. 
The structure of this proof was given in 
Subsection~\ref{SS-sketch-of-the-proof},
and we have spent the previous sections filling in all the difficult details.
In this section
we provide the final and easy bits of the proof 
that the order types of
the reduction relations~$\leqbe$ and~$\leqhp$ 
(see Definition~\ref{defn:reductions})
are depicted correctly as well.

\subsection{\texorpdfstring%
	{$[2]\leqhp[1,0]$ and $[0^{k+1}] \leqhp [0^k]$}
	{[2] headplusreduces to [0k] and [0\{k+1\}] headplusreduces to [0k]}}
\label{SS:part4-hp}
Let us begin with $[0^{k+1}]\leqhp [0^k]$ for given~$k\geq 2$.
It suffices to prove that
\begin{equation*}
[0^{k+1}]\ \leqhp\ [0^2],
\end{equation*}
because one can easily verify that $[0^2]\leqhp[0^k]$.
\begin{lem}
For every $k\geq 2$ we have $[0^{k}]\leqhp[0^2]$.
\end{lem}
\proof 
We need to find a finite family of B\"ohm terms
from $[0^k]$ to~$[0^2]$
which is jointly injective
(see Definition~\ref{defn:reductions}).
Recall that
(see \ref{SSS:inh-0k})
\begin{equation*}
\ts{[0^k]}\,=\,\{U^k_1,\dotsc,U^k_k\}.
\end{equation*}
Given $i,j\in\{1,\dotsc,k\}$ with $i\neq j$,
there is a B\"ohm term $M_{ij} \in \ts{[0^k]\ra [0]}$
which separates the terms $U^k_i$ and $U^k_j$ in the sense that
\begin{equation*}
M_{ij} U^{k}_i \,\eqb\, U^2_1
\qquad\text{and}\qquad
M_{ij} U^{k}_j \,\eqb\, U^2_2.
\end{equation*}
Indeed,
the B\"ohm term
 $M_{ij}\eqdf \lm{\smash{m^{[0^k]}}x_1 x_2} m\, P_1 \dotsb P_k$
does the job,
where
\begin{equation*}
P_\ell \,\eqdf\,
\begin{cases}
\ x_1 & \text{ if }\ell =i \\
\ x_2 & \text{ otherwise.}
\end{cases}
\end{equation*}
Hence the family of terms $\{ M_{ij}\colon i\neq j\}$ 
is jointly injective.\qed

\begin{prop}
\label{P:2leqhp10}
$[2] \leqhp [1,0]$. 
\end{prop}
\proof 
We have shown there is no injective
transformation from the type~$[2]$ to~$[1,0]$ (see Subsection~\ref{SS:wp1nrw}).
However, we will prove that
the following  B\"ohm transformations $\varrho$ and~$\sigma$
from~$[2]$ to type~$[1,0]$ are
jointly injective (and thus $[2]\leqhp [1,0]$).
\begin{equation*}
\varrho_F \eqdf \lm{h^1} fhc
\qquad\qquad \sigma_F \eqdf \lm{h^1} fhfhc 
\end{equation*}
It suffices to show that~$i$ and~$j$ can be recovered
from~$\hat\sigma(\,\left<i,j\right>\,)$ 
and $\hat\varrho(\,\left<i,j\right>\,)$.
This is indeed the case
as one 
we have the following equalities
for $\left<i,j\right>\in\ts{[2]}$.
\begin{equation*}
\hat\varrho (\,\left< i, j\right>\,) = c_i 
\qquad\qquad
   \hat\sigma(\,\left< i, j \right>\,) = c_{2i - j + 1}.
\end{equation*}
We verify the latter equality and leave the other to the reader.
\begin{alignat*}{4}
\hat\sigma(\,\left<i,j\right>\,) 
& \eqbe\  && \sigma_F \lm{x_1^0} \cdots \sigma_F  \lm{x_i^0} x_j  \\
& \eqb\  && \sigma_F\lm{x_1^0} \cdots \sigma_F  \lm{x_j^0} f^{(i-j)}
	x_j \\
& \eqb && \sigma_F\lm{x_1^0} \cdots \sigma_F  \lm{x_{j-1}^0}
	f f^{(i-j)} f  f^{(i-j)} c  \\
& = && \sigma_F\lm{x_1^0} \cdots \sigma_F  \lm{x_{j-1}^0}
	f^{(2+2(i-j))}   c \\
& \eqb && f^{(j-1)} f^{(2+2(i-j))}c = c_{2i - j + 1}\htam{.}
\end{alignat*}
We have proven that
$[2] \leqhp [1,0]$. \qed

\subsection{\texorpdfstring%
	{$[2] \leqbe [1,0]$}
	{[2] be-reduces to [1,0]}}
For the proof of $[2]\leqbe[1,0]$
we need some preparations.
\label{SS:part4-be}
\begin{lem}
\label{L:addmulrepr}
Addition and multiplication on the Church numerals
is definable
in the following sense.
There are closed terms~$M_+,M_\times : [1,0] \ra [1,0] \ra [1,0]$ with
\begin{equation*}
M_+c_mc_n \,\eqbe\, c_{m+n}\quad\text{and}\quad
M_\times c_m c_n \,\eqbe\, c_{m\cdot n}
\qquad(n,m\in\mathbb{N})\htam.
\end{equation*}
\end{lem}
\proof 
It is not hard to see that the terms
\begin{align*}
	M_+ & = \lm{a^{[0,1]} b^{[0,1]} f^1 c^0} af (bfc) \\
        M_\times 
   & = \lm{a^{[0,1]} b^{[0,1]} f^1 c^0} a(bf)c\htam{.}
\end{align*}
do the job.\qed

\begin{cor}
\label{C:10-cantor-pairing}
The Church numerals contain a pairing in the following sense.
There is a term~$M_p:[1,0] \ra [1,0] \ra [1,0]$ such that
\begin{equation*}
M_pc_nc_m \eqbe M_pc_{n'}c_{m'} \quad \implies\quad 
n = n'\ \htam{and}\ m =m'\htam.
\end{equation*}
\end{cor}
\proof 
The map $P\colon (n,m) \mapsto \frac{1}{2}(n+m)(n+m+1)+m$, 
is a well known bijection
between~$\mathbb{N}^2$ and~$\mathbb{N}$, called the Cantor pairing.
Using Lemma~\ref{L:addmulrepr},
we obtain a term~$M_p$ such that~$M_p(c_n,c_m)\eqbe c_{P(n,m)}$
for all~$n,m$.\qed

\begin{prop}
$[2] \leqbe [1,0]$.
\end{prop}
\proof 
We need to find an $R\in\ts{[2]\ra[1,0]}$
such that
\begin{equation*}
RM \,\eqbe\, RN
\quad\implies\quad
M=N\qquad(M,N\in\ts{[2]}.
\end{equation*}
Let $\varrho$ and $\sigma$
be the substitutions from Proposition~\ref{P:2leqhp10}
which form a multi-head reduction from~$[2]$ to~$[1,0]$
and define $R\eqdf\lm{m^{[2]}} M_p\, (m\varrho_F)(m\sigma_F)$.

Let $M,N\in\ts{[2]}$ with $RM\eqbe RN$
be given, to prove $M=N$.
Then
\begin{equation*}
M_p\,(M\varrho_F)(M\sigma_F)\,\eqbe\,M_p\,(N\varrho_F)(N\sigma_F).
\end{equation*}
So $M\varrho_F \eqbe N\varrho_F$
\& $M\sigma_F \eqbe N\sigma_F$
by Corollary~\ref{C:10-cantor-pairing}.
But then~$M=N$
as $\hat\varrho$ and $\hat\sigma$ are jointly injective
(see Proposition~\ref{P:2leqhp10}).\qed

%% file: conclusions.tex
\section{Conclusion}
	\label{S:conclusions}

We have proven Statman's Hierarchy Theorem (see page~\pageref{T:Hierarchy}).
With it we can mechanically determine
for all types~$A$ and~$B$
in $\mathbb{T}^0$
whether $A\leqbe B$, whether $A\leqh B$,
and whether $A\leqhp B$
only by inspecting the syntactic form of~$A$ and~$B$.
Let us make some final remarks.

\subsection{Contributions}
The \emph{calculus of reductions} 
(see Section~\ref{S:intermezzo})
used to prove the existence of reductions 
is new (including the notions of strong reduction and atomic type).
The method to prove
the absense of reductions
from Subsection~\ref{SS:indiscernibility}
is a generalisation of the work of Dekkers in~\cite{Dekkers-1988}.

The Hierarchy Theorem as presented here is slightly stronger than
the one proven by Statman
in that the original version
only completely determined the relations $\leqbe$ and~$\leqhp$,
but not~$\leqh$
(see Theorem~3.4.18 and Corollary~3.4.27 of~\cite{BDS}),
while our version determines the relation~$\leqh$ as well.
For this we had to add one canonical type, namely $[2]$,
and prove (among other things), that~$[2]\nleqh [1,0]$
(see Subsection~\ref{SS:wp1nrw}).

\subsection{Outlook}
The notions and notation
introduced in this paper
are easily adapted
to a setting with multiple base types $\alpha_1,\alpha_2,\dotsc$.
However,
if one tries to determine the  equivalence classes of $\leq_h$
in this setting
one realises
much more work has to be done.
(Indeed, try, for instance, to prove a variant of 
Theorem~\ref{T:inhabitation} for
multiple base types.) 
Perhaps the development of a software tool based 
on the calculation rules for strong reductions
will be of use in such a project.

\subsection{Acknowledgements}
We are grateful that a reviewer spotted an error in
Subsection~\ref{SS:w3nrw2} of a previous version
of this manuscript.